\documentclass[amsmath,superscriptaddress,citeautoscript,twocolumn,
showpacs,floatfix,prb]{revtex4}
\usepackage{amssymb,amsmath}
\usepackage{graphics}
\usepackage{epstopdf}
\usepackage{epsfig}
\usepackage{color}
\usepackage{amsfonts}
\usepackage{bm}
\usepackage{amscd}
\begin{document}

\title{Transient probing of the symmetry and the asymmetry of electron interference}

\author{Matisse Wei-Yuan Tu}
\affiliation{Department of Physics and Center of Theoretical and Computational Physics, University of Hong Kong, Hong Kong}

\author{Amnon Aharony}
\email{aaharonyaa@gmail.com}
\affiliation{Physics Department, Ben Gurion University, Beer Sheva
84105, Israel} \affiliation{Raymond and Beverly Sackler School of Physics and Astronomy, Tel Aviv University,
Tel Aviv 69978, Israel}

\author{Ora Entin-Wohlman}
\affiliation{Physics Department, Ben Gurion University,  Beer Sheva
84105, Israel} \affiliation{Raymond and Beverly Sackler School of Physics and Astronomy, Tel Aviv University,
Tel Aviv 69978, Israel}

\author{Avraham Schiller \footnote{Deceased, June 22nd, 2013}}
\affiliation{Racah Institute of Physics, The Hebrew University, Jerusalem 91904, Israel}

\author{Wei-Min Zhang}
\email{wzhang@mail.ncku.edu.tw}
\affiliation{Department of Physics, National Cheng Kung University, Tainan 70101, Taiwan}

\begin{abstract}
The transient processes of electron transport in nano-scale devices exhibit special phenomena that exist only in the transient regime. Besides how fast the steady states are approached, one interesting aspect of transient transport arises from its strong dependence on the initial state of the system.

Here we address the issue of how the symmetries embedded in the initial state interplay with those of the system structure in the course of transient transports.
We explicitly explore the transient currents arising from various initial occupations in a double-quantum-dot Aharonov-Bohm interferometer. We find symmetry relations between the transient in-tunneling and out-tunneling dynamics for initially empty or full quantum dots when the energy levels in the electrodes are symmetrically distributed with respect to the energy levels in the QDs. This is true for whatever applied fluxes. We also find the flux-even components of the currents and the flux-odd components of the currents exhibit distinct cross-lead symmetric relations.

\end{abstract}



 \maketitle

\section{Introduction}
Coherence of electron propagations in mesoscopic systems is the foundation of developing quantum electronic devices. The study of interference between different electron propagation paths is an important approach to investigate electron coherence. The manifestation of the underlying coherence in physical observables is closely related to the symmetry of the interfering states. By tuning the symmetry of the system, the interference results can be utilized to modulate electron transport properties, essential for device functionalities. In addition to the manifestation of electron coherence via steady-state current-voltage characteristics, time-dependent coherent transport has also attracted much attention. In particular, the availability of time-resolved measurement techniques in nanoelectronics makes it feasible to reveal the special roles played by the transient effects.\cite{Fujisawa01081304,Lu03422,Bylander05361,Fujisawa06759,Feve071169} In this article, we explore the special transport properties, uniquely exhibited in the transient regime, that reveal the underlying symmetries induced by the device geometries, via the use of an Aharonov-Bohm (AB)\cite{Aharonov59485} interferometer with two quantum dots (QDs).

On one hand, frequency-resolved responses to time-dependent periodic driving have been intensively studied. \cite{Tien63647,Buettiker934114,Platero041,Moskalets11book} On the other hand, the transient transport currents in response to pulsed voltages provide useful information for the switching behaviors of electronic devices. \cite{Wingreen938487,Jauho945528,Maciejko06085324} The transient currents induced by optical excitations on molecular transport devices are interesting for their relevance to optical switches.\cite{Listorti113381,Petrov085651,Volkovich11033403,Petrov1253,Petrov13184709} Real-time migrations of electrons between molecules and conduction channels are on the focus in some energy\cite{Hagfeldt00269} and material applications\cite{Wang13205126}.




Particularly for nanoelectronic systems, steering the transient currents can potentially offer versatile resources for timely operating quantum electronic circuits at nanoscale. In general, the transient dynamics is sensitive to the initial state of the system. In some circumstances, the starting point of the transient dynamics is appropriately given by the steady state of the whole transport setup before the turning on of the pulses\cite{Wingreen938487,Jauho945528,Maciejko06085324}. In a broader context of quantum technology, besides how fast one can switch on and off a current, utilization of the whole trajectory of a system from a prepared initial state to a desired later state is also highly relevant. Experimentally, preparation of desired charge states in QDs have been realized.\cite{Hayashi03226804,Fujisawa06759,Hanson071217,Kim1470} Therefore one is naturally motivated to investigate how the transient transport dynamics depends on the initial preparations.\cite{Yang15165403} In addition to this, tuning geometric symmetry has been found useful in modulating the stationary transport properties. For the purpose of exploiting the transient transport properties as potential resources, it is thus important to understand how the symmetry embedded in the initial state interplays with the geometric symmetry of the interference device.



Geometric symmetry of nanostructures can affect interference effects to change the stationary transport currents. For example, working principles for molecular transistors have been proposed based on the interference effects governed by molecular geometries.
\cite{Cardamone062422,Ke083257,Qian08113301} Experimental observations of such effects rely on properly arranging molecule-lead coupling configurations on the molecular scale.\cite{Guedon12305,Vazquez12663,Ballmann12056801} Another widely applied approach to manipulate stationary transport by tuning interference is to thread a magnetic flux through a ring-shaped nanostructure, rendering the AB effect. This was realized earlier with metallic ring interferometry.\cite{Webb852696} With a QD sitting on one arm, making the other arm of the ring as a reference, studies on the resonant tunneling through the QD have been experimentally carried out.\cite{Yacoby954047,Schuster97417,Ji00779,Ji02076601,Avinun-Kalish05529,Kobayashi02256806,Kobayashi03235304,Aikawa04176802} Ring-shaped natural molecules and artificial molecules are useful nanostructures that can host the AB effect and at the same time provide manipulable geometric symmetry. For ring-shaped natural molecules, effective use of magnetic fluxes either requires unrealistically high magnetic fields\cite{Walczak04524} or other special conditions.\cite{Rai112118} Evoking the AB effect with artificial molecules has been realised by putting two QDs on the two arms of a mesoscopic interferometer.\cite{Holleitner01256802,Sigrist06036804,Hatano11076801} This is known as the double-quantum-dot (DQD) AB interferometer. Such setups offer the advantages of incorporating molecular symmetry into the flux-modulated interference effects. How to use the AB oscillations to discern molecular parities has been theoretically\cite{Kang04117} and experimentally studied.\cite{Hatano11076801} By properly combining it with the spin-orbit interaction, mixed actions of charge and spin interference have been shown to give rise to full spin-polarized transport.\cite{Aharony11035323,Tu14165422} The AB interference is also useful to affect the occupation difference between degenerate QDs\cite{Bedkihal13045418} and modulate the current noises.\cite{Urban09165319,Zivkovic01115304}

Various time-dependent aspects of DQD AB interferometers have been tackled. Real-time detection of interference pattern formation has been experimentally attempted.\cite{Gustavsson082547} The flux-dependent decoherent and coherent dynamics of the uncoupled DQD has been tackled by the exact fermion master equation for noninteracting electrons.\cite{Tu11115318,Tu12195403} With inter-dot Coulomb repulsion considered, photoresponse of the coupled DQD in the AB interferometer has been investigated in the weak-tunneling limit.\cite{Dong044303} The reduced density matrix dynamics under the influence of the flux, including also inter-dot Coulomb interaction, has been studied by the use of the exact-numerical-path-integral.\cite{Bedkihal12155324} The latter approach has also been applied to examine the transient magnetotransport in similar systems.\cite{Bedkihal14235411} The effects of Fano resonances on the dynamics of wave scatterings with a coupled DQD molecule in the AB interferometer has been inspected in the time domain.\cite{Romo12085447}

The choice of a simple uncoupled DQD placed in an AB interferometer provides a suitable platform to clearly define the geometric symmetry of the DQD molecule and the symmetry associated with initial occupations in the QDs. In a previous paper, we have comprehended the flux responses of the QD occupations, the transport currents as well as the circulating currents of the DQD AB interferometers in the transient as well as in the steady-state regimes.\cite{Tu12115453} However, we have only worked under the conditions that tunnelings through the two QDs are symmetric and initially the two QDs are not occupied. Here we remove all these restrictions and we show that DQD AB interferometers indeed provide several benefits that diversify the uses of the transient currents. Our main findings are summarized below: (i) Unlike the steady-state currents, which contain only components that are even in the flux (insensitive to the flux direction), the transient currents can exhibit components that are odd in the flux. They behave distinctly from their even counterparts in terms of the manifestation of the geometric symmetry of the device. (ii) Initially empty QDs and initially occupied QDs can induce completely different ways of electron motions, giving rise to distinguishable in-tunneling and out-tunneling transient currents respectively. Their relationship is associated with the symmetry of the energy level distribution. (iii) Asymmetry among different components of the transient currents can be either enhanced or suppressed by properly tuning the device geometry.

The paper is organized as follows: In Sec. \ref{sec_method}, we briefly review the general model for mesoscopic transport systems and the nonequilibrium Green function formalism (NEGF) for calculating the currents. Methodologically, there exist many sophisticated theoretical approaches to study time-dependent quantum transport .\cite{Cini805887,Jauho945528,Stefanucci04195318,Anders05196801,Kurth05035308,Maciejko06085324,Zheng07195127,Schmidt08235110,Muehlbacher08176403,Jin08234703,Tu08235311,Jin10083013,Segal10205323} Most of them are based on NEGF. Since many papers on quantum transport use the scattering formalism,\cite{Datta95book,Imry97book,Nazarov09book} for pedagogical purposes, we also present a connection to the scattering-state approach for deriving the transient time-dependent transport properties. We pay special attention to how the currents depend on the initial occupations in the central nanostructures. In Sec.~\ref{sec_model_transit}, we introduce our target system, an AB interferometer with two QDs. We identify the purely transient components of the tunneling currents and discuss generally their physical meanings. In Sec. \ref{sec_model_transit_fluxindep}, we show how the symmetry and asymmetry of the device parameters can give rise to distinguishable transient dynamics. Section \ref{sec_conclu} contains our conclusions.




\section{Formalism for transient quantum transport}
\label{sec_method}

A class of quantum transport systems is described by the following general Hamiltonian,
\begin{subequations}
\label{gel_H}
\begin{equation}
\mathcal{H}=\mathcal{H}_{\text{S}}+\mathcal{H}_{\text{E}}+\mathcal{H}_{\text{T}},
\end{equation}
where
\begin{align}
\label{gel_Hs}
\mathcal{H}_{\text{S}}=\sum_{i,j\in\text{S}}\varepsilon_{ij}d_{i}^{\dagger}d^{}_{j},
\end{align} is the Hamiltonian for the central scattering region, and
\begin{align}
\label{gel_HE-sum}
\mathcal{H}_{\text{E}}=\sum_{\alpha}\mathcal{H}_{\alpha}=\sum_{\alpha}\sum_{k\in\alpha}\varepsilon_{\alpha k}c_{\alpha k}^{\dagger}c^{}_{\alpha k},
\end{align} is the sum of Hamiltonians of the electrodes, each labeled by $\alpha$.
The tunneling of electrons between the central region and the electrodes is described by
\begin{align}
\label{gel_HT}
\mathcal{H}_{\text{T}}=\sum_{i,\alpha k}\left\{ t^{}_{\alpha ki}c_{\alpha k}^{\dagger}d^{}_{i}+t^{}_{i\alpha k}d^{\dagger}_{i}c_{\alpha k}^{}\right\}.
\end{align}
\end{subequations} Here the subscripts $i,j\in\text{S}=\{1,2,\cdots,D\}$ enumerates the $D$ single-particle levels within the central region and $k\in\alpha$ stands for the continuum levels $k$ within the electrode $\alpha$. The field operator $d_{j}(d^{\dagger}_{j})$ or $c^{}_{\alpha k}(c^{\dagger}_{\alpha k})$ annihilates (creates) an electron on level $j\in\text{S}$  or level $k\in\alpha$. The hopping amplitude between a level $i$ in the central part and a level $k$ in lead $\alpha$ is given by $t^{}_{\alpha ki}=t^{*}_{i\alpha k}$. Since we are targeting at the single-particle interference effect, interactions leading to dephasing of transport are ignored.\cite{Hackenbroich96110,Bruder96114,Hackenbroich01463}





The essential quantity probed by transport measurements is the tunneling current. The current tunneling out of lead $\alpha$ at time $t$ is defined by
\begin{align}
\label{crnt-gnf-def}
&I_{\alpha }\left( t\right)  =-\frac{d}{dt}\left\langle \mathcal{N}^{}_{\alpha}\left(
t\right) \right\rangle =-i\left\langle \left[ \mathcal{H}
 ,\mathcal{N}^{}_{\alpha}\left( t\right) \right] \right\rangle\nonumber\\
 &=-i\sum_{i}\sum_{k\in\alpha}\left(t_{i\alpha k}\langle d^{\dagger}_{i}(t)c^{}_{\alpha k}(t)\rangle-t_{\alpha ki}\langle c^{\dagger}_{\alpha k}(t)d^{}_{i}(t)\rangle\right),
\end{align} where $\mathcal{N}^{}_{\alpha}(t)=\sum_{k\in\alpha}c_{\alpha k}^{\dagger}c^{}_{\alpha k}$ is the total particle number operator in lead $\alpha$ in the Heisenberg representation. We have also set charge unit $e=1$ and $\hbar=1$. The bracket $\langle\cdot\rangle={\rm tr}(\cdot\hat{\rho}(t_0))$ denotes the average over the initial state, $\hat{\rho}(t_0)$, at time $t=t_{0}$, of the total system including the central scattering region plus the electrodes.

The initial state of the total system $\hat{\rho}_{}(t_0)$ is assumed\cite{Leg871,Jauho945528}
to be a product state of the central area and the electrodes, namely,
\begin{subequations}
\label{init_eqthm}
\begin{align}
\hat{\rho}_{}^{}\left(
t_0\right)=\hat{\rho}^{}_{\text{S}}(t_0)\prod_{\alpha}\hat{\rho}^{}_{\alpha}(t_0),
\end{align} where
\begin{align}
\hat{\rho}^{}_{\alpha}(t_0)=\frac{\exp\left[-\left(\mathcal{H}^{}_{\alpha}-\mu_{\alpha}\mathcal{N}^{}_{\alpha}\right)/k_{B}T_{\alpha}\right]}
{\text{tr}_{\text{}}\exp\left[-\left(\mathcal{H}^{}_{\alpha}-\mu_{\alpha}\mathcal{N}^{}_{\alpha}\right)/k_{B}T_{\alpha}\right]},
\end{align}
\end{subequations} represents the thermal equilibrium of the
electrode $\alpha$, each with the chemical potential $\mu_{\alpha}$ and the temperature $T_{\alpha}$. Here $k_{B}$ is the Boltzmann constant. The initial state of the central scattering area is denoted by $\hat{\rho}^{}_{\text{S}}(t_0)$ and is not restricted. We will discuss how the transient currents depend on $\hat{\rho}^{}_{\text{S}}(t_0)$.

\subsection{Scattering-state method}
\label{sec_method-sctstat}

The time developments of the tunneling currents $I_{\alpha }\left( t\right)$ can be calculated via different approaches.\cite{Jauho945528,Jin10083013} Here we introduce the scattering-state method, which tackles the time developments of the Heisenberg field operators $d_{i}^{}(t)$'s and $c^{}_{\alpha k}(t)$'s through the scattering states. It starts with solving the Lippman-Schwinger equation\cite{Lippmann50469} for the scattering states, as the eigenstate of the total Hamiltonian,
\begin{equation}
\left[\psi_{\alpha k}^{\dagger},\mathcal{H}\right]=-\varepsilon_{\alpha k}\psi_{\alpha k}^{\dagger}+i\eta\left(c_{\alpha k}^{\dagger}-\psi_{\alpha k}^{\dagger}\right),\label{Lippman-Schwinger-eq}
\end{equation}
where $\eta\rightarrow0^{+}$. The solution is found to be\cite{Schiller9814978}
\begin{align}
&\psi_{\alpha k}^{\dagger}=\nonumber\\
&c_{\alpha k}^{\dagger}
\!\!+\!\!\sum_{i,j}t_{i\alpha k}\!\!\left(\!\!d_{j}^{\dagger}+\sum_{\beta q}\frac{t_{\beta qj}c^{\dagger}_{\beta q}}{\varepsilon_{\alpha k}-\varepsilon_{\beta q}+i\eta}\right)\!\!{G}_{ji}\left(\varepsilon_{\alpha k}+i\eta\right),\label{scat-sta-sol}
\end{align}
where the Green function in the energy domain, $G_{ij}(z)=[\boldsymbol{G}(z)]_{ij}$,
\begin{equation}
\boldsymbol{G}\left(z\right)\!\!=\!\!\left[z\mathbf{1}_{D}-\boldsymbol{\varepsilon}-\widetilde{\boldsymbol{\Sigma}}\left(z\right)\right]^{-1},\label{Gr_scat-sta}
\end{equation}
has been introduced with the self-energy,
\begin{equation}
\widetilde{\boldsymbol{\Sigma}}\left(z\right)=\int\frac{d\omega}{2\pi}\frac{\boldsymbol{\Gamma}^{}(\omega)}{z-\omega},\label{self-eng_scat-sta}
\end{equation} and the total level-broadening function is,
\begin{align}
&\boldsymbol{\Gamma}^{}(\omega)=\sum_{\alpha}\boldsymbol{\Gamma}^{\alpha}(\omega),~\text{with}
\nonumber\\
&{\Gamma}^{\alpha}_{ij}(\omega)=2\pi\sum_{k\in\alpha}\delta(\omega-\varepsilon^{}_{\alpha k})t_{i\alpha k}t_{\alpha kj}
\label{lvl-brd-ld}
\end{align} where $\left[\boldsymbol{\Gamma}^{\alpha}(\omega)\right]_{ij}=\Gamma^{\alpha}_{ij}(\omega)$.

The states formed from $\psi_{\alpha k}^{\dagger}\vert0\rangle$, where $\vert0\rangle$ denotes the vacuum of the total system, constitute
a complete set in the single-particle space and the relation $\left\{ \psi_{\beta q}^{},\psi_{\alpha k}^{\dagger}\right\} =\delta_{\alpha k,\beta q}$
is satisfied. One can therefore express $d^{}_{j}$
for all $j$ and $c^{}_{\alpha k}$ for all $\alpha k$ in terms of linear combinations of the $\psi_{\alpha k}$'s. The time-dependencies of $d^{}_{j}\left(t\right)$ and $c^{}_{\alpha k}(t)$ are then found through $\psi_{\alpha k}^{}\left(t\right)$ by $\psi_{\alpha k}^{}\left(t\right)=e^{-i\varepsilon_{\alpha k}(t-t_0)}\psi_{\alpha k}^{}$ as $\psi_{\alpha k}^{}$ satisfies the Lippman-Schwinger equation, Eq. (\ref{Lippman-Schwinger-eq}), with the eigenenergy $\varepsilon_{\alpha k}$. Specifically,
\begin{align}
&d^{}_{j}(t)=\sum_{\alpha k}\sum_{i}{G}_{ji}\left(\varepsilon_{\alpha k}+i\eta\right)t_{i\alpha k}e^{-i\varepsilon_{\alpha k}(t-t_0)}\psi_{\alpha k},
\nonumber\\
&c^{}_{\alpha k}(t)\!=\!\!\sum_{\beta q}\!\!\Bigg\{\left(\!\!\delta_{\alpha k,\beta q}\!+\!\!\sum_{i,j}\frac{t_{\alpha kj}{G}_{ji}\left(\varepsilon_{\beta q}+i\eta\right)t_{i\beta q}}{\varepsilon_{\beta q}-\varepsilon_{\alpha k}+i\eta}\!\!\right)\!\!
\nonumber\\&~~~~~~~~~~~~~\times
e^{-i\varepsilon_{\beta q}(t-t_0)}\psi_{\beta q}\Bigg\}.
\label{flopt-expan_byScat-sta}
\end{align} Replacing $\psi_{\alpha k}$ on the right-hand side of Eq.~(\ref{flopt-expan_byScat-sta}) by Eq.~(\ref{scat-sta-sol}), we obtain
\begin{subequations}
\label{dit_hsg}
\begin{align}
&d^{}_{i}\left(t\right)=i\left\{ \sum_{j}G_{ij}^{r}\left(t,t_0\right)d^{}_{j}+\sum_{\alpha k}G_{i,\alpha k}^{r}\left(t,t_0\right)c^{}_{\alpha k}\right\} ,
\nonumber\\
&c^{}_{\alpha k}(t)=i\left\{ \sum_{j}G_{\alpha k,j}^{r}\left(t,t_0\right)d^{}_{j}+\sum_{\beta q}G_{\alpha k,\beta q}^{r}\left(t,t_0\right)c^{}_{\beta q}\right\},
\end{align}
where
\begin{align}
\label{Gr_t_t0}
&G_{ij}^{r}\left(t,t_0\right)=\int_{-\infty}^{\infty}\frac{d\omega}{2\pi}e^{-i\omega (t-t_0)}{G}_{ij}\left(\omega+i\eta\right),
\nonumber\\
&G_{i,\alpha k}^{r}\left(t,t_0\right)=
\int_{-\infty}^{\infty}\frac{d\omega}{2\pi}e^{-i\omega(t-t_0)}\sum_{j}\frac{G_{ij}\left(\omega+i\eta\right)t_{j\alpha k}}{\omega-\varepsilon_{\alpha k}+i\eta},
\nonumber\\
&G_{\alpha k,j}^{r}\left(t,t_0\right)=
\int_{-\infty}^{\infty}\frac{d\omega}{2\pi}e^{-i\omega(t-t_0)}\sum_{l}\frac{t_{\alpha kl}G_{lj}\left(\omega+i\eta\right)}{\omega-\varepsilon_{\alpha k}+i\eta},
\nonumber\\
&G_{\alpha k,\beta q}^{r}\left(t,t_0\right)=\Big\{-i\delta_{\beta q,\alpha k}e^{-i\varepsilon_{\alpha k}(t-t_0)}+
\nonumber\\
&\int_{-\infty}^{\infty}\!\!\frac{d\omega}{2\pi}e^{-i\omega (t-t_0)}\!\!\sum_{jl}\left[\frac{t_{\alpha kj}G_{jl}\left(\omega+i\eta\right)t_{l\beta q}}{\left(\omega-\varepsilon_{\alpha k}+i\eta\right)\left(\omega-\varepsilon_{\beta q}+i\eta\right)}\right]\!\!\Big\}.
\end{align}
\end{subequations}
The result of Eq.~(\ref{dit_hsg}) is identical to that obtained via directly solving the Heisenberg equations.

Using Eq.~(\ref{dit_hsg}) in Eq.~(\ref{crnt-gnf-def}), with the initial state given by Eq.~(\ref{init_eqthm}), we obtain an expression for the time-dependent tunneling current,
\begin{subequations}
\label{dynamic-crnt}
\begin{align}
\label{dynamic-crnt-totformula}
&I^{}_{\alpha }\left( t\right)=
\nonumber\\&
-2{\rm ReTr}\!\!\int_{t_0}^{t}\!\!d\tau\left(\boldsymbol{\Sigma}^{<}_{\alpha}(t,\tau)\boldsymbol{G}^{a}(\tau,t)
+\boldsymbol{\Sigma}^{r}_{\alpha}(t,\tau)\boldsymbol{G}^{<}(\tau,t)\right),
\end{align} where the retarded component of the self-energy is given by,
\begin{align}
\label{retarded-selfeng}
[\boldsymbol{\Sigma}_{\alpha}^{r}\left(t_1,t_2\right)]_{ij}=-i\theta(t_1-t_2)\int\frac{d\omega}{2\pi}\Gamma^{\alpha}_{ij}(\omega)e^{-i\omega(t_1-t_2)},
\end{align} and the lesser component of that self-energy is
\begin{align}
\label{less-selfeng}
[\boldsymbol{\Sigma}^{<}_{\alpha}(t_1,t_2)]_{ij}=i\int\frac{d\omega}{2\pi}f_{\alpha}(\omega)\Gamma^{\alpha}_{ij}(\omega)e^{-i\omega(t_1-t_2)},
\end{align} where $f_{\alpha}(\omega)=1/[e^{(\omega-\mu_{\alpha})/k_{B}T_{\alpha}}+1]$ is the fermi function for electrode $\alpha$.
\end{subequations} In Eq.~(\ref{dynamic-crnt}), the notation $\text{Tr}$ means trace over the indices of the central scattering area.
The advanced Green function is related to the retarded Green function by,
\begin{subequations}
\label{Gr-explicit-Gz}
\begin{align}
\label{Gr-Ga-connect}
{G}^{a}_{ij}(t_1,t_2)=[{G}^{r}_{ji}(t_2,t_1)]^{*},
\end{align} where ${G}^{a}_{ij}(t_1,t_2)=[\boldsymbol{G}^{a}(t_1,t_2)]_{ij}$ and the retarded Green function, defined for arbitrary two times $t_1\ge t_0$ and $t_2\ge t_0$,
\begin{align}
\label{def-Grs}
{G}^{r}_{ij}(t_1,t_2)& =-i\theta\left(t_1-t_2\right)\left.\langle\left\{ d^{}_{i}\left(t_1\right),d_{j}^{\dagger}(t_2)\right\} \rangle\right.,
\end{align} can be computed from Eq.~(\ref{dit_hsg}), resulting in
\begin{align}
{G}^{r}_{ij}(t_1,t_2)& =\theta\left(t_1-t_2\right)\!\!\int_{-\infty}^{\infty}\!\!\frac{d\omega}{2\pi}e^{-i\omega (t_1-t_2)}{G}_{ij}\left(\omega+i\eta\right).
\end{align}
\end{subequations}
\begin{subequations}
\label{G<byGr}
The lesser Green function, defined by
\begin{align}
\label{G<byGr-def}
[\boldsymbol{G}^{<}(t_1,t_2)]_{ij}=i\left.\langle d_{j}^{\dagger}(t_2)d^{}_{i}\left(t_1\right) \rangle\right.,
\end{align}is found via Eq.~(\ref{dit_hsg}) to be
\begin{align}
\boldsymbol{G}^{<}\left( t,t^{\prime }\right) =&\boldsymbol{G}^{r}\left(
t,t_{0}\right) \boldsymbol{G}^{<}\left( t_{0},t_{0}\right)
\boldsymbol{G}^{a}\left( t_{0},t^{\prime }\right)\nonumber\\&
+\int_{t_{0}}^{t}\!\!d\tau \int_{t_{0}}^{t'}\!\!d\tau ^{\prime
}\boldsymbol{G}^{r}\left( t,\tau \right) \boldsymbol{\Sigma}^{<}\left( \tau
,\tau ^{\prime }\right) \boldsymbol{G}^{a}\left( \tau ^{\prime
},t^{\prime }\right), \label{keldysh-eq}
\end{align} where
\begin{align}
\boldsymbol{\Sigma}_{}^{<}\left(\tau,\tau'\right)=\sum_{\alpha}\boldsymbol{\Sigma}_{\alpha}^{<}\left(\tau,\tau'\right),
\end{align} and the lesser Green function at the initial time $t_{0}$ shows an explicit dependence on $\hat{\rho}^{}_{\text{S}}(t_0)$ as
\begin{equation}
\left[ \boldsymbol{G}^{<}\left( t_{0},t_{0}\right) \right]
_{ij}=i\text{tr}\left[ d_{j}^{\dag }d^{}_{i}\hat{\rho}^{}_{\text{S}}(t_0)
\right]
.\label{init-Gless-varrho}
\end{equation}
\end{subequations}Here $\text{tr}$ means tracing over all the degrees of freedom of the whole system (central scattering region plus the electrodes).

\subsection{Transient dependence of the tunneling currents on the initial state of the central area}
\label{sec_method-initstat}

The above result, Eqs.~(\ref{dynamic-crnt},\ref{Gr-explicit-Gz},\ref{G<byGr}), shows that the dependence of the tunneling currents on the initial state of the central scattering region only comes from the second term in Eq.~(\ref{dynamic-crnt-totformula}) through the lesser Green function Eq.~(\ref{G<byGr}). By setting $\hat{\rho}^{}_{\text{S}}(t_0)=\vert0_{\text{S}}\rangle\langle0_{\text{S}}\vert$ (such that $\boldsymbol{G}^{<}\left( t_{0},t_{0}\right)=\boldsymbol{0}$), where $\vert0_{\text{S}}\rangle$ represents the empty state of the central scattering area, or letting $t_0\rightarrow-\infty$, Eq.~(\ref{dynamic-crnt-totformula}) becomes the current obtained in Ref.~[\onlinecite{Jauho945528}]. The expression Eq.~(\ref{keldysh-eq}) can be obtained via the NEGF technique by setting the coupling between the QDs and leads to zero for $t < t_0$.\cite{Stefanucci07195115} One can also obtain Eq.~(\ref{keldysh-eq}) via the influence functional theory.\cite{Jin10083013} Substituting Eq.~(\ref{G<byGr}) into Eq.~(\ref{dynamic-crnt}), the time-dependent tunneling current can thus be separated into two terms,
\begin{subequations}
\label{dynamic-crnt-sep}
\begin{align}
I_{\alpha }\left( t\right)=I^{\text{em.}}_{\alpha }\left( t\right)+I^{\text{occ.}}_{\alpha }\left( t\right),
\end{align} where
\begin{align}
\label{dynamic-crnt-0}
&I^{\text{em.}}_{\alpha }\left( t\right)=-2{\rm ReTr}\!\!\int_{t_0}^{t}\!\!d\tau\left(\boldsymbol{\Sigma}^{<}_{\alpha}(t,\tau)\boldsymbol{G}^{a}(\tau,t)
\right.\nonumber\\
&+\left.\boldsymbol{\Sigma}^{r}_{\alpha}(t,\tau)[\boldsymbol{G}^{<}(\tau,t)]\vert_{0}\right).
\end{align} in which,
\begin{align}
\label{lessG0}
&[\boldsymbol{G}^{<}(\tau,t)]\vert_{0}=\int_{t_{0}}^{\tau}\!\!ds \int_{t_{0}}^{t}\!\!ds ^{\prime
}\boldsymbol{G}^{r}\left( \tau,s \right) \boldsymbol{\Sigma}^{<}\left( s
,s^{\prime }\right) \boldsymbol{G}^{a}\left( s^{\prime
},t\right),
\end{align}
and
\begin{align}
\label{crnt-init-dep}
&I^{\text{occ.}}_{\alpha }\left( t\right)
\nonumber\\&=-2{\rm ReTr}\!\!\int_{t_0}^{t}\!\!d\tau\boldsymbol{\Sigma}^{r}_{\alpha}(t,\tau)
\boldsymbol{G}^{r}\left(
\tau,t_{0}\right) \boldsymbol{G}^{<}\left( t_{0},t_{0}\right)
\boldsymbol{G}^{a}\left( t_{0},t\right).
\end{align}
\end{subequations}
The first term $I^{\text{em.}}_{\alpha}(t)$ describes the tunneling current that is independent of the initial occupation in the central scattering region. The superscript "em" stands for initially "empty" states. The second term $I^{\text{occ.}}_{\alpha}(t)$, where the superscript "occ." abbreviates "initial occupation", explicitly includes this effect. The result Eq.~(\ref{crnt-init-dep}) further indicates that $I^{\text{occ.}}_{\alpha}(t)$ is independent of the initial chemical potentials and temperatures of the electrodes. As long as there are no bound states,\cite{Mahan00book} the Green function, $\boldsymbol{G}^{a}\left( t_{0},t\rightarrow\infty\right)=0$, vanishes in the long-time limit. The current purely induced by initial occupation, $I^{\text{occ.}}_{\alpha }\left( t\right)$, then only survives transiently.


\section{Initial-occupation-dependent transient interferometry with parallel quantum dots}
\label{sec_model_transit}

\begin{figure}[h]
\includegraphics[width=7cm,height=3cm]{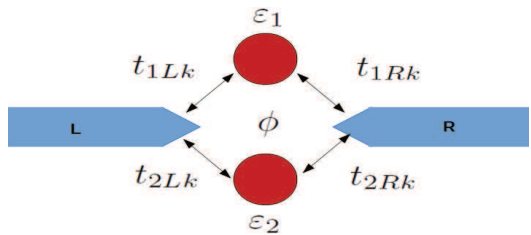} \caption{(color online) A schematic sketch of the DQD
device under consideration. Two single-level QDs with on-site
energies $\varepsilon_{1}$ and $\varepsilon_{2}$ are coupled to the two separate leads with
the tunneling amplitudes $t_{i\alpha k}$. The device is threaded by the
the Aharonov-Bohm flux $\phi$.
} \label{figS}
\end{figure}

The target system in the present work consists of two QDs arranged in parallel between two electrodes. It is schematically shown in Fig.~\ref{figS}.
For simplicity, we assume that each QD contains a single spinless orbital. The Hamiltonian of the central nanostructure Eq.~(\ref{gel_Hs}) is then specified by
\begin{align}
\mathcal{H}_{\text{S}}=\varepsilon_{1}d_{1}^{\dagger}d_{1}+\varepsilon_{2}d_{2}^{\dagger}d_{2}.
\end{align} The two QDs are uncoupled but tunneling through one QD can interfere with tunneling through the other. To reveal and control such interference for the study of transient transport, we utilize a magnetic flux penetrating through the loop formed by the two QDs and the two leads. This implements the AB effect via properly attaching phases to the hopping amplitudes $t_{i\alpha k}=\bar{t}_{i\alpha k}e^{i\phi_{i\alpha}}$ in Eq.~(\ref{gel_HT}) such that the constraint
\begin{align}
\phi_{L}-\phi_{R}=\phi=2\pi\Phi/\Phi_{0},
\end{align}where $\phi_{1\alpha}-\phi_{2\alpha}=\phi_{\alpha}$, is respected. Here $\Phi_{}$ is the applied magnetic flux through the ring and $\Phi_{0}$ is the flux quantum. This enters the Green functions above through the level-broadening functions,
\begin{subequations}
\label{AB_lvbrd}
\begin{align}
\label{with-phase_lvbrd}
\Gamma^{\alpha}_{ij}(\omega)=\bar{\Gamma}^{\alpha}_{ij}(\omega)e^{i(\phi_{i\alpha}-\phi_{j\alpha})},
\end{align} with
\begin{align}
\label{no-phase_lvbrd}
\bar{\Gamma}^{\alpha}_{ij}(\omega)=2\pi\sum_{k\in\alpha}\delta(\omega-\varepsilon^{}_{\alpha k})\bar{t}_{i\alpha k}\bar{t}_{\alpha kj}.
\end{align}
\end{subequations} Without loss of generality, we assume $\bar{\Gamma}^{\alpha}_{ij}(\omega)$ to be real for all $\alpha=L,R$ and $i,j\in\{1,2\}$.

To discuss the transient transport we first extract the components of the tunneling currents that are purely transient. We identify the regime where only these components exist so that the steady-state currents vanish. We elucidate the physical meanings of different transient tunneling currents. This then clarifies the directions for subsequent analysis of how their characters depend on the interferometry parameters.



Here we show the followings. (i) Apart from $I^{\text{occ.}}_{\alpha }\left( t\right)$, purely transient effects in the tunneling currents can also be found from $I^{\text{em.}}_{\alpha}(t)$. These purely transient components can be extracted from the directly observable currents. (ii) Since the steady-state currents vanish in the zero-bias regime, it is convenient to study the transient currents at zero bias. (iii) The currents purely induced by initial occupations can be analyzed by resolving contributions from each QD.


\subsection{transient zero-bias currents}
\label{sec_model_transit_defs}

The present setup is a two-terminal system. The label for reservoir $\alpha$ in Eq.~(\ref{dynamic-crnt-sep}) then takes the value $\alpha=L$ or $\alpha=R$ for a terminal on the left or on the right. In the steady states, charge conservation asserts $I_{\alpha}(t\rightarrow\infty)=-I_{\bar{\alpha}}(t\rightarrow\infty)$, where $\bar{\alpha}$ refers to the opposite of $\alpha$, i.e., $\bar{L}=R$ and $\bar{R}=L$. Before the steady state is reached, due to the time dependence of charge occupation in the central scattering area, one generally has $I_{\alpha}(t)\ne-I_{\bar{\alpha}}(t)$.

Therefore, it is necessary to characterize individually the currents on each side. Generically, we have
\begin{subequations}
\label{dynamic-crnt-0-b0}
\begin{align}
I^{\text{em.}}_{\alpha}(t)=\overline{I}^{\text{em.}}_{\alpha }(t)+\Delta{I}^{\text{em.}}_{\alpha }(t),
\end{align} where
\begin{align}
\label{dynamic-crnt-0-0}
\overline{I}^{\text{em.}}_{\alpha }(t)=\int_{-\infty}^{\infty}\!\!\frac{d\omega}{2\pi}
\bar{f}(\omega)\mathcal{T}^{(+)}_{\alpha}(t,\omega),
\end{align} and
\begin{align}
\label{dynamic-crnt-0-b}
\Delta{I}^{\text{em.}}_{\alpha }(t)=\int_{-\infty}^{\infty}\!\!\frac{d\omega}{2\pi}
\Delta{f}_{\alpha}(\omega)\mathcal{T}^{(-)}_{\alpha}(t,\omega),
\end{align}
are the contributions from the average of the fermi functions,
\begin{align}
&\bar{f}(\omega)=\frac{f_{L}(\omega)+f_{R}(\omega)}{2},
\end{align}
and from the deviation of the fermi distribution in each reservoir from the average,
\begin{align}
\Delta{f}_{\alpha}(\omega)=f_{\alpha}(\omega)-\bar{f}(\omega),
\end{align} respectively. The transmission-like, energy- and time-dependent functions in Eqs.~(\ref{dynamic-crnt-0-0},\ref{dynamic-crnt-0-b}) are
\begin{align}
\label{intftrsm-1}
&\mathcal{T}^{(\pm)}_{\alpha}(t,\omega)=2\text{ImTr}\int_{t_0}^{t}\!\!d\tau
\left\{
\boldsymbol{\Gamma}^{\alpha}(\omega)e^{-i\omega(t-\tau)}\boldsymbol{G}^{a}(\tau,t)\right.
\nonumber\\
&+\boldsymbol{\Sigma}^{r}_{\alpha}(t,\tau)\times\nonumber\\
&\!\!\int_{t_{0}}^{\tau}\!\!\!\!d\tau_{1}\!\!\int_{t_{0}}^{t}\!\!\!\!d\tau_{2}\left.
\!\!\boldsymbol{G}^{r}\left(t,\tau_{1}\right)\!\!
\left[\boldsymbol{\Gamma}^{\alpha}(\omega)\pm\boldsymbol{\Gamma}^{\bar{\alpha}}(\omega)\right]\!\!
e^{-i\omega(\tau_{1}-\tau_{2})}
\boldsymbol{G}^{a}(\tau_{2},t)
\right\}\!,
\end{align}
\end{subequations} obtained by substituting Eq.~(\ref{less-selfeng}) into Eq.~(\ref{dynamic-crnt-0}) and Eq.~(\ref{lessG0}).
In Eq.~(\ref{dynamic-crnt-0-0}), $\mathcal{T}^{(+)}_{\alpha}(t,\omega)$ is the transmission of electrons with energy $\omega$ from reservoir $\alpha$ into the QDs at time $t$.
The second part of the current, $\Delta{I}^{\text{em.}}_{\alpha }(t)$, is nonzero only when $f_{L}(\omega)-f_{R}(\omega)$ is nonzero. It is therefore attributed exclusively to the effect induced by a bias, represented by the difference between the two fermi functions. The other part, $\overline{I}^{\text{em.}}_{\alpha }(t)$, survives even without a bias.

In the steady states, the current, $I=I_{L}(t\rightarrow\infty)=-I_{R}(t\rightarrow\infty)$, is reduced to the well-known expression, namely,\cite{Landauer70863}
\begin{align}
\label{stdyMW}
I=\int_{-\infty}^{\infty}\!\!\frac{d\omega}{2\pi}
(f_{L}(\omega)-f_{R}(\omega))\mathcal{T}^{}_{}(\omega)=\zeta_{\alpha}\Delta{I}^{\text{em.}}_{\alpha }(t\rightarrow\infty),
\end{align}
where
\begin{align}
\mathcal{T}^{}_{}(\omega)\!\!=\!\!\frac{1}{2}\mathcal{T}^{(-)}_{\alpha}(t\rightarrow\infty,\omega)\!\!=\!\!\text{Tr}
\left\{
\boldsymbol{\Gamma}^{\alpha}(\omega)\boldsymbol{G}^{r}(\omega)\boldsymbol{\Gamma}^{\bar{\alpha}}(\omega)
\boldsymbol{G}^{a}(\omega)
\right\},
\end{align} for both $\alpha=L,R$. In Eq.~(\ref{stdyMW}), $\zeta_{L}=1$ and $\zeta_{R}=-1$. Henceforth, from Eqs.~(\ref{stdyMW},\ref{dynamic-crnt-0-b0}) one sees unambiguously that $\overline{I}^{\text{em.}}_{\alpha }(t\rightarrow\infty)=0$ and that the steady-state currents vanish at zero bias.

We thus identify that $\overline{I}^{\text{em.}}_{\alpha }(t)$ is a purely transient component of the current $I^{\text{em.}}_{\alpha}(t)$ for whatever biases.
The effects that are exclusively transients are contained only in $\overline{I}^{\text{em.}}_{\alpha }(t)$ and $I^{\text{occ.}}_{\alpha }\left( t\right)$.
Given the abilities of preparing various initial occupations of the QDs (including the initial empty state),\cite{Hayashi03226804,Fujisawa06759,Hanson071217,Kim1470} the currents purely induced by initial occupations $I^{\text{occ.}}_{\alpha }\left( t\right)$ can be obtained by subtracting $I^{\text{em.}}_{\alpha }\left( t\right)$ from the total current $I_{\alpha}(t)$ starting from various occupations. The other transient component, $\overline{I}^{\text{em.}}_{\alpha }(t)$, can be found from the difference between the currents $I_{\alpha}(t)$ obtained at reversed biases, namely,
\begin{align}
&\left.\overline{I}^{\text{em.}}_{\alpha }(t)\right\vert_{(\mu^{}_{L},\mu^{}_{R})=
(\mu^{}_{A},\mu^{}_{B})}=\frac{1}{2}\times
\nonumber\\&
\left[\left.I^{\text{em.}}_{\alpha}(t)\right\vert_{(\mu^{}_{L},\mu^{}_{R})=(\mu^{}_{A},\mu^{}_{B})}
+\left.I^{\text{em.}}_{\alpha}(t)\right\vert_{(\mu^{}_{L},\mu^{}_{R})=(\mu^{}_{B},\mu^{}_{A})}\right]
\end{align}


\subsection{Resolving level contributions to initial charge induced currents }

To study the effects of initial occupations on the tunneling currents via $I^{\text{occ.}}_{\alpha }\left( t\right)$, one needs to examine various attainable occupations.
For the interferometer with two QDs, particularly interesting initial occupations are the states with one electron either occupying QD1 or QD2, and the fully occupied configuration. We denote $n_{i}(t_0)$ as the initial occupation on QD $i$. These diagonal initial occupations are then described by $\left[ \boldsymbol{G}^{<}\left( t_{0},t_{0}\right) \right]_{ij}=i\delta_{ij}n_{i}(t_0)$, in Eq.~(\ref{crnt-init-dep}) leading  to
\begin{subequations}
\label{crnt-init-diagocc}
\begin{align}
\label{crnt-init-diagocc-0}
I^{\text{occ.}}_{\alpha }\left( t\right)=\sum_{i}n_{i}(t_0)I^{[i]}_{\alpha}(t),
\end{align} with
\begin{align}
\label{crnt-init-diagocc-1}
I^{[i]}_{\alpha}(t)=2\text{Im}\int_{t_0}^{t}\!\!d\tau\left[
\boldsymbol{G}^{a}\left( t_{0},t\right)\boldsymbol{\Sigma}^{r}_{\alpha}(t,\tau)\boldsymbol{G}^{r}\left(
\tau,t_{0}\right)
\right]_{ii},
\end{align}
\end{subequations}
resolving the current specifically induced by an initial occupation on level $i$. Consequently, the current induced by initially occupying both QDs, denoted as $I^{[1+2]}_{\alpha}(t)$, is then
\begin{align}
\label{crnt-1plus2}
I^{[1+2]}_{\alpha}(t)\equiv\left.I^{\text{occ.}}_{\alpha }\left( t\right)\right\vert_{(n_{1}(t_0),n_{2}(t_0))=(1,1)}=\sum_{i=1}^{2}I^{[i]}_{\alpha}(t),
\end{align} according to Eq.~(\ref{crnt-init-diagocc}). The results Eq.~(\ref{dynamic-crnt-sep}) and Eq.~(\ref{crnt-init-diagocc}) imply that one only needs to separately track the dynamics of $I^{[i]}_{\alpha}(t)$ for each level $i$ to see how various initial occupations influence the transient dynamics.





\subsection{Physical meanings of transient currents}
\label{sec_model_transit_defs_physmean}



Having identified distinct transient components of the tunneling currents, we now analyze their corresponding underlying physical processes. Immediately after $t=t_{0}$, the current $I^{\text{em.}}_{\alpha }\left( t\right)$ is contributed solely by the in-tunneling processes (tunneling of electrons from reservoirs into the central scattering area). Since in the beginning there were no electrons in the central part, it is not possible to have electrons tunnel out of the central area into the reservoirs.  On the contrary, $I^{\text{occ.}}_{\alpha }\left( t\right)$ describes currents exclusively induced by out-tunneling processes. The currents are defined as the negative changing rate of the electron numbers in the electrodes. In-tunneling from reservoirs into the QDs decreases the number of electrons in the reservoirs and hence the changing rate of electron number in the reservoirs is negative. Consequently, we anticipate the in-tunneling currents to be positive. On the contrary, we expect the out-tunneling currents to be negative.

Microscopically, the elementary processes underlying the in-tunneling currents are induced by tunneling of electrons from occupied levels below the chemical potentials in the reservoirs to the QDs. The positions of the chemical potentials relative to the energy levels of the QDs determine how much can tunnel into the QDs and how much can tunnel out as well. Therefore the symmetry of energy levels on the QDs relative to the chemical potentials is important for the relationship between the in-tunneling and the out-tunneling currents. In addition to that, the interferometer is also characterized by the geometric symmetry of the device. The individual transient components $I^{[i]}_{\alpha}(t,\phi)$, $\overline{I}^{\text{em.}}_{\alpha }(t,\phi)$ are explicitly associated with the geometric characterization of the interferometer, namely, the up ($i=1$), down ($i=2$), left ($\alpha=L$) and right ($\alpha=R$).
Below we show how to modulate the properties of these transient currents by investigating different components of the transient currents in association with these two symmetries of the system.

\section{Symmetry in the dynamics for the transient currents}
\label{sec_model_transit_fluxindep}

In order to have an unambiguous picture, we explicitly specify the relevant parameters of the system.
We assume the widely applied wide-band approximation. Within this assumption, $\bar{t}_{i\alpha k}=\bar{t}_{i\alpha}$, becomes independent of $k$ in Eq.~(\ref{no-phase_lvbrd}) and consequently
\begin{align}
\label{wb_no-phase_lvbrd}
\bar{\Gamma}^{\alpha}_{ij}(\omega)=\bar{\Gamma}^{\alpha}_{ij}=2\pi\varrho_{\alpha}\bar{t}_{i\alpha}\bar{t}_{j\alpha},
\end{align} where $\varrho_{\alpha}$ is the density-of-state of lead $\alpha$. Following Eq.~(\ref{wb_no-phase_lvbrd}), we have $\bar{\Gamma}^{\alpha}_{12}=2\pi\varrho_{\alpha}\bar{t}_{1\alpha}\bar{t}_{2\alpha}
=\sqrt{(2\pi\varrho_{\alpha}\bar{t}_{1\alpha}\bar{t}_{1\alpha})(2\pi\varrho_{\alpha}\bar{t}_{2\alpha}\bar{t}_{2\alpha})}=\sqrt{\bar{\Gamma}^{\alpha}_{11}\bar{\Gamma}^{\alpha}_{22}}$. The solution to the Green functions and the consequent currents Eqs.~(\ref{dynamic-crnt-0-b0}),(\ref{crnt-init-diagocc}) are explicitly obtained and summarized in Appendix \ref{appx-GrWB}. The parameters of the interferometry are then specified by the four bonds, $\bar\Gamma^{L}_{11}$, $\bar\Gamma^{L}_{22}$, $\bar\Gamma^{R}_{11}$, and $\bar\Gamma^{R}_{22}$, the two on-site energies of the QDs, $\varepsilon_{1}$ and $\varepsilon_{2}$, the two chemical potentials $\mu_{L}$ and $\mu_{R}$ of the reservoirs and the applied flux $\phi$. The currents are functions of both time and flux, namely, $I_{\alpha}(t)\rightarrow I_{\alpha}(t,\phi)$.

The symmetry of the energy level distributions concerns the relative configuration of the four energy references $\varepsilon_{1}$ and $\varepsilon_{2}$, $\mu^{}_{L}$ and $\mu^{}_{R}$.
We denote the energy references by
\begin{align}
\label{ct-dot}
\varepsilon_{0}=(\varepsilon_{1}+\varepsilon_{2})/2,
\end{align} and
\begin{align}
\label{mu-md}
\mu^{}_{0}=(\mu^{}_{L}+\mu^{}_{R})/2.
\end{align}
The geometric symmetry of the interferometer is specified by the distribution of the four bonds and the two on-site energies of the QDs. It is also affected by the applied flux. The geometry of the bonds can be classified according to the geometric symmetry they present: (i) left-right symmetry, namely, $\bar{\Gamma}^{L}_{ii}=\bar{\Gamma}^{R}_{ii}=\bar{\Gamma}_{i}$, and (ii) up-down symmetry, $\bar{\Gamma}^{\alpha}_{11}=\bar{\Gamma}^{\alpha}_{22}=\Gamma^{\alpha}$. We denote the difference between the on-site energies of the two QDs by
\begin{align}
\label{dt-dot}
\delta\varepsilon=\varepsilon_{1}-\varepsilon_{2}.
\end{align} The geometry of the two on-site energies is categorized by being degenerate $\delta\varepsilon=0$ or not $\delta\varepsilon\ne0$, exhibiting up-down symmetry or not. The influences of the flux can be inspected by
generally decomposing any flux-dependent quantity, $Q(\phi)$, into
\begin{subequations}
\label{gel_flux_decps}
\begin{align}
Q(\phi)=Q^{+}(\phi)+Q^{-}(\phi),
\end{align} with
\begin{align}
Q^{\pm}(\phi)=\pm Q^{\pm}(-\phi)
\end{align}
\end{subequations}
representing the even "$+$" and the odd "$-$" responses to the flux.

In Sec.~\ref{sec_model_transit_oddspec}, we illustrate the special transient properties of the odd components of $I^{[i]}_{\alpha}(t,\phi)$, and $\overline{I}^{\text{em.}}_{\alpha }(t,\phi)$. Subsequent discussions reveal that they are not sensitive to left-right asymmetry, unlike their even counterparts.
In Sec.~\ref{sec_model_transit_numcs-1}, we discuss the effects of up-down symmetry and we demonstrate how to affect the left-right asymmetry in the time scales as well as the magnitudes of the currents by the geometry of the system. The case of left-right symmetry is presented in Sec.~\ref{sec_model_fluxindep_sym2} together with geometric factors that influence the up-down asymmetry between $I^{[1]}_{\alpha}(t,\phi)$ and $I^{[2]}_{\alpha}(t,\phi)$. In Sec~\ref{sec_model_fluxindep_sym1} we focus on the symmetry of the energy level distribution. This symmetry is shown to play a key role in the relationship between the main two parts of the purely transient components of the currents, $I^{[i]}_{\alpha}(t,\phi)$ and $\overline{I}^{\text{em.}}_{\alpha }(t,\phi)$. Note that the $I^{[i]}_{\alpha}(t,\phi)$'s do not depend on the chemical potentials. As shown in Eq.~(\ref{dynamic-crnt-0-0}), the geometric factors only enter $\overline{I}^{\text{em.}}_{\alpha}(t,\phi)$ through $\mathcal{T}^{(+)}_{\alpha}(t,\omega,\phi)$, having nothing to do with $\mu_{L}$ and $\mu_{R}$. For simplicity, we set $\mu_{L}=\mu_{R}=\mu_{0}=\varepsilon_{0}$ for inspecting $\overline{I}^{\text{em.}}_{\alpha}(t,\phi)$'s in Sec.~\ref{sec_model_transit_numcs-1} and Sec.~\ref{sec_model_fluxindep_sym2}. Deviations from $\mu_{L}=\mu_{R}=\mu_{0}=\varepsilon_{0}$ are considered in Sec~\ref{sec_model_fluxindep_sym1}.


\subsection{Transient odd components of the tunneling currents }
\label{sec_model_transit_oddspec}

Below we show that the odd components of the currents are insensitive to the left-right asymmetry in the bonds. However, the up-down symmetry in terms of the on-site energies of the QDs is crucial to manifest the odd components.

Substituting explicitly Eq.~(\ref{retardedG-intime}) into Eq.~(\ref{time-dep_Ts_WB}) and Eq.~(\ref{crnt-init-diagocc-WB}) and extracting the parts that are odd in the flux give
\begin{align}
\label{odd-I1}
&I_{\alpha}^{\left[1\right],-}\left(t,\phi\right)  =2\zeta_{\alpha}\bar{\Gamma}_{12}^{\alpha}\bar{\Gamma}_{12}^{\bar{\alpha}}\sin\phi
\nonumber\\
 &\times \left[\text{Im}\left(\frac{b_{-}\left(t\right)}{\Gamma_{g}\left(\phi\right)}b_{+}^{*}\left(t\right)\right)-\delta\varepsilon\left\vert \frac{b_{-}\left(t\right)}{\Gamma_{g}\left(\phi\right)}\right\vert ^{2}\right],
\end{align}
\begin{align}
\label{odd-I2}
&I_{\alpha}^{\left[2\right],-}\left(t,\phi\right)  =2\zeta_{\alpha}\bar{\Gamma}_{12}^{\alpha}\bar{\Gamma}_{12}^{\bar{\alpha}}\sin\phi
 \nonumber\\
 &\times \left[-\text{Im}\left(\frac{b_{-}\left(t\right)}{\Gamma_{g}\left(\phi\right)}b_{+}^{*}\left(t\right)\right)-\delta\varepsilon\left\vert \frac{b_{-}\left(t\right)}{\Gamma_{g}\left(\phi\right)}\right\vert ^{2}\right],
\end{align} and
\begin{align}
\label{odd-Tpm}
&\mathcal{T}_{\alpha}^{\left(\pm\right),-}\left(t,\omega,\phi\right)=\mp4\zeta_{\alpha}\delta\varepsilon\bar{\Gamma}_{12}^{\alpha}\bar{\Gamma}_{12}^{\bar{\alpha}}\sin\phi
\nonumber\\ &\times
\left[\Gamma\left\vert \frac{\tilde{b}_{-}\left(t,\omega\right)}{\Gamma_{g}\left(\phi\right)}\right\vert ^{2}+\text{Re}\left(\frac{\tilde{b}_{-}\left(t,\omega\right)}{\Gamma_{g}\left(\phi\right)}\tilde{b}_{+}^{*}\left(t,\omega\right)\right)\right].
\end{align} In Eqs.~(\ref{odd-I1}),(\ref{odd-I2}),and (\ref{odd-Tpm}), the $\zeta_{\alpha}$'s are defined after Eq.~(\ref{stdyMW}), $\Gamma_{g}(\phi)$ is given by Eq.~(\ref{cmpx-decayrate}), $b_{\pm}\left(t,\phi\right)$ and $\tilde{b}_{\pm}\left(t,\omega\right)$ are defined in Eq.~(\ref{time-amplitudes-cp}) and Eq.~(\ref{time-amplitudes-cp-omega}) respectively. The factor, $\text{Im}\left(\frac{b_{-}\left(t\right)}{\Gamma_{g}\left(\phi\right)}b_{+}^{*}\left(t\right)\right)$, in Eqs.~(\ref{odd-I1},\ref{odd-I2}), vanishes when the geometry of the system satisfies either
\begin{align}
\label{updwn_bdsym}
\bar{\Gamma}^{\alpha}_{11}=\bar{\Gamma}^{\alpha}_{22}=\Gamma^{\alpha},
\end{align} or
\begin{align}
\label{degDQD}
\delta\varepsilon=0.
\end{align} (See Appendix \ref{appx-GrWB-oddbfctr} for a detailed derivation.) The results Eqs.~(\ref{odd-I1}),(\ref{odd-I2}) and (\ref{odd-Tpm}) show that the following two conditions must simultaneously hold for these odd components to be nonzero:
\begin{align}
\label{sinphine0}
\sin(\phi)\ne0.
\end{align} and
\begin{align}
\label{nondegDQD}
\delta\varepsilon\ne0.
\end{align}  The first condition Eq.~(\ref{sinphine0}) explicitly reveals that these odd components arise from the applied flux. The second condition Eq.~(\ref{nondegDQD}) can be used to discern qualitatively if the QDs are degenerate or not by inspecting whether the odd components of the currents are always zero.

Furthermore, Eqs.~(\ref{odd-I1}),(\ref{odd-I2}), and (\ref{odd-Tpm}) also show that regardless of the symmetry of the bonds, the odd components always obey the following symmetry:
\begin{align}
\label{phi-LR_asym}
&I_{L}^{\left[i\right],-}\left(t,\phi\right)=I_{R}^{\left[i\right],-}\left(t,2\pi-\phi\right),
\nonumber\\
&\mathcal{T}_{L}^{\left(\pm\right),-}\left(t,\omega,\phi\right)=\mathcal{T}_{R}^{\left(\pm\right),-}\left(t,\omega,2\pi-\phi\right).
\end{align} Note that Eq.~(\ref{phi-LR_asym}) applies not only to the parts of the currents that are purely transient, namely, $I^{[i]}_{\alpha}(t,\phi)$ and $\overline{I}^{\text{em.}}_{\alpha }(t,\phi)$. It is also true for the part of the current given by $\Delta{I}^{\text{em.}}_{\alpha}(t)$ (see Eq.~(\ref{dynamic-crnt-0-b0}) with Eq.~(\ref{phi-LR_asym})).

The system under consideration is a two-terminal setup.  Therefore, the steady-state currents must obey phase rigidity making only the even components remain.\cite{Yeyati9514360R,Yacoby969583} Taking the steady-state limit, $t\rightarrow\infty$, in Eqs.~(\ref{odd-I1}),(\ref{odd-I2}), and (\ref{odd-Tpm}), we see they all vanish (see Appendix \ref{appx-GrWB-oddbfctr}). Henceforth, the effects which arise from the odd components are exclusively transient. The above analysis then leads us to consider the effect of geometric symmetry for the even and the odd components separately according to whether $\delta\varepsilon=0$ or $\delta\varepsilon\ne0$.

\subsection{Up-down symmetry of the bonds, $\bar{\Gamma}^{\alpha}_{11}=\bar{\Gamma}^{\alpha}_{22}$}

\label{sec_model_transit_numcs-1}

Here we discuss the effects of up-down symmetry realized by Eq.~(\ref{updwn_bdsym}).
Given the equally strong upper and lower bonds, the currents purely induced by occupying one QD shall also exhibit up-down symmetry,\cite{footnote0} namely, $I^{[1]}_{\alpha}(t,\phi)=I^{[2]}_{\alpha}(t,\phi)$ (see Appendix \ref{appx-GrWB-updwnsym} for a detailed proof using Eq.~(\ref{updwn_bdsym}) only without Eq.~(\ref{degDQD})). Intuitively, a large asymmetry between the magnitudes of the currents on the left and that on the right can be realized by $\Gamma^{L}\ll\Gamma^{R}$. More interestingly, we show that the left-right asymmetry between the times of approaching the steady-state limit can be enhanced or suppressed by the choices of $\delta\varepsilon$ in comparison to $\vert\Gamma_{12}(\phi)\vert$.

\subsubsection{degenerate QDs}

\label{sec_model_transit_numcs-1-deg}

Figure \ref{crnts_S38} demonstrates the currents with up-down symmetry in the bonds and with degenerate QD levels, $\delta\varepsilon=0$. In this case, only the even components of the currents are nonzero. The currents on the left and on the right are shown as a function of time at various fluxes on the first two rows of Fig.~\ref{crnts_S38} (plots (a1) for $I^{[1]}_{L}(t,\phi)=I^{[2]}_{L}(t,\phi)$, (a2) for $\overline{I}^{\text{em.}}_{L}(t,\phi)$, (b1) for $I^{[1]}_{R}(t,\phi)=I^{[2]}_{R}(t,\phi)$, (b2) for $\overline{I}^{\text{em.}}_{R}(t,\phi)$).

At degeneracy, the rates of approaching steady states are governed only by the bonds. Therefore, the left-right asymmetry in the bonds would be revealed also in the difference between the times needed for the currents on the left and on the right to reach the steady states. This is illustrated on the first two rows of Fig.~\ref{crnts_S38} (compare the time scales in (a1),(a2) with those in (b1),(b2) in Fig.~\ref{crnts_S38}). When the up-down symmetry is realized simultaneously by Eq.~(\ref{updwn_bdsym}) and Eq.~(\ref{degDQD}), Eqs.~(\ref{retardedG-intime}), (\ref{time-amplitudes-cp}), (\ref{Gr12_updwbdsym}), (\ref{gr11sq2}), (\ref{Ginfeqsym}) and (\ref{I1sym}) yield
\begin{align}
\label{I1_udwnsym_deg}
&I^{[1]}_{\alpha}(t,\phi)=-\Gamma^{\alpha}\left\{\frac{1}{2}\left(1-\frac{\left(\Gamma^{\alpha}+\Gamma^{\bar{\alpha}}\cos\phi\right)}{\delta\Gamma\left(\phi\right)}\right)
e^{-(\Gamma-\delta\Gamma\left(\phi\right))t}\right.
\nonumber\\&\left.
+\frac{1}{2}\left(1+\frac{\left(\Gamma^{\alpha}+\Gamma^{\bar{\alpha}}\cos\phi\right)}{\delta\Gamma\left(\phi\right)}\right)
e^{-(\Gamma+\delta\Gamma\left(\phi\right))t}\right\}.
\end{align} The solution Eq.~(\ref{I1_udwnsym_deg}) shows that the initial-occupation-induced currents on both sides $\alpha=L,~R$ commonly possess two exponential decay terms $e^{-(\Gamma\pm\delta\Gamma\left(\phi\right))t}$, each of which is associated with an amplitude. The corresponding decay rates are plotted in Fig.~\ref{crnts_S38}(c). Equation~(\ref{I1_udwnsym_deg}) says that it is the difference between the amplitudes with $\alpha=L$ and with $\alpha=R$ that is responsible for the asymmetry between $I^{[1]}_{L}(t,\phi)$ and $I^{[1]}_{R}(t,\phi)$. The overall factor $\Gamma^{\alpha}$ in front of the RHS of Eq.~(\ref{I1_udwnsym_deg}) is reflected on the apparent difference in the magnitudes of the left and the right currents (see magnitudes in Fig.~\ref{crnts_S38}(a1) and Fig.~\ref{crnts_S38}(b1)). The difference in the time scales arise from the difference in the relative importance of the amplitudes in front of the faster decay term $e^{-(\Gamma+\delta\Gamma\left(\phi\right))t}$ and the slower decay term $e^{-(\Gamma-\delta\Gamma\left(\phi\right))t}$. Furthermore, the currents on the left shown by Fig.~\ref{crnts_S38}(a1),(a2) are clearly different for different fluxes. The currents on the right show less visible changes for different fluxes (see Fig.~\ref{crnts_S38}(b1),(b2)). These asymmetric features shown by Fig.~\ref{crnts_S38}(a1) and Fig.~\ref{crnts_S38}(b1) can be derived by imposing $\Gamma^{R}\gg\Gamma^{L}$ in Eq.~(\ref{I1_udwnsym_deg}) (see Appendix \ref{appx-AsymLR-deg}).

In Fig.~\ref{crnts_S38}, one also observes a symmetry between $I^{[1]}_{\alpha}(t,\phi)$ and $\overline{I}^{\text{em.}}_{\alpha}(t,\phi)$ showing that $\overline{I}^{\text{em.}}_{\alpha}(t,\phi)=-I^{[1]}_{\alpha}(t,\phi)$. The interpretation of $I^{[i]}_{\alpha}(t,\phi)$ and $\overline{I}^{\text{em.}}_{\alpha}(t,\phi)$ as out-tunneling and in-tunneling currents in Sec.~\ref{sec_model_transit_defs_physmean} is verified by the signs of these quantities (see the negative signs in Fig.~\ref{crnts_S38}(a1),(b1) and the positive signs in Fig.~\ref{crnts_S38}(a2),(b2)). As we show later that as long as $\mu_{0}=\varepsilon_{0}$ with either Eq.~(\ref{updwn_bdsym}) or Eq.~(\ref{degDQD}) being satisfied, one can infer $\overline{I}^{\text{em.}}_{\alpha}(t,\phi)$ from  $I^{[i]}_{\alpha}(t,\phi)$'s (see detailed discussion in Sec.~\ref{sec_model_fluxindep_sym1}). Therefore in Sec.~\ref{sec_model_transit_numcs-1} (where the discussions are all for the case Eq.~(\ref{updwn_bdsym})) only the $I^{[i]}_{\alpha}(t,\phi)$'s are discussed.


\subsubsection{non-degenerate QDs}
\label{sec_model_transit_numcs-1-nondeg}

\begin{figure}[h]
\includegraphics[width=8.0cm,height=9cm]{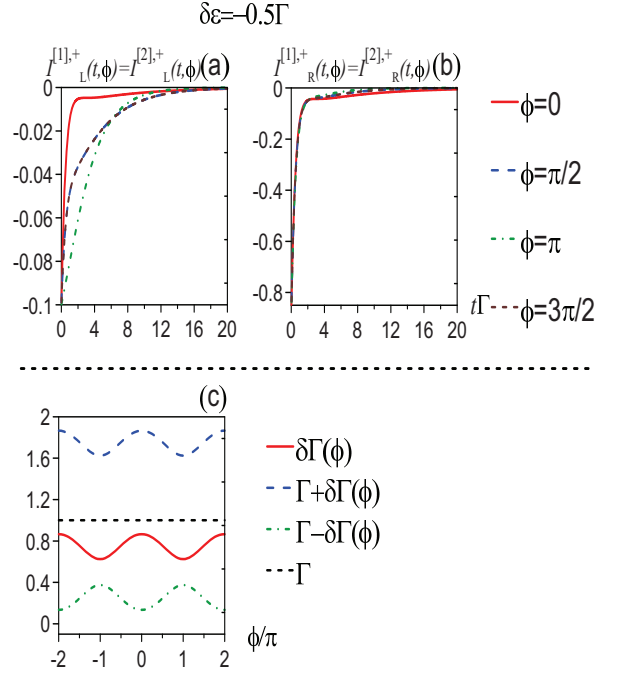} \caption{(color online) The even components of the initial-occupation-induced currents with up-down symmetry, $\bar{\Gamma}^{L}_{11}=\bar{\Gamma}^{L}_{22}=0.1\Gamma$ and $\bar{\Gamma}^{R}_{11}=\bar{\Gamma}^{R}_{22}=0.9\Gamma$ are shown in plots (a) and (b). Here we set $\delta\varepsilon=-0.5\Gamma$ such that $\vert\Gamma_{12}(\phi)\vert^{2}>\delta\varepsilon^{2}$ and the corresponding decay rates are plotted in (c).
} \label{crnts_S49-even}
\end{figure}

\begin{figure}[h]
\includegraphics[width=8.0cm,height=5cm]{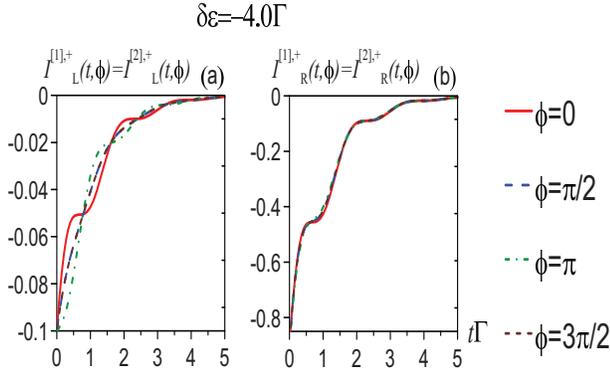} \caption{(color online) The even components of the initial-occupation-induced currents with non-degenerate QDs of on-site energy splitting, $\delta\varepsilon=-4\Gamma$, and with up-down symmetry, $\bar{\Gamma}^{L}_{11}=\bar{\Gamma}^{L}_{22}=0.1\Gamma$ and $\bar{\Gamma}^{R}_{11}=\bar{\Gamma}^{R}_{22}=0.9\Gamma$. Different line styles are for different fluxes as indicated on the right.
} \label{crnts_S50-even}
\end{figure}

\begin{figure}[h]
\includegraphics[width=8.0cm,height=5cm]{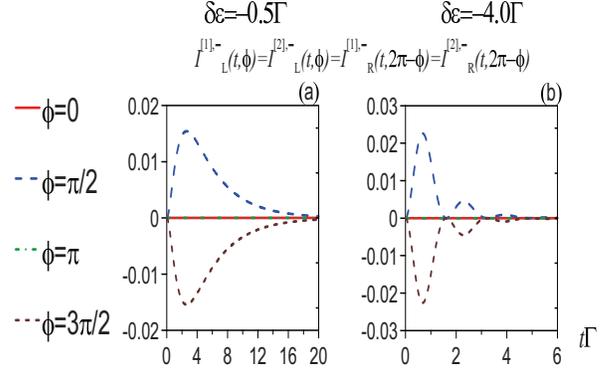} \caption{(color online) The odd components of the initial-occupation-induced currents. Plots (a) and (b) are with the same parameters used in Fig.~\ref{crnts_S49-even} for $\vert\Gamma_{12}(\phi)\vert^{2}>\delta\varepsilon^{2}$ and in Fig.~\ref{crnts_S50-even} for $\vert\Gamma_{12}(\phi)\vert^{2}<\delta\varepsilon^{2}$ respectively.
} \label{crnts_S49_50-odd}
\end{figure}

When the on-site energies of the QDs are not degenerate, $\delta\varepsilon\ne0$, then the odd components of the currents appear when $\sin\phi\ne0$, according to Eqs.~(\ref{odd-I1}),(\ref{odd-I2}), and (\ref{odd-Tpm}).
The time dependence of the currents relies on the time dependence of the Green function $\boldsymbol{G}^{r}\left(t,t_{0}\right)$. The latter, given by Eq.~(\ref{retardedG-intime}), has its time dependence governed by the parameter $\Gamma_{g}(\phi)$ defined in Eq.~(\ref{cmpx-decayrate}). The condition of Eq.~(\ref{updwn_bdsym}) makes $\Gamma_{g}(\phi)=\sqrt{\vert\Gamma_{12}(\phi)\vert^{2}-\delta\varepsilon^{2}}$  become either real, if $\vert\Gamma_{12}(\phi)\vert^{2}>\delta\varepsilon^{2}$, or purely imaginary if $\vert\Gamma_{12}(\phi)\vert^{2}<\delta\varepsilon^{2}$. Since the odd components satisfy Eq.~(\ref{phi-LR_asym}), to focus on the left-right asymmetry we first discuss these two cases of  $\vert\Gamma_{12}(\phi)\vert^{2}>\delta\varepsilon^{2}$ and  $\vert\Gamma_{12}(\phi)\vert^{2}<\delta\varepsilon^{2}$ separately for the even components. The effects of non-degeneracy on the odd components of the currents are discussed later. In all the discussions below for $\delta\varepsilon\ne0$, associated with Figs.~\ref{crnts_S49-even}, \ref{crnts_S50-even}, and \ref{crnts_S49_50-odd}, the equality $I^{[1],\pm}_{\alpha}(t,\phi)=I^{[2],\pm}_{\alpha}(t,\phi)$ under Eq.~(\ref{updwn_bdsym}) is also numerically verified.
\\
\begin{figure}[h]
\includegraphics[width=9.0cm,height=12cm]{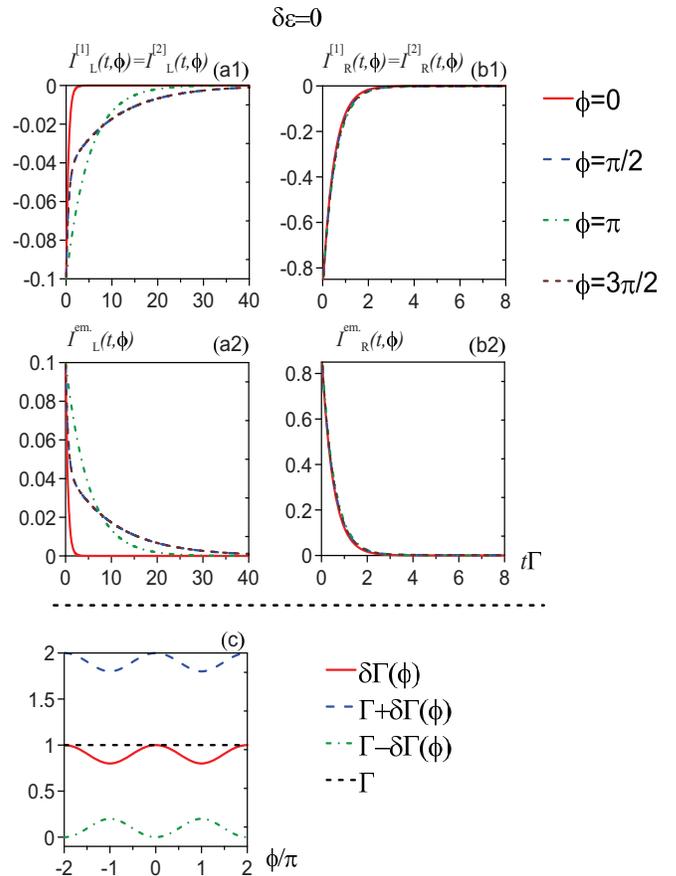} \caption{(color online) Various transient components of the currents with up-down symmetry, $\bar{\Gamma}^{L}_{11}=\bar{\Gamma}^{L}_{22}=0.1\Gamma$ and $\bar{\Gamma}^{R}_{11}=\bar{\Gamma}^{R}_{22}=0.9\Gamma$ at degeneracy $\delta\varepsilon=0$.  In plots (a1),(a2),(b1) and (b2) different line styles are for different fluxes as indicated on the right. In plot (c) we show the decay rates as a function of flux under the bonds and on-site energies specified above.  We have set $\mu_{L}=\mu_{R}=\mu_{0}=\varepsilon_{0}$ in this figure and as well as in Fig.~\ref{crnts_S37} and Fig.~\ref{crnts_S5-even-odd}. The temperature is taken to be $k_{B}T=0.1\Gamma$ for all figures, such that the thermal effects due to finite temperature become unimportant.
} \label{crnts_S38}
\end{figure}

$a.$ \emph{Even components for} $\delta\varepsilon\ne0$ \emph{with} $\vert\Gamma_{12}(\phi)\vert^{2}>\delta\varepsilon^{2}$
\\
\\
We first demonstrate the even components of the transient currents in Fig.~\ref{crnts_S49-even} with $\vert\Gamma_{12}(\phi)\vert^{2}>\delta\varepsilon^{2}\ne0$.
Opening the energy splitting between the QDs introduces another energy scale. This suppresses the sole dominance of the bonds in the rates of approaching steady states. Consequently the difference between the times required by $I^{[1]}_{L}(t,\phi)$ and $I^{[1]}_{R}(t,\phi)$ to reach the steady-state limits displayed in Fig.~\ref{crnts_S49-even}(a) and (b) is not as obvious as those shown by Fig.~\ref{crnts_S38}(a1) and (b1).  The same set of left-right asymmetric bonds is used in both Figs.~\ref{crnts_S38} and \ref{crnts_S49-even}. The even component of the initial-occupation-induced current with $\vert\Gamma_{12}(\phi)\vert^{2}>\delta\varepsilon^{2}\ne0$ now becomes (using the same derivation of Eq.~(\ref{I1_udwnsym_deg}) without applying Eq.~(\ref{degDQD})):
\begin{align}
\label{I1_udwnsym_nondegLS}
&I^{[1],+}_{\alpha}(t,\phi)=-\Gamma^{\alpha}\times
\nonumber\\&
\left\{\frac{1}{2}\left(1-\frac{\left(\Gamma^{\alpha}+\Gamma^{\bar{\alpha}}\cos\phi\right)}{\delta\Gamma\left(\phi\right)}
+\left(\frac{\delta\varepsilon}{\delta\Gamma\left(\phi\right)}\right)^{2}
\right)
e^{-(\Gamma-\delta\Gamma\left(\phi\right))t}\right.
\nonumber\\&\left.
+\frac{1}{2}\left(1+\frac{\left(\Gamma^{\alpha}+\Gamma^{\bar{\alpha}}\cos\phi\right)}{\delta\Gamma\left(\phi\right)}
+\left(\frac{\delta\varepsilon}{\delta\Gamma\left(\phi\right)}\right)^{2}
\right)
e^{-(\Gamma+\delta\Gamma\left(\phi\right))t}\right.
\nonumber\\&\left.
-\left(\frac{\delta\varepsilon}{\delta\Gamma\left(\phi\right)}\right)^{2}e^{-\Gamma t}
\right\},
\end{align} where $\delta\Gamma\left(\phi\right)=\sqrt{\vert\Gamma_{12}(\phi)\vert^{2}-\delta\varepsilon^{2}}>0$ (see Eqs.~(\ref{decay-rates}) and (\ref{cmpx-decayrate})).  In Eq.~(\ref{I1_udwnsym_nondegLS}) in addition to the two decay terms, $e^{-(\Gamma\pm\delta\Gamma\left(\phi\right))t}$, another exponential decay term $e^{-\Gamma t}$ appears. This latter term has its amplitude determined by the non-degeneracy of the QDs' on-site energies. Setting $\delta\varepsilon=0$ in Eq.~(\ref{I1_udwnsym_nondegLS}) reduces it to Eq.~(\ref{I1_udwnsym_deg}). In the degenerate case we have learned that the choice of $\Gamma_{R}\gg\Gamma_{L}$ makes the amplitude in $I^{[1]}_{R}(t,\phi)$ for the slow decay term $e^{-(\Gamma-\delta\Gamma\left(\phi\right))t}$ approach zero. However, when $\delta\varepsilon$ is finite, all the amplitudes in Eq.~(\ref{I1_udwnsym_nondegLS}) are not negligible. This explains why the left-right asymmetry in time scales for reaching the steady-state limits shown in Fig.~\ref{crnts_S49-even} is not as drastic as that shown in Fig.~\ref{crnts_S38}. The left-right asymmetry in the magnitudes of the currents for $\delta\varepsilon\ne0$ is still clearly seen (compare the magnitudes of Fig.~\ref{crnts_S49-even}(a) with those in Fig.~\ref{crnts_S49-even}(b) respectively). This, again, reflects the overall factor $\Gamma^{\alpha}$ in Eq.~(\ref{I1_udwnsym_nondegLS}). The more visible flux dependence in Fig.~\ref{crnts_S49-even}(a), in contrast to the less visible flux dependence in Fig.~\ref{crnts_S49-even}(b), can be explained by the same reason applied to Fig.~\ref{crnts_S38} (see Appendix \ref{appx-AsymLR-deg}). Letting $\delta\varepsilon\ne0$ also raises the rate of the slow decay term by varying the flux (compare the line for $\Gamma-\delta\Gamma(\phi)$ in Fig.~\ref{crnts_S49-even}(c) to that in Fig.~\ref{crnts_S38}(c)). The time required to reach the steady-state limit is thus reduced (compare the time scales in Fig.~\ref{crnts_S49-even}(a) to that in Fig.~\ref{crnts_S38}(a1)).





$b.$ \emph{Even components for} $\delta\varepsilon\ne0$ \emph{with} $\vert\Gamma_{12}(\phi)\vert^{2}<\delta\varepsilon^{2}$
\\
\\
The results of further increasing $\delta\varepsilon$ such that $\vert\Gamma_{12}(\phi)\vert^{2}<\delta\varepsilon^{2}$ are demonstrated in Fig.~\ref{crnts_S50-even}, still with $\Gamma^{R}\gg\Gamma^{L}$. Similar to Fig.~\ref{crnts_S49-even}, increasing $\delta\varepsilon$ decreases the asymmetry between the time scales while the asymmetry in the magnitudes of the left and the right currents are unaffected. For $\vert\Gamma_{12}(\phi)\vert^{2}>\delta\varepsilon^{2}$, Figs.~\ref{crnts_S38} and \ref{crnts_S49-even} show only smooth time evolutions of $I^{[1],+}_{\alpha}(t,\phi)$, described by exponential decays. However, for $\vert\Gamma_{12}(\phi)\vert^{2}<\delta\varepsilon^{2}$, a step-like feature in the time evolutions of $I^{[1],+}_{L}(t,\phi)$ is observed in Fig.~\ref{crnts_S50-even}(a) for $\phi=0$ (see the red solid line) and for $\phi=\pi$ (see the green dashed-dotted line). Asymmetrically for $I^{[1],+}_{R}(t,\phi)$ in Fig.~\ref{crnts_S50-even}(b), this feature exists for all the tested fluxes. Indeed, when $\vert\Gamma_{12}(\phi)\vert^{2}<\delta\varepsilon^{2}$ is obeyed, the even components of the initial-occupation-induced current under Eq.~(\ref{updwn_bdsym}) reads (using the same approach of deriving Eq.~(\ref{I1_udwnsym_nondegLS}) with the condition $\vert\Gamma_{12}(\phi)\vert^{2}<\delta\varepsilon^{2}$)
\begin{align}
\label{I1_udwnsym_nondegGT}
&I^{[1],+}_{\alpha}(t,\phi)
\nonumber\\&
=-\Gamma^{\alpha}e^{-\Gamma t}\left\{
\left[1-\left(\frac{\delta\varepsilon}{\varepsilon_{g}\left(\phi\right)}\right)^{2}\right]\cos\left(\varepsilon_{g}\left(\phi\right)t\right)
\right.\nonumber\\&
-\frac{\Gamma^{\alpha}+\Gamma^{\bar{\alpha}}\cos\phi}{\varepsilon_{g}\left(\phi\right)}
\sin\left(\varepsilon_{g}\left(\phi\right)t\right)
\left.+\left(\frac{\delta\varepsilon}{\varepsilon_{g}\left(\phi\right)}\right)^{2}\right\},
\end{align} where $\varepsilon_{g}\left(\phi\right)=\sqrt{\delta\varepsilon^{2}-\vert\Gamma_{12}(\phi)\vert^{2}}>0$. The result of Eq.~(\ref{I1_udwnsym_nondegGT}) is very different from Eq.~(\ref{I1_udwnsym_nondegLS}). In addition to the exponential decay factor $e^{-\Gamma t}$, time-oscillatory terms, $\cos\left(\varepsilon_{g}\left(\phi\right)t\right)$ and $\sin\left(\varepsilon_{g}\left(\phi\right)t\right)$, also appear. The step-like feature in the time evolution is a result of the combined effect of the exponential decay and the oscillatory terms. The asymmetry between the left and the right currents in showing up the step-like feature in different fluxes is explained by applying $\Gamma^{R}\gg\Gamma^{L}$ to Eq.~(\ref{I1_udwnsym_nondegGT}) (see Appendix \ref{appx-AsymLR-nondeg}).
\\



$c.$ \emph{The odd components of the initial-occupation-induced currents}
\\
\\
The odd components of the transient currents under the same set of parameters used in Fig.~\ref{crnts_S49-even} are displayed in Fig.~\ref{crnts_S49_50-odd}(a) for $\vert\Gamma_{12}(\phi)\vert^{2}>\delta\varepsilon^{2}$. The parameters discussed in Fig.~\ref{crnts_S50-even} are shown in Fig.~\ref{crnts_S49_50-odd}(b) for $\vert\Gamma_{12}(\phi)\vert^{2}<\delta\varepsilon^{2}$. The numerical calculations in Fig.~\ref{crnts_S49_50-odd} verify the odd components obey Eq.~(\ref{phi-LR_asym}). The effects of the splitting between the two QDs' on-site energies are well contrasted by comparing Fig.~\ref{crnts_S49_50-odd}(a) with Fig.~\ref{crnts_S49_50-odd}(b). The time evolution of the odd component of the initial-occupation-induced current under Eq.~(\ref{updwn_bdsym}) and $\vert\Gamma_{12}(\phi)\vert^{2}>\delta\varepsilon^{2}$ is explicitly given by
\begin{align}
\label{I1_udwnsym_odd}
I^{[1],-}_{\alpha}(t,\phi)
=-\zeta_{\alpha}\frac{\Gamma^{L}\Gamma^{R}}{\delta\Gamma^{2}(\phi)}\delta\varepsilon\sin(\phi)(\cosh(\delta\Gamma(\phi)t)-1)
e^{-\Gamma t}.
\end{align} Its time dependence is thus governed by the three exponential decay factors $e^{-(\Gamma\pm\delta\Gamma\left(\phi\right))t}$ and $e^{-\Gamma t}$ and the relative ratios between the amplitudes for these factors are independent of either $\phi$ or $\alpha$. On the other hand, the oscillatory behaviors are observed in Fig.~\ref{crnts_S49_50-odd}(b), corresponding to,
\begin{align}
\label{I1GT_udwnsym_odd}
I^{[1],-}_{\alpha}(t,\phi)
=-\zeta_{\alpha}\frac{\Gamma^{L}\Gamma^{R}}{\varepsilon_{g}^{2}(\phi)}\delta\varepsilon\sin(\phi)(\cos(\varepsilon_{g}(\phi)t)-1)
e^{-\Gamma t},
\end{align} for $\vert\Gamma_{12}(\phi)\vert^{2}<\delta\varepsilon^{2}$. The difference between Fig.~\ref{crnts_S49_50-odd}(a) and Fig.~\ref{crnts_S49_50-odd}(b) in the times of approaching steady-state limit can also be read from Eq.~(\ref{I1_udwnsym_odd}) and Eq.~(\ref{I1GT_udwnsym_odd}). When $\vert\Gamma_{12}(\phi)\vert^{2}>\delta\varepsilon^{2}$, the current is dominated by the slow decay $e^{-(\Gamma-\delta\Gamma\left(\phi\right))t}$ in Eq.~(\ref{I1_udwnsym_odd}). However, when $\vert\Gamma_{12}(\phi)\vert^{2}<\delta\varepsilon^{2}$ the currents exhibit a single decay factor $e^{-\Gamma t}$, which decays faster than $e^{-(\Gamma-\delta\Gamma\left(\phi\right))t}$ [Eq.~(\ref{I1GT_udwnsym_odd})]. Therefore the decay times exhibited in Fig.~\ref{crnts_S49_50-odd}(a) are longer than those found in Fig.~\ref{crnts_S49_50-odd}(b).

\subsection{left-right symmetry $\bar{\Gamma}^{L}_{ii}=\bar{\Gamma}^{R}_{ii}$}
\label{sec_model_fluxindep_sym2}

When the bonds obey left-right symmetry $\bar{\Gamma}^{L}_{ii}=\bar{\Gamma}^{R}_{ii}$, then the even components follow the left-right symmetry, namely, $I^{[i],+}_{L}(t,\phi)=I^{[i],+}_{R}(t,\phi)$ (see later discussions around Eqs.~(\ref{I_lrsym_gel_deg}) and (\ref{I_lrsym_gel_even})) and $\overline{I}^{\text{em.,+}}_{L }(t,\phi)=\overline{I}^{\text{em.,+}}_{R }(t,\phi)$ (see Appendix \ref{appx-Tplusplus} for a proof without assuming up-down symmetry by Eq.~(\ref{updwn_bdsym}) or Eq.~(\ref{degDQD})). Note that the relation between the odd components of the left and the right currents is subjected to Eq.~(\ref{phi-LR_asym}) for whatever setting of the bonds.

Below we show that the up-down asymmetry between $I^{[1]}_{\alpha}(t,\phi)$ and $I^{[2]}_{\alpha}(t,\phi)$, in terms of their time scales for approaching the steady-state limit, can be enhanced by the asymmetry $\bar{\Gamma}_{1}\ll\bar{\Gamma}_{2}$ for both degenerate $\delta\varepsilon=0$ and non-degenerate $\delta\varepsilon\ne0$ cases.

\subsubsection{degenerate QDs}
\label{sec_model_fluxindep_sym2-deg}
\begin{figure}[h]
\includegraphics[width=8.0cm,height=9cm]{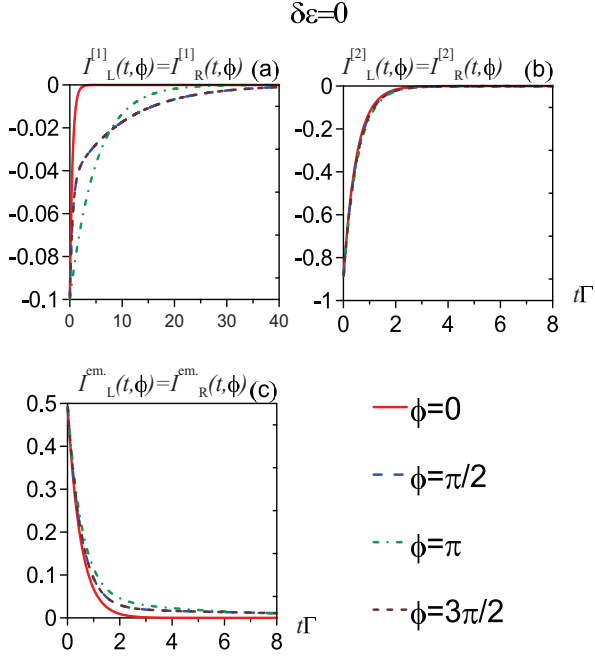} \caption{(color online) Various transient components of the currents with left-right symmetry, $\bar{\Gamma}^{L}_{11}=\bar{\Gamma}^{R}_{11}=0.1\Gamma$ and $\bar{\Gamma}^{L}_{22}=\bar{\Gamma}^{R}_{22}=0.9\Gamma$ at degeneracy $\delta\varepsilon=0$. Different line styles are for different fluxes as indicated on the right of plot (c).
} \label{crnts_S37}
\end{figure}

In Fig.~\ref{crnts_S37} we plot the time evolutions of the purely transient components of the currents under $\bar{\Gamma}^{L}_{ii}=\bar{\Gamma}^{R}_{ii}$ with $\delta\varepsilon=0$.
For an initial charge in QD1 (or QD2), the immediate out-tunneling is taken through the upper bonds (or the lower bonds). Therefore, when up-down asymmetry is present in the bonds, the times for $I^{[1]}_{\alpha}(t,\phi)$ and $I^{[2]}_{\alpha}(t,\phi)$ to reach the steady-state limit as well as the magnitudes of these currents would also show such asymmetry. From Fig.~\ref{crnts_S37}, we see that the up-down asymmetry of the bonds is clearly revealed in the difference between $I^{[1]}_{\alpha}(t,\phi)$ and $I^{[2]}_{\alpha}(t,\phi)$ (compare Fig.~\ref{crnts_S37}(a) with Fig.~\ref{crnts_S37}(b) both in time scales and magnitudes).

Setting $\bar{\Gamma}^{L}_{ii}=\bar{\Gamma}^{R}_{ii}=\bar{\Gamma}^{}_{i}$ and $\delta\varepsilon=0$ [Eq.~(\ref{degDQD})] in Eq.~(\ref{retardedG-intime}) and substituting it further into Eq.~(\ref{crnt-init-diagocc-WB}) yield
\begin{align}
\label{I_lrsym_gel_deg}
I_{\alpha}^{[i]}\left(t,\phi\right) & =I_{\text{f}}^{\left[i\right]}\left(\phi\right)e^{-\left(\Gamma+\delta\Gamma\left(\phi\right)\right)t}+I_{\text{s}}^{\left[i\right]}\left(\phi\right)e^{-\left(\Gamma-\delta\Gamma\left(\phi\right)\right)t}\nonumber \\
 & +I_{0}^{\left[i\right]}\left(\phi\right)e^{-\Gamma t},
\end{align} for $i=1,2$ and here $\delta\Gamma\left(\phi\right)=\sqrt{\vert\Gamma_{12}(\phi)\vert^{2}+\Gamma_{d}^{2}}$, where $\Gamma^{}_{d}$ is defined in Eq.~(\ref{GmSD}). The subscript f in $I_{\text{f}}^{\left[i\right]}\left(\phi\right)$
refers to {}``fast'' decay term and conversely the subscript
s in $I_{\text{s}}^{\left[i\right]}\left(\phi\right)$ refers to {}``slow''
decay term. The explicit expressions of the amplitudes in Eq.~(\ref{I_lrsym_gel_deg}) are given by Eq.~(\ref{amp_I_lrsym_deg}) in Appendix \ref{appx-AsymUDW-deg}. These amplitudes do not depend on $\alpha$ and consequently $I^{[i]}_{L}(t,\phi)=I^{[i]}_{R}(t,\phi)$. Similar to Eq.~(\ref{I1_udwnsym_nondegLS}), Eq.~(\ref{I_lrsym_gel_deg}) also shows the three decay factors, $e^{-\left(\Gamma\pm\delta\Gamma\left(\phi\right)\right)t}$, and $e^{-\Gamma t}$, commonly possed by $I^{[1]}_{\alpha}(t,\phi)$ and $I^{[2]}_{\alpha}(t,\phi)$. With large asymmetry between the upper and the lower bonds, $\bar{\Gamma}_{1}\ll\bar{\Gamma}_{2}$, $I^{[1]}_{\alpha}(t,\phi)$ is governed by the amplitude for the slow decay term while the amplitudes for different decay factors are comparable in $I^{[2]}_{\alpha}(t,\phi)$ (see Appendix \ref{appx-AsymUDW-deg} for more details).

In the present case with $\delta\varepsilon=0$ and $\mu^{}_{0}=\varepsilon^{}_{0}$, we have $\overline{I}^{\text{em.}}_{\alpha }(t,\phi)=-[I^{[1]}_{\alpha}(t,\phi)+I^{[2]}_{\alpha}(t,\phi)]/2$ (see more detailed discussions in Sec.~\ref{sec_model_fluxindep_sym1}). The up-down asymmetry in the bonds is also revealed by comparing Fig.~\ref{crnts_S37}(c) with Fig.~\ref{crnts_S37}(a) and Fig.~\ref{crnts_S37}(b). Since $\left\vert I^{[1]}_{\alpha}(t)\right\vert\ll\left\vert I^{[2]}_{\alpha}(t)\right\vert$, we see that $\overline{I}^{\text{em.}}_{\alpha }(t)$ and $-I^{[2]}_{\alpha}(t)/2$ have about the same magnitudes.
\subsubsection{non-degenerate QDs}
\label{sec_model_fluxindep_sym2-nondeg}

\begin{figure}[h]
\includegraphics[width=8.5cm,height=8.5cm]{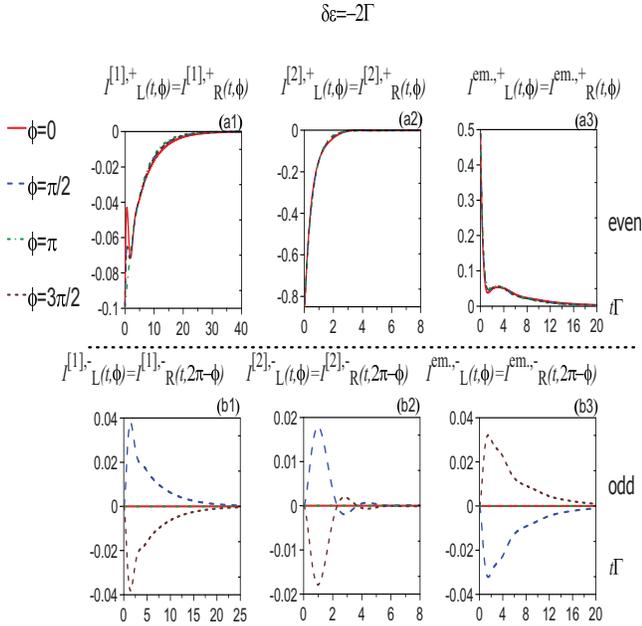} \caption{(color online) Various components of the transient currents with left-right symmetry, $\bar{\Gamma}^{L}_{11}=\bar{\Gamma}^{R}_{11}=0.1\Gamma$ and $\bar{\Gamma}^{L}_{22}=\bar{\Gamma}^{R}_{22}=0.9\Gamma$ at non-degeneracy $\delta\varepsilon=-2\Gamma$. Different line styles are for different fluxes as indicated on the left. The parts that are even in the flux are shown in plots (a1),(a2),(a3) and the corresponding odd parts are shown in plots (b1),(b2) and (b3).
} \label{crnts_S5-even-odd}
\end{figure}

The time evolutions of the purely transient components of the currents under left-right symmetry $\bar{\Gamma}^{L}_{ii}=\bar{\Gamma}^{R}_{ii}$ with non-degeneracy $\delta\varepsilon\ne0$ are demonstrated in Fig.~\ref{crnts_S5-even-odd}.
With $\delta\varepsilon\ne0$, the up-down asymmetry in terms of the difference between the times needed by $I^{[1]}_{\alpha}(t,\phi)$ and $I^{[2]}_{\alpha}(t,\phi)$ to reach the steady-state limit is still obvious (comparing the time scales in Fig.~\ref{crnts_S5-even-odd}(a1),(b1) with those in Fig.~\ref{crnts_S5-even-odd}(a2),(b2)). This is unlike the cases discussed in Sec.~\ref{sec_model_transit_numcs-1}. In this former discussion, letting $\delta\varepsilon\ne0$ has reduced the difference in the times to reach the steady-state limit between $I^{[1]}_{L}(t,\phi)$ and $I^{[1]}_{R}(t,\phi)$ induced by the asymmetry in the bonds ${\Gamma}^{L}\ll{\Gamma}^{R}$. The asymmetry between $I^{[1]}_{\alpha}(t,\phi)$ and $I^{[2]}_{\alpha}(t,\phi)$ due to $\bar{\Gamma}^{}_{1}\ll\bar{\Gamma}^{}_{2}$ is not reduced by letting $\delta\varepsilon\ne0$ as shown by Fig.~\ref{crnts_S5-even-odd}.

Setting $\bar{\Gamma}^{L}_{ii}=\bar{\Gamma}^{R}_{ii}=\bar{\Gamma}^{}_{i}$ in Eq.~(\ref{retardedG-intime}) and substituting it into Eq.~(\ref{crnt-init-diagocc-WB}) yield
\begin{align}
\label{I_lrsym_gel_even}
&I_{\alpha}^{[i],+}\left(t,\phi\right)  =
\nonumber \\&
I_{\text{f}}^{\left[i\right]}\left(\phi\right)e^{-\left(\Gamma+\delta\Gamma\left(\phi\right)\right)t}
+I_{\text{s}}^{\left[i\right]}\left(\phi\right)e^{-\left(\Gamma-\delta\Gamma\left(\phi\right)\right)t}
+
\nonumber \\
& I_{\text{osc}}^{\left[i\right]}\left(\phi\right)e^{-\Gamma t}\cos\left(\varepsilon_{g}\left(\phi\right)t\right)+I_{\text{oss}}^{\left[i\right]}\left(\phi\right)e^{-\Gamma t}\sin\left(\varepsilon_{g}\left(\phi\right)t\right),
\end{align}
for $i=1,2$ in which the superscript $+$ denote the component that
is even in the flux $\phi$. The subscripts osc and oss in $I_{\text{osc}}^{\left[i\right]}\left(\phi\right)$
and $I_{\text{oss}}^{\left[i\right]}\left(\phi\right)$ stand for
oscillation as cosine and oscillation as sine respectively. The details of the amplitudes in Eq.~(\ref{I_lrsym_gel_even}) are found in Eq.~(\ref{amp_lrsym_gel_even-nondeg}) and setting $\delta\varepsilon=0$ reduces Eq.~(\ref{I_lrsym_gel_even}) to Eq.~(\ref{I_lrsym_gel_deg}). Analysis of these amplitudes shows that $\delta\varepsilon\ne0$ does not affect the dominance of the slow decay term in $I^{[1]}_{\alpha}(t,\phi)$ caused by $\bar{\Gamma}_{1}\ll\bar{\Gamma}_{2}$ (see the first paragraph after Eq.~(\ref{I1_nondeg_lrsym_lgupdwasym})). Similar to the discussions in Sec.~\ref{sec_model_transit_numcs-1}, the nonzero $\delta\varepsilon$ gives rise to oscillatory contributions as the last two terms of Eq.~(\ref{I_lrsym_gel_even}). More interestingly, the large asymmetry between the upper and the lower bonds can also render an asymmetry in the manifestation of these oscillatory terms. Figure.~\ref{crnts_S5-even-odd}(a1) for $I_{\alpha}^{[1],+}\left(t,\phi\right)$ displays spike features that can only arise from the contributions of oscillatory terms. However, in Fig.~\ref{crnts_S5-even-odd}(a2) for $I_{\alpha}^{[2],+}\left(t,\phi\right)$ this time-dependent feature is suppressed, showing only smooth monotonic time evolutions. Detailed explanations of how the amplitudes for the oscillatory terms are affected by the asymmetry $\bar{\Gamma}_{1}\ll\bar{\Gamma}_{2}$ is found in the second paragraph after Eq.~(\ref{I1_nondeg_lrsym_lgupdwasym}) in Appendix \ref{appx-AsymUDW-nondeg}.



The odd components of the initial-occupation-induced currents are given by
\begin{align}
\label{I_lrsym_gel_odd}
&I_{\alpha}^{[1/2],-}\left(t,\phi\right)=
\pm\frac{\bar{\Gamma}_{1}\bar{\Gamma}_{2}}{\left\vert \Gamma_{g}\left(\phi\right)\right\vert ^{2}}
\zeta_{\alpha}\sin\phi\times \nonumber
\\
& \left\{ -\frac{\varepsilon_{g}\left(\phi\right)\pm\delta\varepsilon}{2}e^{-\left(\Gamma+\delta\Gamma\left(\phi\right)\right)t}
+\frac{\varepsilon_{g}\left(\phi\right)\mp\delta\varepsilon}{2}e^{-\left(\Gamma-\delta\Gamma\left(\phi\right)\right)t}\right.
\nonumber
\\
& +e^{-\Gamma t}\left[\pm\delta\varepsilon\cos\left(\varepsilon_{g}\left(\phi\right)t\right)
-\delta\Gamma\left(\phi\right)\sin\left(\varepsilon_{g}\left(\phi\right)t\right)\right]
\Bigg\}
,
\end{align}
where the upper sign is for $I_{\alpha}^{[1],-}\left(t,\phi\right)$
and the lower sign is for $I_{\alpha}^{[2],-}\left(t,\phi\right)$. The asymmetry between the odd components, $I_{\alpha}^{[1],-}\left(t,\phi\right)$ and $I_{\alpha}^{[2],-}\left(t,\phi\right)$, is seen by comparing Fig.~\ref{crnts_S5-even-odd}(b1) with Fig.~\ref{crnts_S5-even-odd}(b2). Consistent with the even part, Fig.~\ref{crnts_S5-even-odd}(b1) shows longer times to reach the steady-state limit for $I_{\alpha}^{[1],-}\left(t,\phi\right)$ while Fig.~\ref{crnts_S5-even-odd}(b2) shows much shorter times to reach the steady-state limit for $I_{\alpha}^{[2],-}\left(t,\phi\right)$. Moreover, the oscillatory behavior in Fig.~\ref{crnts_S5-even-odd}(b1) for $I_{\alpha}^{[1],-}\left(t,\phi\right)$ is suppressed while in Fig.~\ref{crnts_S5-even-odd}(b2) clear oscillatory time evolutions are displayed for $I_{\alpha}^{[2],-}\left(t,\phi\right)$. How the asymmetry between the upper and the lower bonds renders asymmetries in the amplitudes between $I_{\alpha}^{[1],-}\left(t,\phi\right)$ and $I_{\alpha}^{[2],-}\left(t,\phi\right)$ is explained in Appendix \ref{appx-AsymUDW-oddnondeg}.


The time evolutions for $\overline{I}^{\text{em.},+}_{\alpha }(t,\phi)$ and $\overline{I}^{\text{em.},-}_{\alpha }(t,\phi)$ are displayed in Fig.~\ref{crnts_S5-even-odd}(a3) and Fig.~\ref{crnts_S5-even-odd}(b3) respectively. Without any up-down symmetry, $\bar{\Gamma}^{\alpha}_{11}\ne\bar{\Gamma}^{\alpha}_{22}$ and $\delta\varepsilon\ne0$, as we will discuss later in the following subsection, $\overline{I}^{\text{em.}}_{\alpha }(t,\phi)$ is not definitely related to $I_{\alpha}^{[1]}\left(t,\phi\right)$ and $I_{\alpha}^{[2]}\left(t,\phi\right)$.  Figures~\ref{crnts_S5-even-odd}(a1),(a2) and (a3) show that the sign of $\overline{I}^{\text{em.},+}_{\alpha }(t,\phi)$ is positive while the signs of $I_{\alpha}^{[1],+}\left(t,\phi\right)$ and $I_{\alpha}^{[2],+}\left(t,\phi\right)$ are negative. The intuitive interpretation of $I^{[i]}_{\alpha}(t,\phi)$ and $\overline{I}^{\text{em.}}_{\alpha}(t,\phi)$ as out-tunneling and in-tunneling currents is thus witnessed also for up-down asymmetry with $\bar{\Gamma}^{\alpha}_{11}\ne\bar{\Gamma}^{\alpha}_{22}$ and $\delta\varepsilon\ne0$ by the even components. For the odd components, Fig.~\ref{crnts_S5-even-odd}(b1),(b2) and (b3) show that the sign of $\overline{I}^{\text{em.},-}_{\alpha }(t,\phi)$ is opposite to that of $I_{\alpha}^{[1],\pm}\left(t,\phi\right)$ and $I_{\alpha}^{[2],\pm}\left(t,\phi\right)$. As revealed by Eqs.~(\ref{odd-I1}),(\ref{odd-I2}) and (\ref{odd-Tpm}), the signs of the odd components are largely determined by the direction of the flux, manifested by the overall common factor $\sin\phi$. Further comparison of the even and the odd components in terms of the relation between $\overline{I}^{\text{em.}}_{\alpha }(t,\phi)$ and $I_{\alpha}^{[i]}\left(t,\phi\right)$'s is discussed later.



\subsection{Symmetry of the energy level distribution}
\label{sec_model_fluxindep_sym1}





 The symmetry of the energy level distribution is characterized by the positions of the chemical potentials relative to the on-site energies of the QDs. Here we show that this symmetry, combined with the geometric up-down symmetry, provides a link between  $\overline{I}^{\text{em.}}_{\alpha }(t,\phi)$ and $I^{[i]}_{\alpha}(t,\phi)$'s. We further show that such a connection can be used to reveal the different mechanisms underlying the even and the odd components of these transient currents.

By Eq.~(\ref{dynamic-crnt-0-0}), we see that both the geometry of the system and the energy configuration involving the chemical potentials determine the transient current, $\overline{I}^{\text{em.}}_{\alpha }(t,\phi)$, obtained under initial empty QDs. These two factors enter $\overline{I}^{\text{em.}}_{\alpha }(t,\phi)$ via the two distinct terms $\mathcal{T}^{(+)}_{\alpha}(t,\omega,\phi)$ and the fermi functions in Eq.~(\ref{dynamic-crnt-0-0}). Note that the transmission-like function, $\mathcal{T}^{(+)}_{\alpha}(t,\omega,\phi)$, given in Eq.~(\ref{intftrsm-1}), does not depend on the chemical potentials.  It is the fermi function in Eq.~(\ref{dynamic-crnt-0-0}) that makes $\overline{I}^{\text{em.}}_{\alpha }(t,\phi)$ depend on the chemical potentials of the reservoirs. The role played by the symmetry of the energy level distribution in $\overline{I}^{\text{em.}}_{\alpha }(t,\phi)$ can be manifested when we have geometric up-down symmetry, realized by either Eq.~(\ref{updwn_bdsym}) or Eq.~(\ref{degDQD}), leading to (see Appendix \ref{appx-Tplusplus} for a detailed proof)
\begin{align}
\label{sym_trsmP}
\mathcal{T}^{(+)}_{\alpha}(t,\varepsilon_{0}+\omega,\phi)=\mathcal{T}^{(+)}_{\alpha}(t,\varepsilon_{0}-\omega,\phi).
\end{align}  If further the chemical potentials (or the on-site energies of the QDs) obey the symmetry,
\begin{align}
\label{mu0eqep0}
\mu^{}_{0}=\varepsilon^{}_{0},
\end{align} (allowing also $\mu^{}_{L}\ne\mu^{}_{R}$), then
the integral in Eq.~(\ref{dynamic-crnt-0-0}) with the use of Eq.~(\ref{sym_trsmP}) becomes (see Appendix \ref{appx-IinIoutrel} for detailed derivation)
\begin{align}
\label{Iem_WBsim}
&\overline{I}^{\text{em.}}_{\alpha }(t,\phi)=\frac{1}{2}\int_{-\infty}^{\infty}\!\!\frac{d\omega}{2\pi}
\mathcal{T}^{(+)}_{\alpha}(t,\omega,\phi).
\end{align} Following from Eq.~(\ref{Iem_WBsim}), an equality between $\overline{I}^{\text{em.}}_{\alpha }(t,\phi)$, $I^{[1]}_{\alpha}(t,\phi)$ and $I^{[2]}_{\alpha}(t,\phi)$ can be established (see Appendix \ref{appx-IinIoutrel} for details):
\begin{align}
\label{p-h_sym_flux-indep}
\overline{I}^{\text{em.}}_{\alpha }(t,\phi)=-\frac{1}{2}I^{[1+2]}_{\alpha}(t,\phi).
\end{align} This result Eq.~(\ref{p-h_sym_flux-indep}) holds for arbitrary fluxes under the symmetry of the energy level configuration, Eq.~(\ref{mu0eqep0}), and the existence of symmetry between the upper and the lower paths, given by Eq.~(\ref{updwn_bdsym}) or Eq.~(\ref{degDQD}). We have mentioned the numerical observation of Eq.~(\ref{p-h_sym_flux-indep}) in Sec.~\ref{sec_model_transit_numcs-1} and Sec.~\ref{sec_model_fluxindep_sym2-deg} at zero bias. In what follows we proceed to discuss the deviations from Eq.~(\ref{mu0eqep0}) and the case of $\mu^{}_{L}\ne\mu^{}_{R}$.

\subsubsection{zero bias $\mu^{}_{L}=\mu^{}_{R}$}
\label{sec_model_fluxindep_sym1_zrbs}

\begin{figure}[h]
\includegraphics[width=8.4cm,height=12cm]{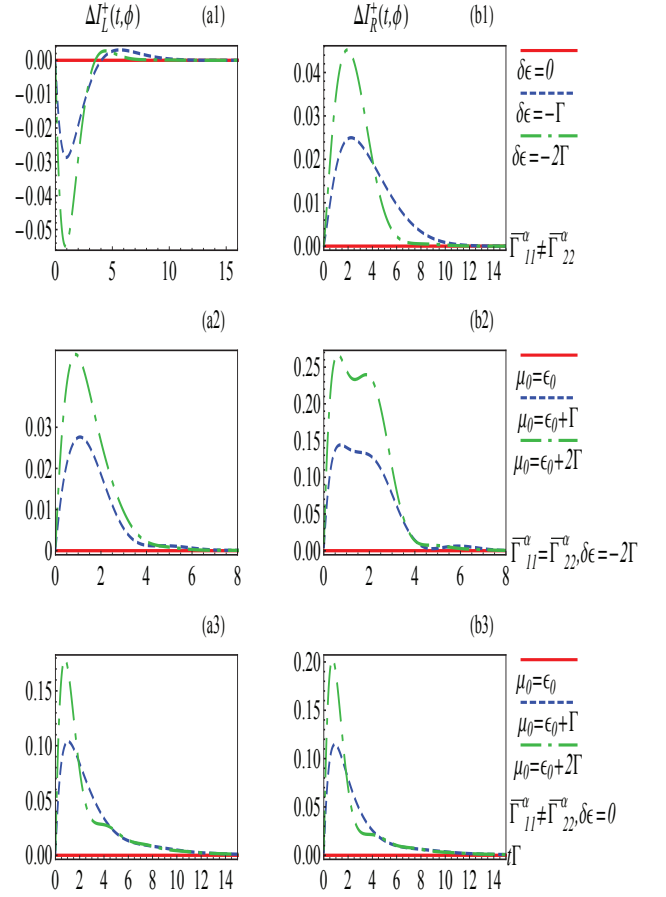} \caption{(color online) The difference between in-tunneling and out-tunneling currents, quantified by the even component of Eq.~(\ref{Del_I_alpha}), $\Delta{I}^{+}_{\alpha}(t,\phi)$. Here we take $\phi=\pi/2$.  In plots (a1),(b1), the bonds are $\bar{\Gamma}^{L}_{11}=0.2$, $\bar{\Gamma}^{R}_{11}=0.6$, $\bar{\Gamma}^{L}_{22}=0.7$, $\bar{\Gamma}^{R}_{22}=0.5$. Different line styles there correspond to different $\delta\varepsilon$ as shown on the right of (b1). In plots (a2,b2), the bonds are $\bar{\Gamma}^{L}_{11}=\bar{\Gamma}^{L}_{22}=0.1\Gamma$ and $\bar{\Gamma}^{R}_{11}=\bar{\Gamma}^{R}_{22}=0.9\Gamma$ and and the on-site energies are $\delta\varepsilon=-2\Gamma$. Different line styles there correspond to different positions of $\mu_{0}$ as indicated to the right of (b2). In plots (a3,b3), we set degenerate QDs with the setting of unequal bonds used in (a1,b1) for all the curves. Different line styles here are for different positions of $\mu_{0}$ as shown on the right of (b3).
} \label{crnts_S6_7_8_3_11_12_6_9_10-even}
\end{figure}

\begin{figure}[h]
\includegraphics[width=8.4cm,height=8cm]{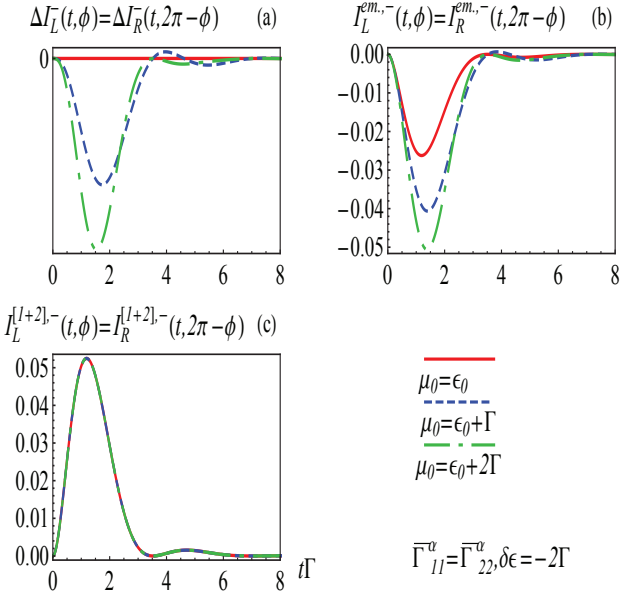} \caption{(color online) The odd components of $\Delta{I}_{\alpha}(t,\phi)$ (plot (a)), and its composition currents, $\overline{I}^{\text{em.}}_{\alpha }(t,\phi)$ (plot (b)) and $I^{[1+2]}_{\alpha}(t,\phi)$ (plot (c)), at $\phi=\pi/2$ with other parameters used in Fig.~\ref{crnts_S6_7_8_3_11_12_6_9_10-even}(a2),(b2).
} \label{crnts_S6_7_8_3_11_12_6_9_10-odd}
\end{figure}

Note that Eq.~(\ref{p-h_sym_flux-indep}) holds for both of the cases $\mu^{}_{L}=\mu^{}_{R}$ and $\mu^{}_{L}\ne\mu^{}_{R}$. At a zero bias $\overline{I}^{\text{em.}}_{\alpha }(t,\phi)=I_{\alpha}(t,\phi)$ and the observed currents are purely transient. For clarity, we first numerically demonstrate Eq.~(\ref{p-h_sym_flux-indep}) with $\mu^{}_{L}=\mu^{}_{R}=\mu^{}_{0}$ in Figs.~\ref{crnts_S6_7_8_3_11_12_6_9_10-even} and \ref{crnts_S6_7_8_3_11_12_6_9_10-odd} for the even and the odd components respectively. We define
\begin{align}
\label{Del_I_alpha}
\Delta{I}_{\alpha}(t,\phi)=\overline{I}^{\text{em.}}_{\alpha }(t,\phi)-\left[-\frac{1}{2}I^{[1+2]}_{\alpha}(t,\phi)\right],
\end{align} to quantify the deviation from Eq.~(\ref{p-h_sym_flux-indep}), due to asymmetry in geometry or energy level distribution.  The roles played by the up-down symmetry in terms of the energy splitting between the upper and the lower QDs is demonstrated in Fig.~\ref{crnts_S6_7_8_3_11_12_6_9_10-even}(a1),(b1). The importance of the symmetry of the energy level distributions, characterized by the deviation of $\mu_{0}$ from $\varepsilon_{0}$, is exemplified in Fig.~\ref{crnts_S6_7_8_3_11_12_6_9_10-even}(a2),(b2) for the bonds exhibiting up-down symmetry and in Fig.~\ref{crnts_S6_7_8_3_11_12_6_9_10-even}(a3),(b3) with degenerate QDs. In the calculations for Fig.~\ref{crnts_S6_7_8_3_11_12_6_9_10-even}, we have verified that $\overline{I}^{\text{em.},+}_{\alpha }(t,\phi)>0$ while $I^{[1+2],+}_{\alpha}(t,\phi)<0$, ensuring the identification of the former as the in-tunneling and of the latter as the out-tunneling currents.
The results in Fig.~\ref{crnts_S6_7_8_3_11_12_6_9_10-even}(a1),(b1) show that as long as $\delta\varepsilon=0$, regardless of the settings of the bonds, the in-tunneling and the out-tunneling currents are equally strong (see the red solid lines which remain at zero all the time in Fig.~\ref{crnts_S6_7_8_3_11_12_6_9_10-even}(a1),(b1)) given that $\mu_{0}=\varepsilon_{0}$. In both Fig.~\ref{crnts_S6_7_8_3_11_12_6_9_10-even}(a2),(b2) and Fig.~\ref{crnts_S6_7_8_3_11_12_6_9_10-even}(a3),(b3) the relative position of $\mu_{0}$ to $\varepsilon_{0}$ has been varied. We show that only when there exists a common energy symmetric point for states in the reservoirs and those in the QDs can the in-tunneling and the out-tunneling dynamics be symmetric to each other  (see the non-vanishing dashed lines in Fig.~\ref{crnts_S6_7_8_3_11_12_6_9_10-even}(a2),(b2),(a3),(b3) in contrast to the red solid lines). Note that although in Fig.~\ref{crnts_S6_7_8_3_11_12_6_9_10-even} we have only shown the results for one value of the flux, the above conclusion holds for all fluxes.

The corresponding odd components of Eq.~(\ref{Del_I_alpha}), along with the underlying currents, $\overline{I}^{\text{em.}}_{\alpha }(t,\phi)$ and $I^{[1+2]}_{\alpha}(t,\phi)$, are shown in Fig.~\ref{crnts_S6_7_8_3_11_12_6_9_10-odd}. By Fig.~\ref{crnts_S6_7_8_3_11_12_6_9_10-odd}(a), we show for the odd components how the deviation from the symmetry of Eq.~(\ref{mu0eqep0}) deviates $\overline{I}^{\text{em.}}_{\alpha }(t,\phi)$ from $-I^{[1+2]}_{\alpha}(t,\phi)/2$. In Fig.~\ref{crnts_S6_7_8_3_11_12_6_9_10-odd}(b) and Fig.~\ref{crnts_S6_7_8_3_11_12_6_9_10-odd}(c), we see how variation in $\mu^{}_{0}$ actually only changes $\overline{I}^{\text{em.}}_{\alpha}(t,\phi)$, leaving $I^{[1+2]}_{\alpha}(t,\phi)$ unaffected, as ensured by Eq.~(\ref{dynamic-crnt-sep}). The higher the $\mu^{}_{0}$ is away from $\varepsilon_{0}$, a larger part of $\mathcal{T}^{(+)}_{\alpha}(t,\omega,\phi)$ is effectively included in the integral of Eq.~(\ref{dynamic-crnt-0-0}) whose upper bound is governed by $\mu^{}_{0}$ (at low temperature). The maximal attainable magnitude of $\overline{I}^{\text{em.}}_{\alpha}(t,\phi)$ in the transient process thus grows with the deviation of $\mu^{}_{0}$ from $\varepsilon_{0}$ (see the changes of the lines in Fig.~\ref{crnts_S6_7_8_3_11_12_6_9_10-odd}(b)). More interestingly, Fig.~\ref{crnts_S6_7_8_3_11_12_6_9_10-odd}(b) and Fig.~\ref{crnts_S6_7_8_3_11_12_6_9_10-odd}(c) exemplify the special properties of the odd components of the currents. It shows that $\overline{I}^{\text{em.},-}_{\alpha }(t,\phi)$ can be negative while $I^{[1+2],-}_{\alpha}(t,\phi)$ can be positive (see also Fig.~\ref{crnts_S5-even-odd}(b1),(b2),(b3) for $\overline{I}^{\text{em.},-}_{\alpha }(t,\phi)$ showing negative values while $I^{[1+2],-}_{\alpha}(t,\phi)$ showing positive values). This is unlike their even counterparts where the signs of the currents are consistent with the in-tunneling and the out-tunneling processes. The distinction between the mechanisms for the even part and the odd part of the transient currents can be further revealed without the restriction of zero bias.

\subsubsection{finite biases $\mu^{}_{L}\ne\mu^{}_{R}$}
\label{sec_model_fluxindep_sym1_arbbs}

At a finite bias, the directly observable current $I^{\text{em.}}_{\alpha }\left( t,\phi\right)$  contains both $\overline{I}^{\text{em.}}_{\alpha }(t,\phi)$, which obeys Eq.~(\ref{p-h_sym_flux-indep}), and $\Delta I^{\text{em.}}_{\alpha }(t,\phi)$, which is not involved in Eq.~(\ref{p-h_sym_flux-indep}). It is thus very interesting to see how the symmetry embedded in Eq.~(\ref{p-h_sym_flux-indep}) can be manifested also at a finite bias from the directly observable current $I^{\text{em.}}_{\alpha }\left( t,\phi\right)$. Below we show that directly from the even and the odd components of $I^{}_{\alpha }\left( t,\phi\right)$,  Eq.~(\ref{p-h_sym_flux-indep}) can be revealed. Complementarily, with the aid of Eq.~(\ref{p-h_sym_flux-indep}) the distinct effects of the even and odd responses of the currents to the flux can be manifested.

Indeed, interchanging the left and the right bonds, the current components on the two sides are directly related to each other as
\begin{subequations}
\label{I_LR_ps_sym}
\begin{align}
\label{I_LR_ps_sym-1}
\overline{I}^{\text{em.},\pm}_{\alpha }(t,\phi)=
\pm\left. \overline{I}^{\text{em.},\pm}_{\bar{\alpha} }(t,\phi)\right\vert_{L\leftrightarrow R},
\end{align} and
\begin{align}
\label{I_LR_ps_sym-2}
\Delta I^{\text{em.},\pm}_{\alpha }(t,\phi)=
\mp\left.\Delta I^{\text{em.},\pm}_{\bar{\alpha} }(t,\phi)\right\vert_{L\leftrightarrow R},
\end{align}
\end{subequations} where $\bar{\alpha}$ is the opposite side of $\alpha$ and $\left.\cdot\right\vert_{L\leftrightarrow R}$ indicates the quantity $\cdot$ on the RHS of Eq.~(\ref{I_LR_ps_sym}) is evaluated by interchanging the left and the right bonds, $\bar{\Gamma}_{ii}^{L}\leftrightarrow\bar{\Gamma}_{ii}^{R}$, from those used in the LHS of the same equation. The superscripts $+$ and $-$ denote the even and the odd flux-dependent parts of these quantities. The result, Eq.~(\ref{I_LR_ps_sym}), is a direct consequence of the geometric properties of the system.\cite{footnote1} It does not require any symmetry among the bonds and it holds with no regard to the positions of $\mu_{L}$ and $\mu_{R}$.
By Eq.~(\ref{dynamic-crnt-sep}) and Eq.~(\ref{dynamic-crnt-0-b0}), the current that starts with initially fully occupied QDs is
\begin{align}
\label{I_d}
&\left.I^{}_{\alpha }\left( t,\phi\right)\right\vert_{n_{1}(t_0)=n_{2}(t_0)=1}
\nonumber\\&=\overline{I}^{\text{em.}}_{\alpha }(t,\phi)+
\Delta I^{\text{em.}}_{\alpha }(t,\phi)
+I_{\alpha}^{[1+2]}(t,\phi).
\end{align} To see how the symmetry, Eq.~(\ref{p-h_sym_flux-indep}), manifests itself, we substitute Eq.~(\ref{p-h_sym_flux-indep}) into Eq.~(\ref{I_d}), yielding
\begin{align}
\label{I_d-2}
\left.I^{}_{\alpha }\left( t,\phi\right)\right\vert_{n_{1}(t_0)=n_{2}(t_0)=1}=-\overline{I}^{\text{em.}}_{\alpha }(t,\phi)+\Delta I^{\text{em.}}_{\alpha }(t,\phi).
\end{align} Combining Eq.~(\ref{I_d-2}) with Eq.~(\ref{I_LR_ps_sym}), we are led to a new relation
\begin{align}
\label{I_d_em_LR_sym}
\left.I^{\pm}_{\alpha}(t,\phi)\right\vert_{n_{1}(t_0)=n_{2}(t_0)=1}^{}
=\mp\left.I^{\pm}_{\bar{\alpha}}(t,\phi)\right\vert_{n_{1}(t_0)=n_{2}(t_0)=0}^{L\leftrightarrow R}.
\end{align}

The result of Eq.~(\ref{I_d_em_LR_sym}) does not require the breaking of the directly observable current $I^{}_{\alpha }\left( t,\phi\right)$ into a purely transient part and a part that would have remained in the steady-state limit. The physical quantities on both sides of Eq.~(\ref{I_d_em_LR_sym}) can be directly obtained from $I^{}_{\alpha }\left( t,\phi\right)$.
The LHS of Eq.~(\ref{I_d_em_LR_sym}) is the current on lead $\alpha$ with the QDs being initially fully occupied, which is dominated by the out-tunneling process in the transient regime and vice versa for the RHS, where initially the QDs are empty. For the even components, the two sides of Eq.~(\ref{I_d_em_LR_sym}) differ by a sign. This reflects the sign difference of the in-tunneling and the out-tunneling currents. In contrast, for the odd components, there is no such a sign difference.

\begin{figure}[h]
\includegraphics[width=8.4cm,height=7cm]{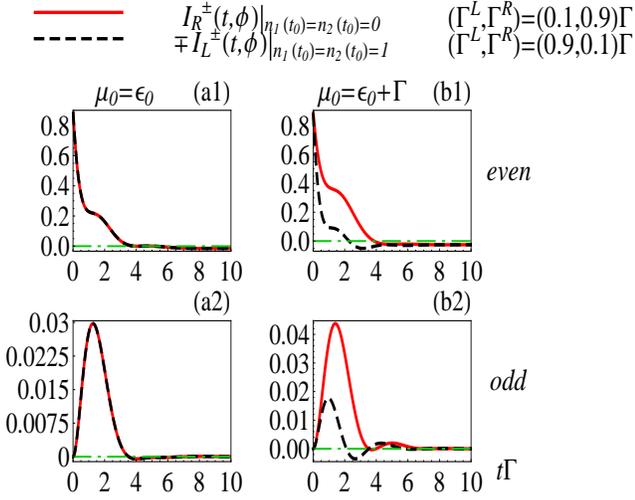} \caption{(color online) The even and the odd components of the directly observable currents that start at different initial occupations. The applied bias is $eV=0.5\Gamma$ and the flux is taken at $\phi=\pi/2$. The plots (a1) and (a2) are with $\mu_{0}=\varepsilon_{0}$ while the plots (b1) and (b2) are with $\mu_{0}=\varepsilon_{0}+\Gamma$. In all the plots we have let $\bar{\Gamma}^{\alpha}_{11}=\bar{\Gamma}^{\alpha}_{22}={\Gamma}^{\alpha}$ with $\delta\varepsilon=-2\Gamma$. The red solid lines in (a1) and (b1) are for the even component $\left.I^{+}_{R}(t,\phi)\right\vert_{n_{1}(t_0)=n_{2}(t_0)=0}$ while those in (a2) and (b2) are for the odd component $\left.I^{-}_{R}(t,\phi)\right\vert_{n_{1}(t_0)=n_{2}(t_0)=0}$ evaluated with $(\Gamma^L,\Gamma^R)=(0.1,0.9)\Gamma$. The black long-dashed lines in (a1) and (b1) are for the even components, $-\left.I^{+}_{L}(t,\phi)\right\vert_{n_{1}(t_0)=n_{2}(t_0)=1}$, and those in (a2) and (b2) are for the odd components, $\left.I^{-}_{L}(t,\phi)\right\vert_{n_{1}(t_0)=n_{2}(t_0)=1}$, evaluated with interchanged bonds as $(\Gamma^L,\Gamma^R)=(0.9,0.1)\Gamma$. The green dash-dot lines are drawn to indicate the zero value. One sees that in the long-time limit, the even components (a1) and (b1) are away from zero while the odd components (a2) and (b2) overlap with zero.
} \label{figev}
\end{figure}

We demonstrate  Eq.~(\ref{I_d_em_LR_sym}) numerically in Fig.~\ref{figev} with $eV=\mu_{L}-\mu_{R}\ne0$.
Heed that, at a finite bias, the breaking of the equality Eq.~(\ref{I_d_em_LR_sym}) when Eq.~(\ref{p-h_sym_flux-indep}) does not hold is a purely transient phenomena. In the steady-state limit the purely transient components vanish, $\overline{I}^{\text{em.}}_{\alpha }(t\rightarrow\infty,\phi)=0$, $I^{[1+2]}_{\alpha}(t\rightarrow\infty,\phi)=0$, leading to $I_{\alpha}(t\rightarrow\infty,\phi)=\Delta I^{\text{em.}}_{\alpha }(t\rightarrow\infty,\phi)$ and Eq.~(\ref{I_LR_ps_sym-2}) immediately becomes Eq.~(\ref{I_d_em_LR_sym}). One can see in Fig.~\ref{figev}(a1) that $-\left.I^{+}_{L}(t,\phi)\right\vert_{n_{1}(t_0)=n_{2}(t_0)=1}^{}=\left.I^{+}_{R}(t,\phi)\right\vert_{n_{1}(t_0)=n_{2}(t_0)=0}^{L\leftrightarrow R}$ and in Fig.~\ref{figev}(a2) that $\left.I^{-}_{L}(t,\phi)\right\vert_{n_{1}(t_0)=n_{2}(t_0)=1}^{}=\left.I^{-}_{R}(t,\phi)\right\vert_{n_{1}(t_0)=n_{2}(t_0)=0}^{L\leftrightarrow R}$, confirming the symmetry behind Eq.~(\ref{p-h_sym_flux-indep}). In Fig.~\ref{figev}(b1),(b2), in which we have raised $\mu_{0}$ away from $\varepsilon_{0}$, then Eq.~(\ref{I_d_em_LR_sym}) is only transiently broken. However, both sides of Eq.~(\ref{I_d_em_LR_sym}) for the even part can remain nonzero to the steady states, depending on the bias applied (see captions of Fig.~\ref{figev}). This testifies that the effect caused by the symmetry behind Eq.~(\ref{p-h_sym_flux-indep}) is exclusively transient.


\section{conclusion}
\label{sec_conclu}


In summary, we have identified several effects which are uniquely present in the transient regime for the AB interferometer with two QDs:

1. The odd components of the currents arise as a purely transient effect (see the definition of even and odd components in Eq.~(\ref{gel_flux_decps})). The fixed relation between the odd components of the currents on the left and on the right, given by Eq.~(\ref{phi-LR_asym}) for whatever bonding geometry, witnesses the intrinsically transient effect brought up by the flux-induced interference.

2. The geometric asymmetry of the interferometer is manifested by the asymmetry between individual even components of transient currents. This contrasts the steady-state limit, where the current on the left is always related to that on the right by $I_{L}(t\rightarrow\infty,\phi)=-I_{R}(t\rightarrow\infty,\phi)$ for whatever distributions of the bonds, on-site energies and chemical potentials. The currents purely induced by initially occupying either the upper or the lower QD do not survive in the steady-state limit. This restricts the detection and utilization of up-down asymmetry of the bonds in the steady-state limit.

Specifically we find that the left-right asymmetry in the times of approaching the steady-state limit, caused by ${\Gamma}^{L}_{}\ne{\Gamma}^{R}_{}$, can be reduced by a larger up-down asymmetry of the QDs' on-site energies. In contrast, the non-degeneracy of the two QDs' on-site energies does not affect the up-down asymmetry in the times of approaching steady-state limit, caused by $\bar{\Gamma}^{}_{1}\ne\bar{\Gamma}^{}_{2}$. These results suggest a way to modulate the anisotropy of transient transport by the combination of tuning the on-site energies between the two QDs and the relative strengths among the bonds.


3. The two main purely transient components of the total tunneling currents, namely, $\overline{I}^{\text{em.}}_{\alpha }(t,\phi)$ and $I^{[1+2]}_{\alpha}(t,\phi)$,  manifest the underlying symmetry among the energy levels in a unique way. This is given by Eq.~(\ref{p-h_sym_flux-indep}), which holds when the system obeys the symmetry of the energy levels as Eq.~(\ref{mu0eqep0}) with the up-down geometric symmetry given by Eq.~(\ref{updwn_bdsym}) or Eq.~(\ref{degDQD}). The distinct mechanisms underlying the even part and the odd part of the transient currents can be revealed by the aid of Eq.~(\ref{I_d_em_LR_sym}). Both sides of Eq.~(\ref{I_d_em_LR_sym}) are directly observable at finite biases. The even component, whose existence does not rely on non-vanishing flux unlike the odd component, clearly reveals the distinction between the in-tunneling and the out-tunneling transient processes carried out by $\overline{I}^{\text{em.},+}_{\alpha }(t,\phi)$ and $I^{[i],+}_{\alpha}(t,\phi)$ respectively.



The above results are obtained the under wide-band approximation, which allows no bound states. If the system possesses bound states, then the stationary currents can depend on the initial occupations of the QDs.\cite{Yang15165403} The effects associated with the initial-occupation-induced currents are expected to remain not only in the transient regime. The wide-band approximation is appropriate near the band center where the density-of-states is flat. For the level-broadening functions other than the wide-band approximation, the specific line-shape of the band on each of the leads should be taken into account for the geometric symmetry. The density-of-states in each of the leads should also be considered as a part of the factors that determine the symmetry of the energy level distribution. This could affect the relationship between the in-tunneling and the out-tunneling currents.

Although our calculations are based on a specific interferometer, our analysis, in the following senses, is not restricted to this particular example. First, the separation of the tunneling current into the two contributions that are respectively independent and dependent of initial occupations in the central scattering region, namely, Eq.~(\ref{dynamic-crnt-sep}), is a result of the general Hamiltonian Eq.~(\ref{gel_H}) and the initial state Eq.~(\ref{init_eqthm}). It holds regardless of the number of reservoirs. Second, the separation of a purely transient component from the tunneling current starting from empty central scattering region, given by Eq.~(\ref{dynamic-crnt-0-b0}), is common to any two-terminal device. This can be generalized to multi-terminal setups by rewriting the fermi functions in Eq.~(\ref{less-selfeng}) (and henceforth appear in Eq.~(\ref{dynamic-crnt-0}) and Eq.~(\ref{lessG0})) into the sum of the average of all fermi functions and the deviation from it. Third, the interpretation of the transient components of the currents as the in-tunneling and the out-tunneling currents are based on an intuitive understanding of the corresponding situation described by these components. Such an interpretation is not limited to this specific interferometer. Fourth, the manifestation of the asymmetry of the bonds through the asymmetry in the transient currents is intuitively anticipated. This is also applicable to other transport systems.


We have pointed out how the transient components of the tunneling currents can be extracted from directly observable currents.
Given the feasibilities of preparing various electron occupations in QDs in experiments\cite{Hayashi03226804,Fujisawa06759,Hanson071217,Kim1470}  and performing time-resolved current measurements\cite{Fujisawa01081304,Lu03422,Bylander05361,Fujisawa06759,Feve071169}, our theoretical results are relevant to experimental tests.
Various examples have demonstrated how to infer the parameters of the underlying electronic structures from the transient currents. These include the fermi level positions in pulsed scenarios,\cite{Wingreen938487,Jauho945528,Maciejko06085324} the inter-dot hopping strengths in arrays of coupled QDs,\cite{Taranko12023711} the spin-sensitive energy splittings\cite{Souza07205315,Perfetto08155301} and difference in frequency between two driving fields\cite{Trocha10115320}. The correspondence between the symmetry (asymmetry) of geometric factors of the system and the symmetry (asymmetry) among different components of the transient currents discussed here also provides a way to infer the symmetry (asymmetry) of the parameters of the interferometer. The transient existence of the odd components is a direct probe of the non-degeneracy of the QDs and the nonzero flux. Quantum electronics has taken the advantage of the temporal coherence of electron tunnelings for making switches. Here we have shown that the transient current at a terminal can be separated into distinct components and these different components show distinguishable characters. The transient current at one terminal can also differ nontrivially from that at the other terminal. They can all be modulated by tuning the geometry of the system. In addition to the switches, the diversified ways of these different transient currents could potentially be exploited as alternative resources for operating electronic devices.








\begin{acknowledgements}
This work was supported by the Israeli
Science Foundation (ISF) and by the infrastructure program
of Israel Ministry of Science and Technology under
contract 3-11173. This research was
supported by the Ministry of Science and Technology of ROC under Contract No.
NSC-102-2112-M-006-016-MY3, and the National Center for Theoretical Science of Taiwan. It is also supported in part by the Headquarters of University Advancement at the National Cheng Kung University,
which is sponsored by the Ministry of Education of ROC. We acknowledge the facility of High Performance Computing Cluster of National Cheng Kung University.
\end{acknowledgements}

\appendix

\section{Explicit solutions under wide-band approximation}
\label{appx-GrWB}

By the use of Eqs.~(\ref{wb_no-phase_lvbrd},\ref{with-phase_lvbrd}) in Eqs.~(\ref{Gr_scat-sta},\ref{self-eng_scat-sta}), the retarded Green function, as the first line of Eq.~(\ref{Gr_t_t0}) is found to be,
\begin{subequations}
\label{retardedG-intime}
\begin{align}
\boldsymbol{G}^{r}\left(t,t_{0}\right)=\boldsymbol{\mathcal{A}}^{+}_{}(\phi)c^{}_{+}(t,\phi)+\boldsymbol{\mathcal{A}}^{-}_{}(\phi)c^{}_{-}(t,\phi),
\end{align}  where
\begin{align}
\boldsymbol{\mathcal{A}}^{\pm}_{}(\phi)=-\frac{i}{2}\left(\begin{array}{cc}
\left(1\pm\frac{\Gamma_{d}+i\delta\varepsilon}{\Gamma_{g}(\phi)}\right)&\pm\Gamma_{12}(\chi,\phi)/\Gamma_{g}(\phi)\\
\pm\Gamma_{21}(\chi,\phi)/\Gamma_{g}(\phi) & \left(1\mp\frac{\Gamma_{d}+i\delta\varepsilon}{\Gamma_{g}(\phi)}\right)
\end{array}\right)
\end{align} and
\begin{align}
\label{time-amplitudes}
c^{}_{\pm}(t,\phi)=e^{-i(\varepsilon_{0}\pm\varepsilon_{g}(\phi)/2)(t-t_{0})}e^{-(\Gamma\pm\delta\Gamma_{}(\phi))(t-t_{0})/2}.
\end{align}
\end{subequations}
The total broadening and the asymmetry between the broadenings of the two QDs are characterized by
\begin{align}
\label{GmSD}
\Gamma_{}=(\Gamma_{11}+\Gamma_{22})/2,~\Gamma_{d}=(\Gamma_{11}-\Gamma_{22})/2,
\end{align} with
\begin{align}
\Gamma_{ii}=\sum_{\alpha=L,R}\bar{\Gamma}^{\alpha}_{ii},
\end{align} for $i=1,2$. The indirect tunnel coupling between the two QDs via the leads is
\begin{align}
\Gamma_{12}(\chi,\phi)=e^{i\chi/2}(\bar{\Gamma}^{L}_{12}e^{i\phi/2}+\bar{\Gamma}^{R}_{12}e^{-i\phi/2}),
\end{align} where
\begin{align}
\chi=\phi_{L}+\phi_{R},
\end{align}
is the gauge phase and $\Gamma_{21}(\chi,\phi)=[\Gamma_{12}(\chi,\phi)]^{*}$. The time-dependence of the retarded Green function is characterized by
\begin{align}
\label{decay-rates}
\delta\Gamma_{}(\phi)=\text{Re}\Gamma_{g}(\phi),~\varepsilon_{g}(\phi)=\text{Im}\Gamma_{g}(\phi)
\end{align} where
\begin{align}
\label{cmpx-decayrate}
\Gamma_{g}(\phi)=\sqrt{\vert\Gamma_{12}(\phi)\vert^{2}+(\Gamma_{d}+i\delta\varepsilon)^{2}},
\end{align}
is a complex quantity whose real part gives the difference in decay rates and the imaginary part shows the phase difference between the time evolving amplitudes $c_{+}(t,\phi)$ and $c_{-}(t,\phi)$ in Eq.~(\ref{retardedG-intime}). Note that $\vert\Gamma_{12}(\phi)\vert^{2}=\vert\Gamma_{12}(\chi,\phi)\vert^{2}$ is independent of $\chi$.

Under the wide-band approximation, we can simplify Eq.~(\ref{crnt-init-dep}) to
\begin{align}
I^{\text{occ.}}_{\alpha}(t,\phi)=i\text{Tr}\left[
\boldsymbol{\Gamma}^{\alpha}\boldsymbol{G}^{r}(t,t_{0})\boldsymbol{G}^{<}(t_{0},t_{0})\boldsymbol{G}^{a}(t_{0},t)\right],
\end{align} and Eq.~(\ref{dynamic-crnt-0-b0}) with
\begin{subequations}
\label{time-dep_Ts_WB}
\begin{align}
&\mathcal{T}^{(\pm)}_{\alpha}(t,\omega,\phi)=\text{Tr}\left\{
\boldsymbol{\Gamma}^{\alpha}\left[i\left(\boldsymbol{\bar{G}}^{r}(t,\omega)-\boldsymbol{\bar{G}}^{a}(t,\omega)\right)
\right.\right.\nonumber\\
&\left.\left.-
\boldsymbol{\bar{G}}^{r}(t,\omega)(\boldsymbol{\Gamma}^{\alpha}\pm\boldsymbol{\Gamma}^{\bar{\alpha}})\boldsymbol{\bar{G}}^{a}(t,\omega)\right]\right\},
\end{align} where
\begin{align}
&\boldsymbol{\bar{G}}^{r}(t,\omega)=\int_{t_0}^{t}d\tau e^{i\omega(t-\tau)}\boldsymbol{G}^{r}(t,\tau),
\nonumber\\
&\boldsymbol{\bar{G}}^{a}(t,\omega)=\int_{t_0}^{t}d\tau e^{-i\omega(t-\tau)}\boldsymbol{G}^{a}(\tau,t).
\end{align}
\end{subequations} Here $\mathcal{T}^{(\pm)}_{\alpha}(t,\omega,\phi)=\mathcal{T}^{(\pm)}_{\alpha}(t,\omega)$ are the transmission-like functions defined in Eq.~(\ref{dynamic-crnt-0-b0}) with the additional argument $\phi$ emphasizing its flux dependence.
The current induced by an initial occupation on level $i$, Eq.~(\ref{crnt-init-diagocc-1}), becomes,
\begin{align}
\label{crnt-init-diagocc-WB}
I^{[i]}_{\alpha}(t,\phi)=-\left[
\boldsymbol{G}^{a}(t_{0},t)\boldsymbol{\Gamma}^{\alpha}\boldsymbol{G}^{r}(t,t_{0})\right]_{ii}.
\end{align}

The solution Eq.~(\ref{retardedG-intime}) reduces to those used in Ref.~[\onlinecite{Tu12115453}] by setting $\bar{\Gamma}^{\alpha}_{11}=\bar{\Gamma}^{\alpha}_{22}$ , for $\alpha=L,R$.

\subsection{Properties of the odd components}
\label{appx-GrWB-oddbfctr}
In Eqs.~(\ref{odd-I1}),(\ref{odd-I2}),(\ref{odd-Tpm}), we have defined
\begin{align}
\label{time-amplitudes-cp}
b_{\pm}\left(t,\phi\right)=\frac{1}{2}\left[c_{+}\left(t,\phi\right)\pm c_{-}\left(t,\phi\right)\right],
\end{align} and
\begin{align}
\label{time-amplitudes-cp-omega}
\tilde{b}_{\pm}\left(t,\omega\right)=\int_{t_0}^{t}d\tau e^{i\omega(t-\tau)}b_{\pm}\left(\tau,\phi\right).
\end{align}
The condition Eq.~(\ref{updwn_bdsym}) makes $\Gamma_{g}\left(\phi\right)$ of Eq.~(\ref{cmpx-decayrate}) become either purely real or purely imaginary. The condition Eq.~(\ref{degDQD}) ensures that $\Gamma_{g}\left(\phi\right)$ is purely real.

When $\Gamma_{g}\left(\phi\right)$ is real, then by Eq.~(\ref{time-amplitudes})
\begin{equation}
c_{+}\left(t,\phi\right)c_{-}^{*}\left(t,\phi\right)=e^{-\Gamma t},
\end{equation}
 and consequently
\begin{align}
&\text{Im}\left[\frac{b_{-}\left(t,\phi\right)}{\Gamma_{g}\left(\phi\right)}b_{+}^{*}\left(t,\phi\right)\right]
=\frac{\text{Im}\left[c_{+}\left(t,\phi\right)c_{-}^{*}\left(t,\phi\right)\right]/2}{\Gamma_{g}\left(\phi\right)}\nonumber\\
&=\frac{\text{Im}\left[e^{-\Gamma t}\right]/2}{\Gamma_{g}\left(\phi\right)}=0.
\end{align}
When $\Gamma_{g}\left(\phi\right)$ is purely imaginary, denoted as $\Gamma_{g}\left(\phi\right)=i\varepsilon_{g}\left(\phi\right)$,
where $\varepsilon_{g}\left(\phi\right)$ is real, then
\begin{align}
\left\vert c_{\pm}\left(t,\phi\right)\right\vert ^{2}
=\left\vert e^{-\left[i\varepsilon_{0}+\frac{1}{2}\left(\Gamma\pm i\varepsilon_{g}\left(\phi\right)\right)\right]t}\right\vert ^{2}
= e^{-\Gamma t},
\end{align} and consequently we have
\begin{align}
&\text{Im}\left[\frac{b_{-}\left(t,\phi\right)}{\Gamma_{g}\left(\phi\right)}b_{+}^{*}\left(t,\phi\right)\right]
\!\!=\!\!\text{Im}\!\!\left[\frac{\frac{1}{4}\left[\!\left\vert c_{+}\left(t,\phi\right)\right\vert ^{2}-\left\vert c_{-}\left(t,\phi\right)\right\vert ^{2}\!\!\right]}{i\varepsilon_{g}\left(\phi\right)}\right]
\nonumber\\&=0.
\end{align}
Therefore, we conclude that whenever Eq.~(\ref{updwn_bdsym}) or Eq.~(\ref{degDQD}) holds,
we are led to
\begin{align}
\text{Im}\left[\frac{b_{-}\left(t,\phi\right)}{\Gamma_{g}\left(\phi\right)}b_{+}^{*}\left(t,\phi\right)\right]=0.\label{zero-identity-1}
\end{align}  The result of Eq.~(\ref{odd-Tpm}) under the condition Eq.~(\ref{updwn_bdsym}) reduces to the transient breaking of phase rigidity we found in Ref.~[\onlinecite{Tu12115453}] for the current $I_{\alpha}(t,\phi)$ starting from empty QDs with up-down symmetric bonds.

As $t\rightarrow\infty$, $c^{}_{\pm}(t,\phi)$ in Eq.~(\ref{time-amplitudes}) approaches zero (so does $b_{\pm}^{}(t,\phi)$ in Eq.~(\ref{time-amplitudes-cp})) and therefore the expressions in Eq.~(\ref{odd-I1}) and Eq.~(\ref{odd-I2}) also approach zero.
By explicitly substituting Eq.~(\ref{time-amplitudes}) into Eq.~(\ref{odd-Tpm}) and taking $t\rightarrow\infty$, the result also vanishes. This shows that the odd components are nonzero only in the transient regime.

\subsection{Prove $I^{[1]}_{\alpha}(t,\phi)=I^{[2]}_{\alpha}(t,\phi)$ when $\bar{\Gamma}_{11}^{\alpha}=\bar{\Gamma}_{22}^{\alpha}$}
\label{appx-GrWB-updwnsym}

Under the condition Eq.~(\ref{updwn_bdsym}), the currents induced by initially occupying one QD,
Eq.~(\ref{crnt-init-diagocc-WB}), become
\begin{align}
\label{I1sym}
&I_{\alpha}^{\left[1\right]}\left(t\right)=-\Gamma^{\alpha}\times\nonumber\\
&\!\!\left(\left\vert G^{r}_{11}\left(t\right)\right\vert ^{2}\!\!+\left\vert G^{r}_{21}\left(t\right)\right\vert ^{2}\!\!+2\text{Re}\left[e^{i\phi_{\alpha}}G^{r}_{21}\left(t\right)(G^{r}_{11}\left(t\right))^{*}\right]\right)
\end{align} and
\begin{align}
\label{I2sym}
&I_{\alpha}^{\left[2\right]}\left(t\right)=-\Gamma^{\alpha}\times\nonumber\\
&\!\!\left(\left\vert G^{r}_{12}\left(t\right)\right\vert ^{2}\!\!+\left\vert G^{r}_{22}\left(t\right)\right\vert ^{2}\!\!+2\text{Re}\left[e^{-i\phi_{\alpha}}G^{r}_{12}\left(t\right)(G^{r}_{22}\left(t\right))^{*}\right]\right),
\end{align} where $G^{r}_{ij}(t)=[\boldsymbol{G}^{r}\left(t,t_{0}\right)]_{ij}$ for $i,j\in\{1,2\}$.
From Eq.~(\ref{retardedG-intime}), we have,
\begin{align}
\label{Gr12_updwbdsym}
&G^{r}_{12}\left(t\right)=i\Gamma_{12}\left(\chi,\phi\right)\frac{b_{-}\left(t,\phi\right)}{\Gamma_{g}\left(\phi\right)},
\nonumber\\
&G^{r}_{21}\left(t\right)=i\Gamma_{12}^{*}\left(\chi,\phi\right)\frac{b_{-}\left(t,\phi\right)}{\Gamma_{g}\left(\phi\right)},
\end{align} and therefore
\begin{equation}
\label{G12sq2eqsym}
\left\vert G^{r}_{12}\left(t\right)\right\vert ^{2}=\left\vert G^{r}_{21}\left(t\right)\right\vert ^{2}.
\end{equation} On the other hand, Eq.~(\ref{retardedG-intime}) with Eq.~(\ref{time-amplitudes-cp}) gives
\begin{align}
\label{gr11sq2}
&\left\vert G^{r}_{11}\left(t\right)\right\vert^{2} =\nonumber \\
&\left\vert b_{+}\left(t,\phi\right)\right\vert^{2}+\delta\varepsilon^{2}\left\vert \frac{b_{-}\left(t,\phi\right)}{\Gamma_{g}\left(\phi\right)}\right\vert ^{2}-2\delta\varepsilon\text{Im}\left[\frac{b_{-}\left(t,\phi\right)}{\Gamma_{g}\left(\phi\right)}b_{+}^{*}\left(t,\phi\right)\right],
\end{align} and
\begin{align}
\label{gr22sq2}
&\left\vert G^{r}_{22}\left(t\right)\right\vert^{2} =\nonumber \\
&\left\vert b_{+}\left(t,\phi\right)\right\vert^{2}+\delta\varepsilon^{2}\left\vert \frac{b_{-}\left(t,\phi\right)}{\Gamma_{g}\left(\phi\right)}\right\vert ^{2}+2\delta\varepsilon\text{Im}\left[\frac{b_{-}\left(t,\phi\right)}{\Gamma_{g}\left(\phi\right)}b_{+}^{*}\left(t,\phi\right)\right].
\end{align} Apply Eq.~(\ref{zero-identity-1}) to Eq.~(\ref{gr11sq2}) and Eq.~(\ref{gr22sq2}) then yield,
\begin{align}
\label{eqG11G22sq2}
\left\vert G^{r}_{11}\left(t\right)\right\vert ^{2}=\left\vert G^{r}_{22}\left(t\right)\right\vert ^{2}.
\end{align} Utilizing Eq.~(\ref{zero-identity-1}) to the last terms of Eq.~(\ref{I1sym}) and Eq.~(\ref{I2sym}), we find
\begin{align}
\label{Ginfeqsym}
 & \text{Re}\left[e^{i\phi_{\alpha}}G^{r}_{21}\left(t\right)(G^{r}_{11}\left(t\right))^{*}\right]
\nonumber \\
&=  \left(\Gamma^{\alpha}+\Gamma^{\bar{\alpha}}\cos\phi\right)\text{Re}\left[\frac{b_{-}\left(t,\phi\right)}{\Gamma_{g}\left(\phi\right)}b_{+}^{*}\left(t,\phi\right)\right]
\nonumber \\&
+\zeta_{\alpha}\Gamma^{\bar{\alpha}}\delta\varepsilon\sin\phi\left\vert \frac{b_{-}\left(t,\phi\right)}{\Gamma_{g}\left(\phi\right)}\right\vert ^{2}
\nonumber \\&= \text{Re}\left[e^{-i\phi_{\alpha}}G^{r}_{12}\left(t\right)(G^{r}_{22}\left(t\right))^{*}\right].
\end{align} Substituting Eq.~(\ref{G12sq2eqsym}), Eq.~(\ref{eqG11G22sq2}) and Eq.~(\ref{Ginfeqsym}) into Eq.~(\ref{I1sym}) and Eq.~(\ref{I2sym}) results in
$I^{[1]}_{\alpha}(t,\phi)=I^{[2]}_{\alpha}(t,\phi)$.

\section{Limits of large asymmetries}

\subsection{Large left-right asymmetry under up-down symmetric bonds}
\label{appx-AsymLR}

\subsubsection{degenerate QDs $\delta\varepsilon=0$}
\label{appx-AsymLR-deg}

Here we present more detailed discussions of the initial-occupation-induced currents under degeneracy with up-down symmetry in the bonds, with large left-right asymmetry.
To reveal the left-right asymmetry, we have let $\Gamma^{R}\gg\Gamma^{L}$ in Fig.~\ref{crnts_S38}. Setting $\alpha=R$ (therefore $\bar{\alpha}=L$) in Eq.~(\ref{I1_udwnsym_deg}) with $\Gamma^{R}\gg\Gamma^{L}$ we find $\delta\Gamma(\phi)\approx\Gamma^{R}$ by Eq.~(\ref{decay-rates}) and Eq.~(\ref{cmpx-decayrate}) and subsequently $\frac{\left(\Gamma^{L}+\Gamma^{R}\cos\phi\right)}{\delta\Gamma\left(\phi\right)}\approx\cos\phi$, $\frac{\left(\Gamma^{R}+\Gamma^{L}\cos\phi\right)}{\delta\Gamma\left(\phi\right)}\approx1$, $\Gamma+\delta\Gamma\left(\phi\right)\lesssim2\Gamma$ and $0<\Gamma-\delta\Gamma\left(\phi\right)\ll\Gamma$. Under such circumstances, the initial-occupation-induced currents, Eq.~(\ref{I1_udwnsym_deg}), become approximately
\begin{align}
\label{I1_udwnsym_deg-LllR}
&I^{[1]}_{L}(t,\phi)\approx-\Gamma^{L}\left\{\frac{1}{2}\left(1-\cos\phi\right)
e^{-(\Gamma-\delta\Gamma\left(\phi\right))t}\right.
\nonumber\\&\left.
+\frac{1}{2}\left(1+\cos\phi\right)
e^{-(\Gamma+\delta\Gamma\left(\phi\right))t}\right\},
\nonumber\\&
I^{[1]}_{R}(t,\phi)\approx-\Gamma^{R}e^{-(\Gamma+\delta\Gamma\left(\phi\right))t}.
\end{align} The amplitude for the slow decay term $e^{-(\Gamma-\delta\Gamma\left(\phi\right))t}$ in $I^{[1]}_{R}(t,\phi)$ becomes very small and the amplitude for the fast decay term $e^{-(\Gamma+\delta\Gamma\left(\phi\right))t}$ remains finite. The dynamics of $I^{[1]}_{R}(t,\phi)$ is dominated by the fast decay term as shown by the last line of Eq.~(\ref{I1_udwnsym_deg-LllR}). On the other hand, setting $\alpha=L$ in Eq.~(\ref{I1_udwnsym_deg}) leads to comparable amplitudes for both of the exponential decay terms (see the first two lines of Eq.~(\ref{I1_udwnsym_deg-LllR})). After the fast decay term becomes effectively zero, the slow decay term still remains visible for $I^{[1]}_{L}(t,\phi)$ and dominates its approach to steady states.

At large left-right asymmetry Eq.~(\ref{I1_udwnsym_deg-LllR}) also reveals two distinct situations in terms of the dependence on the flux. The amplitude in $I^{[1]}_{R}(t,\phi)$ is independent of the flux while the amplitudes in $I^{[1]}_{L}(t,\phi)$ are clearly flux-dependent via $\cos\phi$. The decay rates, plotted in Fig.~\ref{crnts_S38}(c), only weakly depend on the flux. They do not alter the main flux-dependence given by the amplitudes. Besides, the red solid lines for $\phi=0$ in Fig.~\ref{crnts_S38}(a1),(a2) show a particular fast saturation, in comparison to other lines for other values of the flux. Indeed, setting $\phi=0$ in Eq.~(\ref{I1_udwnsym_deg}) further reduces it to
\begin{align}
\label{I1_udwnsym_deg-zeroflux}
&I^{[1]}_{\alpha}(t,\phi=0)=-\Gamma^{\alpha}e^{-2\Gamma t}.
\end{align} The initial-occupation-induced current $I^{[1]}_{\alpha}(t,\phi)$ at $\phi=0$ saturates to its steady-state value with a maximal rate $2\Gamma$ attainable from $\Gamma+\delta\Gamma\left(\phi\right)$ by varying $\phi$, as shown by Fig.~\ref{crnts_S38}(c).

\subsubsection{non-degenerate QDs $\delta\varepsilon\ne0$}
\label{appx-AsymLR-nondeg}

The step-like feature and its asymmetry between the left and the right currents shown in Fig.~\ref{crnts_S50-even} are detailed here.
For $I^{[1],+}_{L}(t,\phi)$ one sees that the curves for different flux cross at some points of time (see Fig.~\ref{crnts_S50-even}(a)). For $I^{[1],+}_{R}(t,\phi)$, different values of the flux follow similar evolution trajectories exhibiting step-like features (see Fig.~\ref{crnts_S50-even}(b)). The currents on the left and on the right differ, besides the overall factor $\Gamma^{\alpha}$ in Eq.~(\ref{I1_udwnsym_nondegGT}), only by the factor $\Gamma^{\alpha}+\Gamma^{\bar{\alpha}}\cos\phi$ in the amplitude in front of the oscillating term $\sin\left(\varepsilon_{g}\left(\phi\right)t\right)$. Applying $\Gamma^{L}\ll\Gamma^{R}$ to Eq.~(\ref{I1_udwnsym_nondegGT}) with $\alpha=L$ leads $\Gamma^{\alpha}+\Gamma^{\bar{\alpha}}\cos\phi\approx\Gamma^{R}\cos\phi$. In Fig.~\ref{crnts_S50-even}(a), the values of the lines with $\phi=\pi/2$ and $\phi=3\pi/2$ (the overlapping blue dashed and the brown short-dashed lines) are between the values of the lines at $\phi=0$ (the red solid line) and $\phi=\pi$ (the green dash-dotted line) at times that these different lines do not cross. This coincides with the dominance of $\Gamma^{R}\cos\phi$ in the amplitude before $\sin\left(\varepsilon_{g}\left(\phi\right)t\right)$, showing $\cos(\phi=0)>\cos(\phi=\pi/2)=\cos(\phi=3\pi/2)>\cos(\phi=\pi)$. This signifies that the term $\sin\left(\varepsilon_{g}\left(\phi\right)t\right)$ is important for this feature.
Note that the left-right asymmetry $\Gamma^{L}\ll\Gamma^{R}$ yields $\varepsilon_{g}\left(\phi\right)\approx\sqrt{\delta\varepsilon^{2}-(\Gamma^{R})^{2}}$ and therefore the main dependence on $\phi$ relies on the amplitude $\Gamma^{R}\cos\phi$. The crossing points correspond to the times satisfying $\sin\left(\varepsilon_{g}\left(\phi\right)t\right)=0$. On the other hand, applying $\Gamma^{L}\ll\Gamma^{R}$ to Eq.~(\ref{I1_udwnsym_nondegGT}) with $\alpha=R$ yields $\Gamma^{\alpha}+\Gamma^{\bar{\alpha}}\cos\phi\approx\Gamma^{R}$. Therefore, the step-like feature of $I^{[1],+}_{R}(t,\phi)$ in Fig.~\ref{crnts_S50-even}(b) appears for all these values of flux since the amplitude before $\sin\left(\varepsilon_{g}\left(\phi\right)t\right)$ is not as sensitive to $\phi$ as it is in the case of $I^{[1],+}_{L}(t,\phi)$.

\subsection{Large up-down asymmetry with left-right symmetry}

\subsubsection{degenerate QDs}
\label{appx-AsymUDW-deg}

The amplitudes in Eq.~(\ref{I_lrsym_gel_deg}), obtained with left-right symmetry and degenerate QDs, explicitly read
\begin{subequations}
\label{amp_I_lrsym_deg}
\begin{align}
\label{I1_deg_lrsym_ampfs}
 & I_{\text{f}/\text{s}}^{\left[1\right]}\left(\phi\right)\nonumber \\
= & -\frac{\bar{\Gamma}_{1}}{2}\left\{ \frac{\left(1\pm\Gamma_{d}/\delta\Gamma\left(\phi\right)\right)^{2}}{2}\right.
\nonumber \\
 & \left.+\frac{\left(1+\cos\phi\right)}{\delta\Gamma\left(\phi\right)^{2}}
 \left[\left(\bar{\Gamma}_{2}\right)^{2}\pm\bar{\Gamma}_{2}\left(\delta\Gamma\left(\phi\right)\pm\Gamma_{d}\right)\right]\right\},
\end{align}
\begin{align}
&I_{0}^{\left[1\right]}\left(\phi\right)=-\bar{\Gamma}_{1}\times\nonumber\\&\left\{ \frac{1-\left(\Gamma_{d}/\delta\Gamma\left(\phi\right)\right)^{2}}{2}-\frac{\left(1+\cos\phi\right)}{\delta\Gamma^{2}\left(\phi\right)}\left[\left(\bar{\Gamma}_{2}\right)^{2}+\bar{\Gamma}_{2}\Gamma_{d}\right]\right\} ,\label{I1_deg_lrsym_amp0}
\end{align}
and
\begin{align}
 & I_{\text{f}/\text{s}}^{\left[2\right]}\left(\phi\right)\nonumber \\
= & -\frac{\bar{\Gamma}_{2}}{2}\left\{ \frac{\left(1\mp\Gamma_{d}/\delta\Gamma\left(\phi\right)\right)^{2}}{2}\right.\nonumber \\
 & \left.+\frac{\left(1+\cos\phi\right)}{\delta\Gamma\left(\phi\right)^{2}}\left[\left(\bar{\Gamma}_{1}\right)^{2}\pm\bar{\Gamma}_{1}\left(\delta\Gamma\left(\phi\right)\mp\Gamma_{d}\right)\right]\right\} ,\label{I2_deg_lrsym_ampfs}
\end{align}
\begin{align}
&I_{0}^{\left[2\right]}\left(\phi\right)=-\bar{\Gamma}_{2}\times
\nonumber\\&
\left\{ \frac{1-\left(\Gamma_{d}/\delta\Gamma\left(\phi\right)\right)^{2}}{2}-\frac{\left(1+\cos\phi\right)}{\delta\Gamma^{2}\left(\phi\right)}\left[\left(\bar{\Gamma}_{1}\right)^{2}-\bar{\Gamma}_{1}\Gamma_{d}\right]\right\} .\label{I2_deg_lrsym_amp0}
\end{align}
\end{subequations}

In Eq.~(\ref{I1_deg_lrsym_ampfs}) and Eq.~(\ref{I2_deg_lrsym_ampfs}) the upper (lower) sign is for the f (s) amplitude. The up-down asymmetry can be manifested by applying $\bar{\Gamma}_{1}\ll\bar{\Gamma}_{2}$ to Eqs.~(\ref{I_lrsym_gel_deg}) and (\ref{amp_I_lrsym_deg}). This leads to $\Gamma_{d}\approx-\bar{\Gamma}_{2},\delta\Gamma\left(\phi\right)\approx\bar{\Gamma}_{2}$ and consequently
\begin{subequations}
\label{I_lrsym_lg12asym_deg}
\begin{equation}
\label{I_lrsym_lg12asym_deg-1s}
I_{\text{s}}^{\left[1\right]}\left(\phi\right)\approx-\frac{\bar{\Gamma}_{1}}{2}\left(1-\cos\phi\right),
\end{equation}
\begin{equation}
\label{I_lrsym_lg12asym_deg-1f}
I_{\text{f}}^{\left[1\right]}\left(\phi\right)\approx-\frac{\bar{\Gamma}_{1}}{2}\left(1+\cos\phi\right),
\end{equation}
\begin{equation}
\label{I_lrsym_lg12asym_deg-10}
I_{0}^{\left[1\right]}\left(\phi\right)\approx0.
\end{equation}
and
\begin{equation}
\label{I_lrsym_lg12asym_deg-2s}
I_{\text{s}}^{\left[2\right]}\left(\phi\right)\approx-\frac{\bar{\Gamma}_{2}}{2}\left(1+\cos\phi\right)\frac{\bar{\Gamma}_{1}^{2}}{\bar{\Gamma}_{2}^{2}},
\end{equation}
\begin{equation}
\label{I_lrsym_lg12asym_deg-2f}
I_{\text{f}}^{\left[2\right]}\left(\phi\right)\approx-\frac{\bar{\Gamma}_{2}}{2}\left[2+\left(1+\cos\phi\right)\frac{\bar{\Gamma}_{1}^{2}+2\bar{\Gamma}_{1}\bar{\Gamma}_{2}}{\bar{\Gamma}_{2}^{2}}\right],
\end{equation}
\begin{equation}
\label{I_lrsym_lg12asym_deg-20}
I_{0}^{\left[2\right]}\left(\phi\right)\approx\bar{\Gamma}_{2}\left(1+\cos\phi\right)\frac{\bar{\Gamma}_{1}^{2}+\bar{\Gamma}_{1}\bar{\Gamma}_{2}}{\bar{\Gamma}_{2}^{2}}.
\end{equation}
\end{subequations}

In Eq.~(\ref{I_lrsym_lg12asym_deg}) for the part of $I_{\alpha}^{[1]}\left(t,\phi\right)$, the two amplitudes for the fast and slow decay terms, $I_{\text{f}}^{\left[1\right]}\left(\phi\right)$ and $I_{\text{s}}^{\left[1\right]}\left(\phi\right)$ respectively given by Eq.~(\ref{I_lrsym_lg12asym_deg-1f}) and Eq.~(\ref{I_lrsym_lg12asym_deg-1s}), are comparable to each other. However, for $I_{\alpha}^{[2]}\left(t,\phi\right)$, the large up-down asymmetry has led to $ \left\vert I_{\text{s}}^{\left[2\right]}\left(\phi\right)\right\vert \ll \left\vert I_{\text{f}}^{\left[2\right]}\left(\phi\right)\right\vert$ (comparing Eq.~(\ref{I_lrsym_lg12asym_deg-2s}) with Eq.~(\ref{I_lrsym_lg12asym_deg-2f}) under $\bar{\Gamma}^{}_{1}/\bar{\Gamma}^{}_{2}\ll1$). Therefore, $I_{\alpha}^{[2]}\left(t,\phi\right)$ is governed by the fast decay term $e^{-(\Gamma+\delta\Gamma(\phi))t}$, showing a faster approach to the steady-state limit (see the time scales of Fig.~\ref{crnts_S37}(b)), in contrast to the slow approach exhibited by $I_{\alpha}^{[1]}\left(t,\phi\right)$ (see the time scales of Fig.~\ref{crnts_S37}(a)). Furthermore, the factor $(1+\cos\phi)$ in Eq.~(\ref{I_lrsym_lg12asym_deg-2s}), Eq.~(\ref{I_lrsym_lg12asym_deg-2f}) and Eq.~(\ref{I_lrsym_lg12asym_deg-20}) are all multiplied by numbers that scale with $\bar{\Gamma}^{}_{1}/\bar{\Gamma}^{}_{2}\ll1$. Comparing Eq.~(\ref{I_lrsym_lg12asym_deg-1s}), Eq.~(\ref{I_lrsym_lg12asym_deg-1f}) with Eq.~(\ref{I_lrsym_lg12asym_deg-2s}), Eq.~(\ref{I_lrsym_lg12asym_deg-2f}) and Eq.~(\ref{I_lrsym_lg12asym_deg-20}) with $\bar{\Gamma}^{}_{1}/\bar{\Gamma}^{}_{2}\ll1$ then explains why $I_{\alpha}^{[1]}\left(t,\phi\right)$ (of Fig.~\ref{crnts_S37}(a)) shows a clear flux dependence while $I_{\alpha}^{[2]}\left(t,\phi\right)$ (of Fig.~\ref{crnts_S37}(b)) does not.

\subsubsection{non-degenerate QDs for the even components}
\label{appx-AsymUDW-nondeg}

The amplitudes in Eq.~(\ref{I_lrsym_gel_even}) for $I_{\alpha}^{[1],+}\left(t,\phi\right)$, obtained with left-right symmetry but non-degenerate QDs, are explicitly given
by
\begin{subequations}
\label{amp_lrsym_gel_even-nondeg}
\begin{align}
\label{I1_nondeg_lrsym_ampfs}
 & I_{\text{f}/\text{s}}^{\left[1\right]}\left(\phi\right)= -\frac{\bar{\Gamma}_{1}}{2}\left\{ \frac{\left(1+\left\vert r_{g}\left(\phi\right)\right\vert ^{2}\pm2\text{Re}\left(r_{g}\left(\phi\right)\right)\right)}{2}\right.
\nonumber \\
 & \left.+\frac{\left(1+\cos\phi\right)}{\left\vert \Gamma_{g}\left(\phi\right)\right\vert ^{2}}\left[\left(\bar{\Gamma}_{2}\right)^{2}\pm\bar{\Gamma}_{2}\left(\delta\Gamma\left(\phi\right)\pm\Gamma_{d}\right)\right]\right\} ,
\end{align}
\begin{align}
\label{I1_nondeg_lrsym_ampCos}
I_{\text{osc}}^{\left[1\right]}\left(\phi\right)&=-\bar{\Gamma}_{1}\left\{ \frac{1-\left\vert r_{g}\left(\phi\right)\right\vert ^{2}}{2}\right.
\nonumber\\&
\left.-\frac{\left(1+\cos\phi\right)}{\left\vert \Gamma_{g}\left(\phi\right)\right\vert ^{2}}\left[\left(\bar{\Gamma}_{2}\right)^{2}+\bar{\Gamma}_{2}\Gamma_{d}\right]\right\} ,
\end{align}
and
\begin{align}
\label{I1_nondeg_lrsym_ampSin}
I_{\text{oss}}^{\left[1\right]}\left(\phi\right)=-\bar{\Gamma}_{1}\left\{ \text{Im}\left(r_{g}\left(\phi\right)\right)-\frac{\left(1+\cos\phi\right)}{\left\vert \Gamma_{g}\left(\phi\right)\right\vert ^{2}}\bar{\Gamma}_{2}\varepsilon_{g}\left(\phi\right)\right\} .
\end{align}
The amplitudes for $I_{\alpha}^{[2],+}\left(t,\phi\right)$ are given
by
\begin{align}
\label{I2_nondeg_lrsym_ampfs}
 & I_{\text{f}/\text{s}}^{\left[2\right]}\left(\phi\right)
=  -\frac{\bar{\Gamma}_{2}}{2}\left\{ \frac{\left(1+\left\vert r_{g}\left(\phi\right)\right\vert ^{2}\mp2\text{Re}\left(r_{g}\left(\phi\right)\right)\right)}{2}\right.\nonumber \\
 & \left.+\frac{\left(1+\cos\phi\right)}{\left\vert \Gamma_{g}\left(\phi\right)\right\vert ^{2}}\left[\left(\bar{\Gamma}_{1}\right)^{2}\pm\bar{\Gamma}_{1}\left(\delta\Gamma\left(\phi\right)\mp\Gamma_{d}\right)\right]\right\} ,
\end{align}
\begin{align}
\label{I2_nondeg_lrsym_ampCos}
I_{\text{osc}}^{\left[2\right]}\left(\phi\right)=-\bar{\Gamma}_{2}\left\{ \frac{1-\left\vert r_{g}\left(\phi\right)\right\vert ^{2}}{2}-\frac{\left(1+\cos\phi\right)}{\left\vert \Gamma_{g}\left(\phi\right)\right\vert ^{2}}\left[\left(\bar{\Gamma}_{1}\right)^{2}-\bar{\Gamma}_{1}\Gamma_{d}\right]\right\} ,
\end{align}
\begin{align}
\label{I2_nondeg_lrsym_ampSin}
I_{\text{oss}}^{\left[2\right]}\left(\phi\right)=-\bar{\Gamma}_{2}\left\{ -\text{Im}\left(r_{g}\left(\phi\right)\right)-\frac{\left(1+\cos\phi\right)}{\left\vert \Gamma_{g}\left(\phi\right)\right\vert ^{2}}\bar{\Gamma}_{1}\varepsilon_{g}\left(\phi\right)\right\} .
\end{align}
\end{subequations} In Eq.~(\ref{I1_nondeg_lrsym_ampfs}) and Eq.~(\ref{I2_nondeg_lrsym_ampfs}) the upper (lower) sign is for fast (slow) decay term. The notation $r_{g}\left(\phi\right)$ is defined by
\begin{align}
\label{rg}
r_{g}\left(\phi\right)=\frac{\Gamma_{d}+i\delta\varepsilon}{\Gamma_{g}\left(\phi\right)}.
\end{align}

The up-down asymmetry induced by $\bar{\Gamma}_{1}\ll\bar{\Gamma}_{2}$, leading to $\Gamma_{d} \approx-\bar{\Gamma}_{2}$, $\Gamma_{g}\left(\phi\right)\approx\bar{\Gamma}_{2}-i\varepsilon_{g}\left(\phi\right)$, and consequently $r_{g}\left(\phi\right)\approx-1$, has rendered the amplitudes Eq.~(\ref{amp_lrsym_gel_even-nondeg}) to
\begin{subequations}
\label{I1_nondeg_lrsym_lgupdwasym}
\begin{align}
\label{I1_nondeg_lrsym_amps_lgupdwasym}
I_{\text{s}}^{\left[1\right]}\left(\phi\right)\approx
-\frac{\bar{\Gamma}_{1}}{2}\left[2-\left(1+\cos\phi\right)\frac{\bar{\Gamma}_{2}^{2}}{\bar{\Gamma}_{2}^{2}+\delta\varepsilon^{2}}\right],
\end{align}
\begin{align}
\label{I1_nondeg_lrsym_ampf_lgupdwasym}
I_{\text{f}}^{\left[1\right]}\left(\phi\right)\approx
-\frac{\bar{\Gamma}_{1}}{2}\left(1+\cos\phi\right)\frac{\bar{\Gamma}_{2}^{2}}{\bar{\Gamma}_{2}^{2}+\delta\varepsilon^{2}},
\end{align}
\begin{align}
\label{I1_nondeg_lrsym_ampcos_lgupdwasym}
I_{\text{osc}}^{\left[1\right]}\left(\phi\right)\approx0,
\end{align}
\begin{align}
\label{I1_nondeg_lrsym_ampsin_lgupdwasym}
I_{\text{oss}}^{\left[1\right]}\left(\phi\right)\approx
-\bar{\Gamma}_{1}\left(1+\cos\phi\right)\frac{\bar{\Gamma}_{2}\delta\varepsilon}{\bar{\Gamma}_{2}^{2}+\delta\varepsilon^{2}},
\end{align}
and
\begin{align}
\label{I2_nondeg_lrsym_amps_lgupdwasym}
I_{\text{s}}^{\left[2\right]}\left(\phi\right)\approx
-\frac{\bar{\Gamma}_{2}}{2}\left(1+\cos\phi\right)\frac{\bar{\Gamma}_{1}^{2}}{\bar{\Gamma}_{2}^{2}+\delta\varepsilon^{2}},
\end{align}
\begin{align}
\label{I2_nondeg_lrsym_ampf_lgupdwasym}
I_{\text{f}}^{\left[2\right]}\left(\phi\right)\approx
-\frac{\bar{\Gamma}_{2}}{2}\left[2+\left(1+\cos\phi\right)\frac{\bar{\Gamma}_{1}^{2}+2\bar{\Gamma}_{1}\bar{\Gamma}_{2}}{\bar{\Gamma}_{2}^{2}+\delta\varepsilon^{2}}\right],
\end{align}
\begin{align}
\label{I2_nondeg_lrsym_ampcos_lgupdwasym}
I_{\text{osc}}^{\left[2\right]}\left(\phi\right)\approx
\bar{\Gamma}_{2}\left(1+\cos\phi\right)\frac{\bar{\Gamma}_{1}^{2}+\bar{\Gamma}_{1}\bar{\Gamma}_{2}}{\bar{\Gamma}_{2}^{2}+\delta\varepsilon^{2}},
\end{align}
\begin{align}
\label{I2_nondeg_lrsym_ampsin_lgupdwasym}
I_{\text{oss}}^{\left[2\right]}\left(\phi\right)\approx
-\bar{\Gamma}_{2}\left(1+\cos\phi\right)\frac{\bar{\Gamma}_{1}\delta\varepsilon}{\bar{\Gamma}_{2}^{2}+\delta\varepsilon^{2}}.
\end{align}
\end{subequations}

Similar to the discussions in Sec.~\ref{sec_model_fluxindep_sym2-deg} and Appendix \ref{appx-AsymLR-nondeg}, Eq.~(\ref{I1_nondeg_lrsym_lgupdwasym}) shows that when $\bar{\Gamma}_1<<\bar{\Gamma}_2$, then  $\left\vert I_{\text{s}}^{\left[2\right]}\left(\phi\right)\right\vert \ll \left\vert I_{\text{f}}^{\left[2\right]}\left(\phi\right)\right\vert$ while $I_{\text{s}}^{\left[1\right]}\left(\phi\right)$ and $I_{\text{f}}^{\left[1\right]}\left(\phi\right)$ are comparable. Figure~\ref{crnts_S5-even-odd}(a1) for $I_{\alpha}^{[1]}\left(t,\phi\right)$ for $t<5\Gamma^{-1}$ shows more visible variations with the flux in contrast to Fig.~\ref{crnts_S5-even-odd}(a2) for $I_{\alpha}^{[2]}\left(t,\phi\right)$. This is also explained by Eq.~(\ref{I1_nondeg_lrsym_lgupdwasym}) using similar reasonings already described in Sec.~\ref{sec_model_fluxindep_sym2-deg}.

The up-down asymmetry in terms of showing the spike feature or not is explained in the followings. Comparing Eq.~(\ref{I1_nondeg_lrsym_ampsin_lgupdwasym}) with Eq.~(\ref{I1_nondeg_lrsym_amps_lgupdwasym}) and Eq.~(\ref{I1_nondeg_lrsym_ampf_lgupdwasym}), we find that the amplitude for the oscillation $I_{\text{oss}}^{\left[1\right]}\left(\phi\right)$ is comparable to the amplitudes for pure decays $I_{\text{s}}^{\left[1\right]}\left(\phi\right)$ and $I_{\text{f}}^{\left[1\right]}\left(\phi\right)$ in $I_{\alpha}^{[1],+}\left(t,\phi\right)$. However, for $I_{\alpha}^{[2],+}\left(t,\phi\right)$, by comparing the amplitudes for oscillations Eq.~(\ref{I2_nondeg_lrsym_ampcos_lgupdwasym}) and Eq.~(\ref{I2_nondeg_lrsym_ampsin_lgupdwasym}) with the dominant amplitude for the pure decay Eq.~(\ref{I2_nondeg_lrsym_ampf_lgupdwasym}), we find that $\left\vert I_{\text{osc}}^{\left[2\right]}\left(\phi\right)\right\vert \ll \left\vert I_{\text{f}}^{\left[2\right]}\left(\phi\right)\right\vert$ and $\left\vert I_{\text{oss}}^{\left[2\right]}\left(\phi\right)\right\vert \ll \left\vert I_{\text{f}}^{\left[2\right]}\left(\phi\right)\right\vert$ due to the asymmetry $\bar{\Gamma}_{1}\ll\bar{\Gamma}_{2}$. Henceforth, the oscillatory feature can be manifested before it is fully damped in $I_{\alpha}^{[1],+}\left(t,\phi\right)$, as shown by Fig~\ref{crnts_S5-even-odd}(a1). But the amplitudes of oscillations are negligible in comparisons to the pure decays in $I_{\alpha}^{[2],+}\left(t,\phi\right)$, as shown by Fig~\ref{crnts_S5-even-odd}(a2).

\subsubsection{non-degenerate QDs for the odd components}
\label{appx-AsymUDW-oddnondeg}

Applying $\bar{\Gamma}_{1}\ll\bar{\Gamma}_{2}$ to Eq.~(\ref{I_lrsym_gel_odd}) results in
\begin{align}
\label{I1odd_lrsym_lgudasym}
&I_{\alpha}^{[1],-}\left(t,\phi\right)\approx\frac{\bar{\Gamma}_{1}\bar{\Gamma}_{2}}{\left\vert \Gamma_{g}\left(\phi\right)\right\vert ^{2}}\zeta_{\alpha}\sin\phi\times,\nonumber \\
 & \left\{ -\delta\varepsilon e^{-\left(\Gamma-\delta\Gamma\left(\phi\right)\right)t}\right.\left.+e^{-\Gamma t}\left[\delta\varepsilon\cos\left(\delta\varepsilon t\right)+\bar{\Gamma}_{2}\sin\left(\delta\varepsilon t\right)\right]\right\} ,
\end{align} and
\begin{align}
\label{I2odd_lrsym_lgudasym}
&I_{\alpha}^{[2],-}\left(t,\phi\right)\approx\frac{\bar{\Gamma}_{1}\bar{\Gamma}_{2}}{\left\vert \Gamma_{g}\left(\phi\right)\right\vert ^{2}}\zeta_{\alpha}\sin\phi\times\nonumber \\
 & \left\{ -\delta\varepsilon e^{-\left(\Gamma+\delta\Gamma\left(\phi\right)\right)t}\right.\left.+e^{-\Gamma t}\left[\delta\varepsilon\cos\left(\delta\varepsilon t\right)-\bar{\Gamma}_{2}\sin\left(\delta\varepsilon t\right)\right]\right\} ,
\end{align} Comparing Eq.~(\ref{I2odd_lrsym_lgudasym}) with Eq.~(\ref{I1odd_lrsym_lgudasym}), we find that $I_{\alpha}^{[2],-}\left(t,\phi\right)$ does not have the slowest decay term $e^{-\left(\Gamma-\delta\Gamma\left(\phi\right)\right)t}$ as $I_{\alpha}^{[1],-}\left(t,\phi\right)$ does. This explains why $I_{\alpha}^{[1],-}\left(t,\phi\right)$ approach the steady-state limit also much slower than $I_{\alpha}^{[2],-}\left(t,\phi\right)$ does, as in the case for the even components.

The decay factor that damps the oscillations, the second term of Eq.~(\ref{I1odd_lrsym_lgudasym}), is given by $e^{-\Gamma t}$. It decays faster than $e^{-\left(\Gamma-\delta\Gamma\left(\phi\right)\right)t}$. Therefore the oscillation has been damped before a period is visible while the slow decay has not completely vanished. In Eq.~(\ref{I2odd_lrsym_lgudasym}), the pure decay factor is given by $e^{-\left(\Gamma+\delta\Gamma\left(\phi\right)\right)t}$ which decays faster than the damping of the amplitudes for the oscillations, $e^{-\Gamma t}$. This describes the visible oscillations whose amplitudes decay with time as seen in Fig.~\ref{crnts_S5-even-odd}(b2) for $I_{\alpha}^{[2],-}\left(t,\phi\right)$.
\\
\section{Properties of the even part of the transient transmission-like function $\mathcal{T}^{(+)}_{\alpha}(t,\omega)$.}
\label{appx-Tplusplus}
The purely transient component of the current that depends on the chemical potentials of the reservoirs, $\overline{I}^{\text{em.}}_{\alpha }(t,\phi)$, is given by Eq.~(\ref{dynamic-crnt-0-0}). The transmission-like function under the wide-band approximation can be explicitly analyzed by Eq.~(\ref{time-dep_Ts_WB}) with the aid of Eq.~(\ref{retardedG-intime}). Below we show that when $\bar{\Gamma}_{ii}^{L}=\bar{\Gamma}_{ii}^{R}$ holds, then $\overline{I}^{\text{em.},+}_{L}(t,\phi)=\overline{I}^{\text{em.},+}_{R }(t,\phi)$. We also prove that when Eq.~(\ref{updwn_bdsym}) or Eq.~(\ref{degDQD}) hold, then Eq.~(\ref{sym_trsmP}) is satisfied.

Explicitly, without assuming any symmetry among the bonds or  the on-site energies of the QDs, Eq.~(\ref{time-dep_Ts_WB}) for $\mathcal{T}_{\alpha}^{\left(+\right)}\left(t,\omega\right)$ can be written as,
\begin{subequations}
\label{Tplus_gel}
\begin{align}
\label{Tplus1+2}
\mathcal{T}_{\alpha}^{\left(+\right)}\left(t,\omega\right)=\left\{ \mathcal{T}_{\alpha1}^{\left(+\right)}\left(t,\omega\right)+\mathcal{T}_{\alpha2}^{\left(+\right)}\left(t,\omega\right)\right\} ,
\end{align}
where
\begin{align}
\label{Tplus1}
\mathcal{T}_{\alpha1}^{\left(+\right)}\left(t,\omega\right) & =\text{Tr}\left\{
\boldsymbol{\Gamma}^{\alpha}\left[i\left(\boldsymbol{\bar{G}}^{r}(t,\omega)-\boldsymbol{\bar{G}}^{a}(t,\omega)\right)
\right]\right\}\nonumber \\
 & =\mathcal{T}_{\alpha1,d}^{\left(+\right)}\left(t,\omega\right)+\mathcal{T}_{\alpha1,f}^{\left(+\right)}\left(t,\omega\right),
\end{align}
in which
\begin{align}
\label{Tplus1_d}
&\mathcal{T}_{\alpha1,d}^{\left(+\right)}\left(t,\omega\right)=
 2\left(\bar{\Gamma}_{11}^{\alpha}+\bar{\Gamma}_{22}^{\alpha}\right)\text{Re}\left[\tilde{b}_{+}\left(t,\omega\right)\right]
 \nonumber \\&
+2\left(\bar{\Gamma}_{11}^{\alpha}-\bar{\Gamma}_{22}^{\alpha}\right)\text{Re}\left[r_{g}\left(\phi\right)\tilde{b}_{-}\left(t,\omega\right)\right],
\end{align} and
\begin{align}
\label{Tplus1_f}
\mathcal{T}_{\alpha1,f}^{\left(+\right)}\left(t,\omega\right)   =4\bar{\Gamma}_{12}^{\alpha}\left[\bar{\Gamma}_{12}^{\alpha}+\bar{\Gamma}_{12}^{\bar{\alpha}}\cos\phi\right]
\text{Re}\left(\frac{\tilde{b}_{-}\left(t,\omega\right)}{\Gamma_{g}\left(\phi\right)}\right).
\end{align} The second term in Eq.~(\ref{Tplus1+2}) reads
\begin{align}
\label{Tplus2}
\mathcal{T}_{\alpha2}^{\left(+\right)}\left(t,\omega\right) & =-\text{Tr}\left\{
\boldsymbol{\Gamma}^{\alpha}\left[
\boldsymbol{\bar{G}}^{r}(t,\omega)(\boldsymbol{\Gamma}^{L}+\boldsymbol{\Gamma}^{R})\boldsymbol{\bar{G}}^{a}(t,\omega)\right]\right\}\nonumber \\
 & =\mathcal{T}_{\alpha2,d}^{\left(+\right)}\left(t,\omega\right)+\mathcal{T}_{\alpha2,f}^{\left(+\right)}\left(t,\omega\right),
\end{align}
in which
\begin{widetext}
\begin{align}
\label{Tplus2_d}
\mathcal{T}_{\alpha2,d}^{\left(+\right)}\left(t,\omega\right) & =
\frac{\left(\bar{\Gamma}_{11}^{\alpha}+\bar{\Gamma}_{22}^{\alpha}\right)}{2}\left\{ \left(\Gamma_{11}+\Gamma_{22}\right)\left[\left\vert \tilde{b}_{+}\left(t,\omega\right)\right\vert ^{2}+\left(\left\vert r_{g}\left(\phi\right)\right\vert ^{2}+\left\vert \frac{\Gamma_{12}\left(\phi\right)}{\Gamma_{g}\left(\phi\right)}\right\vert ^{2}\right)\left\vert \tilde{b}_{-}\left(t,\omega\right)\right\vert ^{2}\right]\right.\nonumber \\
 & \left.+2\left(\Gamma_{11}-\Gamma_{22}\right)\text{Re}\left(\tilde{b}_{+}\left(t,\omega\right)r_{g}^{*}\left(\phi\right)\tilde{b}_{-}^{*}\left(t,\omega\right)\right)+4\left\vert \Gamma_{12}\left(\phi\right)\right\vert ^{2}\text{Re}\left(\tilde{b}_{+}\left(t,\omega\right)\frac{\tilde{b}_{-}^{*}\left(t,\omega\right)}{\Gamma_{g}^{*}\left(\phi\right)}\right)\right\} \nonumber \\
+ & \frac{\left(\bar{\Gamma}_{11}^{\alpha}-\bar{\Gamma}_{22}^{\alpha}\right)}{2}\left\{ \left(\Gamma_{11}-\Gamma_{22}\right)\left[\left\vert \tilde{b}_{+}\left(t,\omega\right)\right\vert ^{2}+\left(\left\vert r_{g}\left(\phi\right)\right\vert ^{2}-\left\vert \frac{\Gamma_{12}\left(\phi\right)}{\Gamma_{g}\left(\phi\right)}\right\vert ^{2}\right)\left\vert \tilde{b}_{-}\left(t,\omega\right)\right\vert ^{2}\right]\right.\nonumber \\
 & \left.+2\left(\Gamma_{11}+\Gamma_{22}\right)\text{Re}\left(\tilde{b}_{+}\left(t,\omega\right)r_{g}^{*}\left(\phi\right)\tilde{b}_{-}^{*}\left(t,\omega\right)\right)+4\left\vert \Gamma_{12}\left(\phi\right)\right\vert ^{2}\left\vert \tilde{b}_{-}\left(t,\omega\right)\right\vert ^{2}\text{Re}\left(\frac{r_{g}\left(\phi\right)}{\Gamma_{g}^{*}\left(\phi\right)}\right)\right\} ,
\end{align} and
\begin{align}
\label{Tplus2_f}
\mathcal{T}_{\alpha2,f}^{\left(+\right),+}\left(t,\omega,\phi\right)
= & 2\bar{\Gamma}_{12}^{\alpha}\left(\bar{\Gamma}_{12}^{\alpha}+\bar{\Gamma}_{12}^{\bar{\alpha}}\cos\phi\right)\times\nonumber \\
 & \left\{ \left\vert \tilde{b}_{+}\left(t,\omega\right)\right\vert ^{2}+(\Gamma_{11}+\Gamma_{22})\text{Re}\left(\tilde{b}_{+}\left(t,\omega\right)\frac{\tilde{b}_{-}^{*}\left(t,\omega\right)}{\Gamma_{g}^{*}\left(\phi\right)}\right)+\right.\nonumber \\
 & \left.+\left(\left(\Gamma_{11}-\Gamma_{22}\right)\text{Re}\left[\frac{r_{g}\left(\phi\right)}{\Gamma_{g}^{*}\left(\phi\right)}\right]+\left\vert \frac{\Gamma_{12}\left(\phi\right)}{\Gamma_{g}\left(\phi\right)}\right\vert ^{2}-\left\vert r_{g}\left(\phi\right)\right\vert ^{2}\right)\left\vert \tilde{b}_{-}\left(t,\omega\right)\right\vert ^{2}\right\} .
\end{align}
\end{widetext}
\end{subequations} Here $\tilde{b}_{\pm}\left(t,\omega\right)$ are defined in Eq.~(\ref{time-amplitudes-cp-omega}), supplemented by Eq.~(\ref{time-amplitudes-cp}) and Eq.~(\ref{retardedG-intime}) and $r_{g}\left(\phi\right)$ is defined in Eq.~(\ref{rg}).
Equations (\ref{Tplus1}) and (\ref{Tplus2_d}) are all even in the flux $\phi$. The only part that is odd in the flux comes from the second term on the second line of Eq.~(\ref{Tplus2}), namely, $\mathcal{T}_{\alpha2,f}^{\left(+\right)}\left(t,\omega,\phi\right)$. This odd term has already been discussed in Sec.~\ref{sec_model_transit_oddspec}, given by Eq.~(\ref{odd-Tpm}), as $\mathcal{T}_{\alpha2,f}^{\left(+\right),-}\left(t,\omega,\phi\right)=\mathcal{T}_{\alpha}^{\left(+\right),-}\left(t,\omega,\phi\right)$. In this Appendix we show only its even part, $\mathcal{T}_{\alpha2,f}^{\left(+\right),+}\left(t,\omega,\phi\right)$, given by Eq.~(\ref{Tplus2_f}). Taking $\bar{\Gamma}_{ii}^{L}=\bar{\Gamma}_{ii}^{R}$ in Eq.~(\ref{Tplus_gel}) one obtains
$\mathcal{T}_{L}^{\left(+\right),+}\left(t,\omega,\phi\right)=\mathcal{T}_{R}^{\left(+\right),+}\left(t,\omega,\phi\right)$ and therefore
$\overline{I}^{\text{em.},+}_{L}(t,\phi)=\overline{I}^{\text{em.},+}_{R }(t,\phi)$ by Eq.~(\ref{dynamic-crnt-0-0}).

Next we discuss the symmetry property of $\mathcal{T}_{\alpha}^{\left(+\right)}\left(t,\omega,\phi\right)$ in terms of its distribution in $\omega$, which is determined by $\tilde{b}_{\pm}^{}(t,\omega)$. Under the condition that Eq.~(\ref{updwn_bdsym}) holds, the second term of Eq.~(\ref{Tplus1_d}), the first term on the second line and the last two lines of Eq.~(\ref{Tplus2_d}) all vanish. Therefore, as long as $\text{Re}\left[\tilde{b}_{+}\left(t,\omega\right)\right]$
, $\left\vert \tilde{b}_{\pm}\left(t,\omega\right)\right\vert ^{2}$
and $\text{Re}\left(\tilde{b}_{+}\left(t,\omega\right)\frac{\tilde{b}_{-}^{*}\left(t,\omega\right)}{\Gamma_{g}^{*}\left(\phi\right)}\right)$
are symmetric with respect to $\omega=\varepsilon_{0}$, then Eq.~(\ref{sym_trsmP}) is satisfied for both the even and the odd components (the same set of functions is involved for the odd component). Note that when Eq.~(\ref{updwn_bdsym}) holds, then $\Gamma_{g}(\phi)$ is either purely real or purely imaginary. Therefore, we only have to discuss the above functions under these two cases of $\Gamma_{g}(\phi)$ being purely real and purely imaginary.

On the other hand, when Eq.~(\ref{degDQD}) holds then $\Gamma_{g}(\phi)$ is purely real and consequently $r_{g}(\phi)$ also becomes real (see Eq.~(\ref{cmpx-decayrate}) and Eq.~(\ref{rg})). When Eq.~(\ref{degDQD}) holds but with $\bar{\Gamma}_{11}^{\alpha}\ne\bar{\Gamma}_{22}^{\alpha}$, then all the following quantities, $\text{Re}\left[\tilde{b}_{\pm}\left(t,\omega\right)\right]$,
$\left\vert \tilde{b}_{\pm}\left(t,\omega\right)\right\vert ^{2}$, and
$\text{Re}\left(\tilde{b}_{+}\left(t,\omega\right)\tilde{b}_{-}^{*}\left(t,\omega\right)\right)$
have to be symmetric with respect to $\omega=\varepsilon_{0}$ in order that Eq.~(\ref{sym_trsmP}) is fulfilled.

We define,
\begin{align}
\label{time-amplitudes-cf}
\tilde{c}_{\pm}\left(t,\omega\right)=\int_{0}^{t}d\tau e^{i\omega\tau}c^{}_{\pm}\left(\tau,\phi\right),
\end{align} where $c^{}_{\pm}\left(\tau,\phi\right)$ is given in Eq.~(\ref{time-amplitudes}). The $\omega$ dependent terms in Eq.~(\ref{Tplus_gel}) and Eq.~(\ref{odd-Tpm}) that are relevant to the present discussions for Eq.~(\ref{sym_trsmP}) are conveniently expressed in terms of $\tilde{c}_{\pm}\left(t,\omega\right)$ as
\begin{align}
\label{bpmabsq}
&\left\vert \tilde{b}_{\pm}\left(t,\omega\right)\right\vert ^{2}=
\nonumber\\
&\frac{\left\vert \tilde{c}_{+}\left(t,\omega\right)\right\vert ^{2}+\left\vert \tilde{c}_{-}\left(t,\omega\right)\right\vert ^{2}\pm2\text{Re}\left[\tilde{c}_{+}\left(t,\omega\right)\tilde{c}_{-}^{*}\left(t,\omega\right)\right]}{4},
\end{align}
\begin{align}
\label{bplusminsta-0}
\text{Re}\left(\tilde{b}_{+}\left(t,\omega\right)\tilde{b}_{-}^{*}\left(t,\omega\right)\right)
=\frac{\left\vert \tilde{c}_{+}\left(t,\omega\right)\right\vert ^{2}-\left\vert \tilde{c}_{-}\left(t,\omega\right)\right\vert ^{2}}{4},
\end{align}
\begin{align}
\label{bplusminsta}
&\text{Re}\left(\frac{\tilde{b}_{+}\left(t,\omega\right)\tilde{b}_{-}^{*}\left(t,\omega\right)}{\Gamma_{g}^{*}(\phi)}\right)=
\nonumber\\&
\left\{
\begin{array}{cc}
\frac{\left\vert \tilde{c}_{+}\left(t,\omega\right)\right\vert ^{2}-\left\vert \tilde{c}_{-}\left(t,\omega\right)\right\vert ^{2}}{4\delta\Gamma(\phi)},
&~\text{if}~\Gamma_{g}(\phi)=\delta\Gamma(\phi)
\\
\frac{\text{Im}\left(\tilde{c}_{+}\left(t,\omega\right)\tilde{c}_{-}^{*}\left(t,\omega\right)\right)}{2\varepsilon_{g}(\phi)},
&~\text{if}~\Gamma_{g}(\phi)=i\varepsilon_{g}(\phi)
\end{array}
\right.
\end{align}


\subsection{The cases for $\Gamma_{g}\left(\phi\right)=\delta\Gamma\left(\phi\right)$}

When $\Gamma_{g}\left(\phi\right)$ is real, then by Eq.~(\ref{time-amplitudes}) and Eq.~(\ref{time-amplitudes-cf}), we have
\begin{widetext}
\begin{align}
\tilde{c}_{\pm}\left(t,\omega\right) & =\frac{e^{-\left(\Gamma\pm\delta\Gamma\left(\phi\right)\right)t/2}\left(\omega-\varepsilon_{0}\right)\sin\left[\left(\omega-\varepsilon_{0}\right)t\right]
-\left[\left(\Gamma\pm\delta\Gamma\left(\phi\right)\right)/2\right]e^{-\left(\Gamma\pm\delta\Gamma\left(\phi\right)\right)t/2}
\left(\cos\left[\left(\omega-\varepsilon_{0}\right)t\right]-1\right)}
{\left(\omega-\varepsilon_{0}\right)^{2}+\left[\left(\Gamma\pm\delta\Gamma\left(\phi\right)\right)/2\right]^{2}}
\nonumber
\\
 & -i\frac{e^{-\left(\Gamma\pm\delta\Gamma\left(\phi\right)\right)t/2}\left[\left(\Gamma\pm\delta\Gamma\left(\phi\right)\right)/2\right]\sin\left[\left(\omega-\varepsilon_{0}\right)t\right]+\left(\omega-\varepsilon_{0}\right)\left(e^{-\left(\Gamma\pm\delta\Gamma\left(\phi\right)\right)t/2}\cos\left[\left(\omega-\varepsilon_{0}\right)t\right]-1\right)}{\left(\omega-\varepsilon_{0}\right)^{2}+\left[\left(\Gamma\pm\delta\Gamma\left(\phi\right)\right)/2\right]^{2}}.
\end{align}
\end{widetext}
Consequently,
\begin{align}
\label{csym-1}
&\text{Re}\left[\tilde{c}_{\pm}\left(t,\varepsilon_{0}+\omega\right)\right]=\text{Re}\left[\tilde{c}_{\pm}\left(t,\varepsilon_{0}-\omega\right)\right],
\nonumber\\
&\text{Im}\left[\tilde{c}_{\pm}\left(t,\varepsilon_{0}+\omega\right)\right]=-\text{Im}\left[\tilde{c}_{\pm}\left(t,\varepsilon_{0}-\omega\right)\right].
\end{align} Using Eq.~(\ref{csym-1}), we find that $\left\vert \tilde{c}_{\pm}\left(t,\omega\right)\right\vert ^{2}=(\text{Re}\left[\tilde{c}_{\pm}\left(t,\omega\right)\right])^{2}+(\text{Im}\left[\tilde{c}_{\pm}\left(t,\omega\right)\right])^{2}$  and $\text{Re}\left[\tilde{c}_{+}\left(t,\omega\right)\tilde{c}_{-}^{*}\left(t,\omega\right)\right]=\left(\text{Re}\left[\tilde{c}_{+}\left(t,\omega\right)\right]\text{Re}\left[\tilde{c}_{-}\left(t,\omega\right)\right]
 +\text{Im}\left[\tilde{c}_{+}\left(t,\omega\right)\right]\text{Im}\left[\tilde{c}_{-}\left(t,\omega\right)\right]\right)$ are both symmetric in $\omega$ with respect to $\omega=\varepsilon_{0}$. Therefore, by Eq.~(\ref{time-amplitudes-cp-omega}), Eq.~(\ref{time-amplitudes-cp}), Eq.~(\ref{time-amplitudes}) and Eq.~(\ref{time-amplitudes-cf}) with Eq.~(\ref{csym-1}), it is found that
$\text{Re}\left[\tilde{b}_{\pm}\left(t,\omega\right)\right]$ is symmetric in $\omega$ with respect to $\omega=\varepsilon_{0}$.
By Eq.~(\ref{bpmabsq}), Eq.~(\ref{bplusminsta-0}) and Eq.~(\ref{bplusminsta}) with Eq.~(\ref{csym-1}), we find that $\left\vert \tilde{b}_{\pm}\left(t,\omega\right)\right\vert ^{2}$ and $\text{Re}\left(\tilde{b}_{+}\left(t,\omega\right)\tilde{b}_{-}^{*}\left(t,\omega\right)\right)$
are both symmetric in $\omega$ with respect to $\omega=\varepsilon_{0}$. This finishes the proof of Eq.~(\ref{sym_trsmP}) for the case that Eq.~(\ref{updwn_bdsym}) holds with $\vert\Gamma_{12}(\phi)\vert^{2}>\delta\varepsilon^{2}$ and the case Eq.~(\ref{degDQD}) holds, leading to $\Gamma_{g}(\phi)$ being real.

\subsection{The cases for $\Gamma_{g}\left(\phi\right)=i\varepsilon_{g}(\phi)$}
When $\Gamma_{g}\left(\phi\right)$ is purely imaginary, then by Eq.~(\ref{time-amplitudes}) and Eq.~(\ref{time-amplitudes-cf}), we have
\begin{align}
\tilde{c}_{\pm}\left(t,\omega\right)=C_{1}\left(\Omega_{\pm}\left(\omega\right),t\right)+iC_{2}\left(\Omega_{\pm}\left(\omega\right),t\right),
\end{align}
where
\begin{align}
&C_{1}\left(\Omega_{\pm}\left(\omega\right),t\right)=
\nonumber\\&
\frac{e^{-\Gamma t/2}\!\Omega_{\pm}\left(\omega\right)\!\sin\left[\Omega_{\pm}\left(\omega\right)t\right]\!\!-\!\!\left(\Gamma/2\right)\left(e^{-\Gamma t/2}\!\!\cos\left[\Omega_{\pm}\left(\omega\right)t\right]\!\!-\!\!1\right)}{\Omega_{\pm}\left(\omega\right)^{2}+\left[\Gamma/2\right]^{2}},
\end{align}
and
\begin{align}
&C_{2}\left(\Omega_{\pm}\left(\omega\right),t\right)=(-1)\times
\nonumber\\&
\frac{e^{-\Gamma t/2}\left[\Gamma/2\right]\sin\left[\Omega_{\pm}\left(\omega\right)t\right]
\!\!+\!\!\Omega_{\pm}\left(\omega\right)\!\!\left(e^{-\Gamma t/2}\!\cos\left[\Omega_{\pm}\left(\omega\right)t\right]\!\!-1\!\right)}{\Omega_{\pm}\left(\omega\right)^{2}+\left[\Gamma/2\right]^{2}}
\end{align}
in which
\begin{align}
\Omega_{\pm}\left(\omega\right)=\omega-\left(\varepsilon_{0}\pm\varepsilon_{g}\left(\phi\right)/2\right).
\end{align}
One immediately sees
\begin{align}
\label{csym-2}
&C_{1}\left(\omega,t\right)=C_{1}\left(-\omega,t\right),
\nonumber\\
&C_{2}\left(\omega,t\right)=-C_{2}\left(-\omega,t\right).
\end{align}
We examine the symmetric property of the function $\text{Re}\left[\tilde{b}_{+}\left(t,\omega\right)\right]=C_{1}\left(\Omega_{+}\left(\omega\right),t\right)+C_{1}\left(\Omega_{-}\left(\omega\right),t\right)$. Evaluating it at $\varepsilon_{0}\pm\omega$ yields,
\begin{align}
\label{Bplusupw}
&\text{Re}\left[\tilde{b}_{+}\left(t,\varepsilon_{0}+\omega\right)\right]=
\nonumber\\&
C_{1}\left(\omega-\varepsilon_{g}\left(\phi\right)/2,t\right)+C_{1}\left(\omega+\varepsilon_{g}\left(\phi\right)/2,t\right),
\end{align} and
\begin{align}
\label{Bplusdwnw}
&\text{Re}\left[\tilde{b}_{+}\left(t,\varepsilon_{0}-\omega\right)\right]
\nonumber\\&
=C_{1}\left(-\omega-\varepsilon_{g}\left(\phi\right)/2,t\right)+C_{1}\left(-\omega+\varepsilon_{g}\left(\phi\right)/2,t\right)
\nonumber\\
&=C_{1}\left(\omega+\varepsilon_{g}\left(\phi\right)/2,t\right)+C_{1}\left(\omega-\varepsilon_{g}\left(\phi\right)/2,t\right).
\end{align} We have applied Eq.~(\ref{csym-2}) in Eq.~(\ref{Bplusdwnw}) to obtain the last line.  Comparing Eq.~(\ref{Bplusupw}) with Eq.~(\ref{Bplusdwnw}), we certify
\begin{align}
\label{BplusSym}
\text{Re}\left[\tilde{b}_{+}\left(t,\varepsilon_{0}-\omega\right)\right]=\text{Re}\left[\tilde{b}_{+}\left(t,\varepsilon_{0}+\omega\right)\right].
\end{align} Next we examine the quantity,
\begin{align}
&\left\vert \tilde{c}_{+}\left(t,\omega\right)\right\vert ^{2}+\left\vert \tilde{c}_{-}\left(t,\omega\right)\right\vert ^{2}
=
\nonumber \\
&C_{1}^{2}\left(\Omega_{+}\left(\omega\right),t\right)+C_{2}^{2}\left(\Omega_{+}\left(\omega\right),t\right)
\nonumber \\
 &+C_{1}^{2}\left(\Omega_{-}\left(\omega\right),t\right)+C_{2}^{2}\left(\Omega_{-}\left(\omega\right),t\right).
\end{align} Evaluating it with $\varepsilon_{0}\pm\omega$ yields
\begin{align}
\label{cabs-1}
 & \left\vert \tilde{c}_{+}\left(t,\varepsilon_{0}+\omega\right)\right\vert ^{2}+\left\vert \tilde{c}_{-}\left(t,\varepsilon_{0}+\omega\right)\right\vert ^{2}
\nonumber\\
= & C_{1}^{2}\left(\omega-\varepsilon_{g}\left(\phi\right)/2,t\right)+C_{2}^{2}\left(\omega-\varepsilon_{g}\left(\phi\right)/2,t\right)
\nonumber\\&+C_{1}^{2}\left(\omega+\varepsilon_{g}\left(\phi\right)/2,t\right)+C_{2}^{2}\left(\omega+\varepsilon_{g}\left(\phi\right)/2,t\right)
\end{align}
and
\begin{align}
\label{cabs-2}
 & \left\vert \tilde{c}_{+}\left(t,\varepsilon_{0}-\omega\right)\right\vert ^{2}+\left\vert \tilde{c}_{-}\left(t,\varepsilon_{0}-\omega\right)\right\vert ^{2}
\nonumber\\
= & C_{1}^{2}\left(-\omega-\varepsilon_{g}\left(\phi\right)/2,t\right)+C_{2}^{2}\left(-\omega-\varepsilon_{g}\left(\phi\right)/2,t\right)
\nonumber\\
&+C_{1}^{2}\left(-\omega+\varepsilon_{g}\left(\phi\right)/2,t\right)+C_{2}^{2}\left(-\omega+\varepsilon_{g}\left(\phi\right)/2,t\right)
\nonumber\\
= & C_{1}^{2}\left(\omega+\varepsilon_{g}\left(\phi\right)/2,t\right)+C_{2}^{2}\left(\omega+\varepsilon_{g}\left(\phi\right)/2,t\right)
\nonumber\\&+C_{1}^{2}\left(\omega-\varepsilon_{g}\left(\phi\right)/2,t\right)+C_{2}^{2}\left(\omega-\varepsilon_{g}\left(\phi\right)/2,t\right).
\end{align} The last two lines of Eq.~(\ref{cabs-2}) are obtained using Eq.~(\ref{csym-2}). Comparing Eq.~(\ref{cabs-1}) with Eq.~(\ref{cabs-2}), we find
\begin{align}
\label{cabs-3}
 & \left\vert \tilde{c}_{+}\left(t,\varepsilon_{0}+\omega\right)\right\vert ^{2}+\left\vert \tilde{c}_{-}\left(t,\varepsilon_{0}+\omega\right)\right\vert ^{2}
\nonumber\\
&=  \left\vert \tilde{c}_{+}\left(t,\varepsilon_{0}-\omega\right)\right\vert ^{2}+\left\vert \tilde{c}_{-}\left(t,\varepsilon_{0}-\omega\right)\right\vert ^{2}.
\end{align} A similar approach is applied to
\begin{align}
&\text{Re}\left[\tilde{c}_{+}\left(t,\omega\right)\tilde{c}_{-}^{*}\left(t,\omega\right)\right]=
\nonumber\\
&C_{1}\left(\Omega_{+}\left(\omega\right),t\right)C_{1}\left(\Omega_{-}\left(\omega\right),t\right)
\nonumber\\&
+C_{2}\left(\Omega_{+}\left(\omega\right),t\right)C_{2}\left(\Omega_{-}\left(\omega\right),t\right),
\end{align} yielding
\begin{align}
 & \text{Re}\left[\tilde{c}_{+}\left(t,\varepsilon_{0}+\omega\right)\tilde{c}_{-}^{*}\left(t,\varepsilon_{0}+\omega\right)\right]
\nonumber\\
= & C_{1}\left(\omega-\varepsilon_{g}\left(\phi\right)/2,t\right)C_{1}\left(\omega+\varepsilon_{g}\left(\phi\right)/2,t\right)
\nonumber\\&
+C_{2}\left(\omega-\varepsilon_{g}\left(\phi\right)/2,t\right)C_{2}\left(\omega+\varepsilon_{g}\left(\phi\right)/2,t\right),
\end{align}
and
\begin{align}
 & \text{Re}\left[\tilde{c}_{+}\left(t,\varepsilon_{0}-\omega\right)\tilde{c}_{-}^{*}\left(t,\varepsilon_{0}-\omega\right)\right]
 \nonumber\\
= & C_{1}\left(-\omega-\varepsilon_{g}\left(\phi\right)/2,t\right)C_{1}\left(-\omega+\varepsilon_{g}\left(\phi\right)/2,t\right)
\nonumber\\&
+C_{2}\left(-\omega-\varepsilon_{g}\left(\phi\right)/2,t\right)C_{2}\left(-\omega+\varepsilon_{g}\left(\phi\right)/2,t\right)
 \nonumber\\
= & C_{1}\left(\omega+\varepsilon_{g}\left(\phi\right)/2,t\right)C_{1}\left(\omega-\varepsilon_{g}\left(\phi\right)/2,t\right)
\nonumber\\&
+\left(-1\right)^{2}C_{2}\left(\omega+\varepsilon_{g}\left(\phi\right)/2,t\right)C_{2}\left(\omega-\varepsilon_{g}\left(\phi\right)/2,t\right).
\end{align} Consequently,
\begin{align}
\label{cabs-4}
&\text{Re}\left[\tilde{c}_{+}\left(t,\varepsilon_{0}+\omega\right)\tilde{c}_{-}^{*}\left(t,\varepsilon_{0}+\omega\right)\right]
\nonumber\\&
=\text{Re}\left[\tilde{c}_{+}\left(t,\varepsilon_{0}-\omega\right)\tilde{c}_{-}^{*}\left(t,\varepsilon_{0}-\omega\right)\right].
\end{align} By the same token, for
\begin{align}
&\text{Im}\left[\tilde{c}_{+}\left(t,\omega\right)\tilde{c}_{-}^{*}\left(t,\omega\right)\right]=  \nonumber\\&
C_{2}\left(\Omega_{+}\left(\omega\right)\!,\!t\right)C_{1}\left(\Omega_{-}\left(\omega\right)\!,t\!\right)
\!\!-\!\!C_{1}\left(\Omega_{+}\left(\omega\right)\!,\!t\right)C_{2}\left(\Omega_{-}\left(\omega\right)\!,\!t\right),
\end{align} we have
\begin{align}
 & \text{Im}\left[\tilde{c}_{+}\left(t,\varepsilon_{0}+\omega\right)\tilde{c}_{-}^{*}\left(t,\varepsilon_{0}+\omega\right)\right]
 \nonumber\\
= & C_{2}\left(\omega-\varepsilon_{g}\left(\phi\right)/2,t\right)C_{1}\left(\omega+\varepsilon_{g}\left(\phi\right)/2,t\right)
 \nonumber\\&
-C_{1}\left(\omega-\varepsilon_{g}\left(\phi\right)/2,t\right)C_{2}\left(\omega+\varepsilon_{g}\left(\phi\right)/2,t\right),
\end{align}
\begin{align}
 & \text{Im}\left[\tilde{c}_{+}\left(t,\varepsilon_{0}-\omega\right)\tilde{c}_{-}^{*}\left(t,\varepsilon_{0}-\omega\right)\right]
 \nonumber\\&
 =C_{2}\left(-\omega-\varepsilon_{g}\left(\phi\right)/2,t\right)C_{1}\left(-\omega+\varepsilon_{g}\left(\phi\right)/2,t\right)
  \nonumber\\&
 -C_{1}\left(-\omega-\varepsilon_{g}\left(\phi\right)/2,t\right)C_{2}\left(-\omega+\varepsilon_{g}\left(\phi\right)/2,t\right)
 \nonumber\\&
= -C_{2}\left(\omega+\varepsilon_{g}\left(\phi\right)/2,t\right)C_{1}\left(\omega-\varepsilon_{g}\left(\phi\right)/2,t\right)
\nonumber\\&
+C_{1}\left(\omega+\varepsilon_{g}\left(\phi\right)/2,t\right)C_{2}\left(\omega-\varepsilon_{g}\left(\phi\right)/2,t\right)
\end{align} and therefore
\begin{align}
\label{cabs-5}
&\text{Im}\left[\tilde{c}_{+}\left(t,\varepsilon_{0}-\omega\right)\tilde{c}_{-}^{*}\left(t,\varepsilon_{0}-\omega\right)\right]
 \nonumber\\&=
\text{Im}\left[\tilde{c}_{+}\left(t,\varepsilon_{0}+\omega\right)\tilde{c}_{-}^{*}\left(t,\varepsilon_{0}+\omega\right)\right].
\end{align}
Using Eqs.~(\ref{cabs-3}),(\ref{cabs-4}), (\ref{cabs-5}), (\ref{bpmabsq}) and the case for $\Gamma_{g}\left(\phi\right)=i\varepsilon_{g}(\phi)$ in Eq.~(\ref{bplusminsta}), we thus conclude that $\left\vert \tilde{b}_{\pm}\left(t,\omega\right)\right\vert ^{2}$ and $\text{Re}\left(\tilde{b}_{+}\left(t,\omega\right)\frac{\tilde{b}_{-}^{*}\left(t,\omega\right)}{\Gamma_{g}^{*}\left(\phi\right)}\right)$
are also symmetric in $\omega$ with respect to $\omega=\varepsilon_{0}$. This with Eq.~(\ref{BplusSym}) completes the proof of Eq.~(\ref{sym_trsmP}) for the case that Eq.~(\ref{updwn_bdsym}) holds with $\vert\Gamma_{12}(\phi)\vert^{2}<\delta\varepsilon^{2}$, leading to $\Gamma_{g}\left(\phi\right)=i\varepsilon_{g}(\phi)$.


\section{Derivations of Eq.~(\ref{Iem_WBsim}) and Eq.~(\ref{p-h_sym_flux-indep})}
\label{appx-IinIoutrel}

Using the zero-bias assumption $f_{L}(\omega)=f_{R}(\omega)=1/[e^{(\omega-\mu_{0})/k_{B}T}+1]\equiv f(\bar{\omega})$, where $\bar{\omega}=\omega-\mu_{0}$, and the property
\begin{align}
f(\bar{\omega})+f(-\bar{\omega})=1,
\end{align}
the integral in Eq.~(\ref{dynamic-crnt-0-b}) can be rewritten to,
\begin{align}
\overline{I}^{\text{em.}}_{\alpha }(t,\phi)
&=\int_{-\infty}^{+\infty}d\omega \bar{f}(\omega)\mathcal{T}^{(+)}_{\alpha}(t,\omega,\phi)
\nonumber\\&
=\int_{-\infty}^{+\infty}d\bar{\omega} f(\bar{\omega})\mathcal{T}^{(+)}_{\alpha}(t,\bar{\omega}+\mu_{0},\phi)
\nonumber\\&
=\frac{1}{2}\int_{-\infty}^{+\infty}d\omega\mathcal{T}^{(+)}_{\alpha}(t,\omega,\phi)
\end{align}
provided that $\mathcal{T}^{(+)}_{\alpha}(t,\omega,\phi)$ is symmetrically distributed around $\omega=\mu_{0}=\varepsilon_{0}$ (see Eq.~(\ref{sym_trsmP})).  This then derives Eq.~(\ref{Iem_WBsim}) at zero bias.

A similar derivation of Eq.~(\ref{Iem_WBsim}) can be extended to the case with a finite bias under the condition $\mu_{0}=\varepsilon_{0}$. The integral expression for the zero-bias current, $\overline{I}^{\text{em.}}_{\alpha }(t,\phi)$, in Eq.~(\ref{dynamic-crnt-0-b}), can be rewritten at zero temperature to $\int_{-\infty}^{+\infty}d\omega \bar{f}(\omega)\mathcal{T}^{(+)}_{\alpha}(t,\omega,\phi)=\left[\int_{-\infty}^{\mu_{0}}+\frac{1}{2}\left(\int_{\mu_{0}}^{\mu_{0}+eV/2}-\int_{\mu_{0}-eV/2}^{\mu_{0}}\right)\right]d\omega \mathcal{T}^{(+)}_{\alpha}(t,\omega,\phi)$, where $eV=\mu_{L}-\mu_{R}$. Obviously when $\mathcal{T}^{(+)}_{\alpha}(t,\omega,\phi)$ is symmetrically distributed around $\omega=\mu_{0}$, we again have $\overline{I}^{\text{em.}}_{\alpha }(t,\phi)=\int_{-\infty}^{+\infty}d\omega\bar{f}(\omega)\mathcal{T}^{(+)}_{\alpha}(t,\omega,\phi)=\frac{1}{2}\int_{-\infty}^{+\infty}d\omega\mathcal{T}^{(+)}_{\alpha}(t,\omega,\phi)$.

 Substituting Eq.~(\ref{time-dep_Ts_WB}) to Eq.~(\ref{Iem_WBsim}) leads to
\begin{align}
\label{Iem_WBsim-1}
&\overline{I}^{\text{em.}}_{\alpha }(t,\phi)=\frac{1}{2}\int_{-\infty}^{\infty}\!\!\frac{d\omega}{2\pi}
\mathcal{T}^{(+)}_{\alpha}(t,\omega)\nonumber\\
&=\frac{1}{2}\text{Tr}\left[\boldsymbol{\Gamma}^{\alpha}-\boldsymbol{\Gamma}^{\alpha}\int_{t_0}^{t}d\tau
\boldsymbol{G}^{r}(\tau,t_0)\boldsymbol{\Gamma}\boldsymbol{G}^{a}(t_0,\tau)\right],
\end{align} where $\boldsymbol{\Gamma}=\boldsymbol{\Gamma}^{L}+\boldsymbol{\Gamma}^{R}$. Note that the flux is embedded in the off-diagonals of $\boldsymbol{\Gamma}^{\alpha}$.

On the other hand, with Eq.~(\ref{crnt-init-diagocc-WB}), the combined contributions from initially occupying each of the QDs read
\begin{align}
\label{I1p2_WBsim}
I^{[1+2]}_{\alpha}(t,\phi)=-\text{Tr}\left[\boldsymbol{\Gamma}^{\alpha}\boldsymbol{G}^{r}(t,t_0)\boldsymbol{G}^{a}(t_0,t)\right].
\end{align} The initial current at $t=t_{0}$ from $\overline{I}^{\text{em.}}_{\alpha }(t,\phi)$ is directly seen by $\overline{I}^{\text{em.}}_{\alpha }(t_0,\phi)=\frac{1}{2}\text{Tr}\left(\boldsymbol{\Gamma}^{\alpha}\right)>0$ and similarly for the initial current from $I^{[1+2]}_{\alpha}(t_0,\phi)=-\text{Tr}\left[\boldsymbol{\Gamma}^{\alpha}\right]<0$. This in turn gives $\overline{I}^{\text{em.}}_{\alpha }(t_0,\phi)=-\frac{1}{2}I^{[1+2]}_{\alpha}(t_0,\phi)$. By the use of Eq.~(\ref{retardedG-intime}), which satisfies $-i\partial\boldsymbol{G}^{r}(t,t_0)/\partial t+(\boldsymbol{\varepsilon}-i\boldsymbol{\Gamma}/2)\boldsymbol{G}^{r}(t,t_0)=0$ for $t>t_0$, one then finds from Eq.~(\ref{Iem_WBsim-1}) and Eq.~(\ref{I1p2_WBsim}) that
$\partial \overline{I}^{\text{em.}}_{\alpha }(t,\phi)/\partial t=-\frac{1}{2}\partial I^{[1+2]}_{\alpha}(t,\phi)/\partial t$. This then derives Eq.~(\ref{p-h_sym_flux-indep}).






\begin{thebibliography}{999}
\bibitem{footnote0}Note that the out-tunneling currents from the initially occupied QDs do not depend on the chemical potentials in the reservoirs (see Eq.~(\ref{crnt-init-dep}) and Eq.~(\ref{crnt-init-diagocc})).
\bibitem{footnote1}
Observing that $\bar{\Gamma}_{ii}^{\alpha}=\left.\bar{\Gamma}_{ii}^{\bar{\alpha}}\right\vert_{L\leftrightarrow R}$, one finds the evaluation of $\Gamma_{12}^{\alpha}=\bar{\Gamma}_{12}^{\alpha}e^{i\phi_{\alpha}}$ at $\alpha=L$ at one set of bonds and the evaluation of it at $\alpha=R$ at left-right exchanged bonds only differ by the phase factor $\phi_{L}\rightarrow\phi_{R}$. Therefore, $\mathcal{T}^{(\pm)}_{\alpha}(t,\omega,[\phi_{L}=\phi_{A},\phi_{R}=\phi_{B}])=\left.\mathcal{T}^{(\pm)}_{\bar{\alpha}}(t,\omega,[\phi_{L}=\phi_{B},\phi_{R}=\phi_{A}])\right\vert_{L\leftrightarrow R}$. As the transmission-like functions depend on the phases $\phi_{L}$ and $\phi_{R}$ only via $\phi=\phi_{L}-\phi_{R}$, we have $\mathcal{T}^{(\pm)}_{\alpha}(t,\omega,[\phi_{L}=\phi_{A},\phi_{R}=\phi_{B}])=\mathcal{T}^{(\pm)}_{\alpha}(t,\omega,\phi_{A}-\phi_{B})
=\left.\mathcal{T}^{(\pm)}_{\bar{\alpha}}(t,\omega,\phi_{B}-\phi_{A})\right\vert_{L\leftrightarrow R}$ and subsequently $\mathcal{T}^{(\pm)}_{\alpha}(t,\omega,[\phi_{L}=\phi_{A},\phi_{R}=\phi_{B}])=\mathcal{T}^{(\pm)}_{\alpha}(t,\omega,\phi)
=\left.\mathcal{T}^{(\pm)}_{\bar{\alpha}}(t,\omega,-\phi)\right\vert_{L\leftrightarrow R}$. This then leads to
(without requiring any symmetry among the bonds) $\mathcal{T}^{(\pm),+}_{\alpha}(t,\omega,\phi)
=+\left.\mathcal{T}^{(\pm),+}_{\bar{\alpha}}(t,\omega,\phi)\right\vert_{L\leftrightarrow R}$ and $\mathcal{T}^{(\pm),-}_{\alpha}(t,\omega,\phi)
=-\left.\mathcal{T}^{(\pm),-}_{\bar{\alpha}}(t,\omega,\phi)\right\vert_{L\leftrightarrow R}$. Combining this result with Eq.~(\ref{dynamic-crnt-0-0}) and Eq.~(\ref{dynamic-crnt-0-b}), we then obtain Eq.~(\ref{I_LR_ps_sym-1}) and Eq.~(\ref{I_LR_ps_sym-2}) respectively.
\end{thebibliography}
\end{document}